\documentclass[twocolumn,
showpacs,
showkeys,
preprintnumbers,
nofootinbib,
superscriptaddress,
amsmath,
amssymb,
floatfix,
secnumarabic,
aps,
pra,
a4paper,
notitlepage,
final,
]{revtex4-1}%

\usepackage{subfiles}

\usepackage[colorlinks=true,urlcolor=blue]{hyperref}
\usepackage{graphicx} 
\graphicspath{{images/}}
\usepackage{epsfig}
\usepackage{subfigure}
\usepackage[notcite,notref,color]{showkeys}
\usepackage{tikz}
\usetikzlibrary{decorations.pathreplacing}
\usetikzlibrary{decorations.pathmorphing}
\usepackage{multirow}
\usepackage{dcolumn}
\newcolumntype{.}{D{.}{.}{-1}}
\newcolumntype{i}[1]{D{.}{.}{#1}}

\newcommand{\beq}{\begin{equation}}
\newcommand{\eeq}{\end{equation}}
\newcommand{\beqar}{\begin{eqnarray}}
\newcommand{\eeqar}{\end{eqnarray}}
\newcommand{\bal}{\begin{aligned}}
\newcommand{\eal}{\end{aligned}}

\def\ham{\hbox{$\cal H$}}
\def\dalam{\hbox
{\vrule\vbox{\hrule\hbox to 1ex{ \hfill}\kern 1 ex\hrule}\vrule}}

\def\1/2{\hbox{$ {1 \over 2}$ }}

\def\h{\hbar}
\def\i/h{{i \over \h}}

\def\inf{\infty}
\def\pd{\partial} 
\def\v{\vec}

\def\cvE{\hbox{ $ \vec {\cal { E }}$}} \def\cvH{\hbox{$ \vec {\cal { H}}$}}

\def\a{\alpha}  
\def\b{\beta}  
 
\def\g{\gamma} \def\G{\Gamma} 
\def\d{\delta} \def\D{\Delta} 
\def\l{\lambda}  
\def\e{\epsilon} \def\E{\hbox{$\cal E $}}

\def\s{\sigma}
\def\r{\rho} \def\vr{\varrho}
\def\x{\xi}
\def\c{\chi} 
\def\vf{\varphi}

\def\p{\psi} 
\def\bp{\bar \psi}

\def\m{\mu}
\def\n{\nu}

\def\z{\zeta}

\def\w{\omega}\def\W{\Omega}
\def\tt{\theta} \def\vt{\vartheta}

\def\<{\langle}
\def\>{\rangle}

\def\({\left(}
\def\[{\left[}
\def\){\right)}
\def\]{\right]}

\newcommand{\myfrac}[2]{{\ifmmode{}^{#1}\!/_{\!#2}\else${}^{#1}\!/_{\!#2}$\fi}}

\date{\today}

\begin{document}

\title{Magnetic vacuum polarization effects in the supercritical QED: spontaneous generation of a stable highly magnetized phase of the vacuum state}

\author{K. A. Sveshnikov}
\affiliation{Department of Physics, Moscow State University, Leninsky Gory, 119991, Moscow, Russian Federation}
\author{E. S. Polshikova}
\email{polshikovaes@gmail.com}
\affiliation{Department of Physics, Moscow State University, Leninsky Gory, 119991, Moscow, Russian Federation}
\author{S. A. Artiukova}
\affiliation{Skolkovo Institute of Science and Technology, Bolshoy Boulevard, 30, bld.1, 121205, Moscow, Russia}
\author{M. A. Boitsov}
\affiliation{Department of Physics, Moscow State University, Leninsky Gory, 119991, Moscow, Russian Federation}
\author{P. A. Grashin}
\affiliation{Department of Physics, Moscow State University, Leninsky Gory, 119991, Moscow, Russian Federation}
\begin{abstract}
    
The spontaneous generation of axial vacuum current and corresponding magnetic field with multipole structure, caused by the supercritical Coulomb source with charge $Z$ and size $R$, is explored in essentially non-perturbative approach  with emphasis on the vacuum energy $\E_{VP}$, considered as a function of $R$ with fixed $Z$. The properties of such highly magnetized vacuum phase are studied in detail. It is shown that the arising this way magnetic component of the supercritical vacuum polarization leads to a significant decrease of the total energy of the system, which could provide the formation of a stable extra-heavy nuclear cluster. Numerical calculations proving the latter in case of the specific spherically symmetric Coulomb source with $Z = 828$ are presented.

\end{abstract}

\maketitle

\section{Introduction}\label{Sec_Intro}

\href{www.overleaf.com/learn/latex/Multi-file_LaTeX_projects}{}

So far, the behavior of the QED-vacuum exposed to a supercritical EM-source is the subject of an active research~\cite{Rafelski2016,Davydov2017,Sveshnikov2017,
Popov2018,*Novak2018,*Maltsev2018,Roenko2018,Maltsev2019,*Maltsev2020,Grashin2022a,Krasnov2022,*Krasnov2022a}.
Of the main interest is the assumption that in such external fields there should take place a deep vacuum state reconstruction, caused by discrete levels diving into the lower continuum and accompanied by such nontrivial effects as spontaneous positron emission combined with vacuum  shells formation (see e.g., Refs.~\cite{Greiner1985a,Plunien1986,Greiner2012,Ruffini2010,Rafelski2016} and citations therein). In the case of its reliable registration, spontaneous emission would be a signal of new physics in strong fields. In 3+1 D QED,  such effects are expected for Coulomb sources of nucleus size with charges $Z>Z_{cr,1} \simeq 170$, which are large enough for direct  observation and could probably be created in low-energy heavy ion collisions. However, in  previous investigations of those at  GSI (Darmstadt, Germany), rechecked later by Argonne Lab. (USA), no evidence of spontaneous emission was found ~\cite{Mueller1994}.

Recently, this problem has been explored in terms of the vacuum polarization energy (throughout the paper --- VP-energy) $\E_{VP}$ ~\cite{Grashin2022a,Krasnov2022,*Krasnov2022a,Grashin2023a}. $\E_{VP}$ plays an essential role in  the region of supercriticality, especially for spontaneous emission, since the latter should be provided solely by the VP-effects without any other channels of energy transfer. In particular, it is indeed the decrease of $\E_{VP}$, caused by the spectrum reconstruction, which provides the spontaneous positrons with corresponding energy for emission. Considered as a function of $Z$, with increasing $Z$ VP-energy reveals a pronounced decline into the negative range, accompanied with negative jumps, exactly equal to the electron rest mass, which occur each time the  discrete level dives into the lower continuum~\cite{Grashin2022a}.  At the same time, as a function of $R$ --- spatial parameter to regulate the cluster size by --- with fixed $Z$,  $\E_{VP}(Z,R)$ simulates the non-perturbative VP-effects in  slow heavy ion collisions in a quite reasonable way. The  point here is that in the supercritical region the behavior of  $\E_{VP}(Z,R)$ turns out to be entirely different in dependence on the value of the Coulomb  charge $Z$. In particular, for $Z<250$ it continuously grows with decreasing $R$ and thus creating serious problems for spontaneous emission in a system of approaching ions with such total $Z$. The activation of positron emission in such a system turns out to be not less than $Z \simeq 250-260$, but actually it will produce a noticeable effect  only from $Z \sim 300$~\cite{Grashin2023a}. With further growth of $Z$ the lowering of $ \E_{VP}(Z,R)$ into the negative range with decreasing $R$  becomes more and more pronounced, especially  in the range $300 < Z < 600$~\cite{Grashin2022a,*Krasnov2022a}, and so in this range the spontaneous emission  has every chance to be detected against the conversion pairs. However, positron emission implies that the corresponding positive lepton number should be left as  a density, concentrated in vacuum shells. Otherwise, either in such processes the lepton number conservation must be broken, or the positron emission must be prohibited. But there are no indications that the lepton number can exist in the form of a spatially dependent density. So the lepton number can create an absolute ban on spontaneous emission. In view of recent attempts in this field of interest
~\cite{Rafelski2016,Popov2018,*Novak2018,*Maltsev2018,
Roenko2018,Maltsev2019,*Maltsev2020,Grashin2022a,Krasnov2022,*Krasnov2022a,Grashin2023a,FAIR2009,Ter2015,MA2017169}, these circumstances require a special study.

One of the possible ways out is based on the proposal that in the supercritical region with increasing charge of the external Coulomb source other nonlinear VP-effects should enter the game, including those associated with the magnetic component of the vacuum state. In particular, it would be natural to assume that for a spherical source there should take place the  spontaneous generation of the axial vacuum current and the corresponding magnetic field, and hence, of a highly magnetized phase of the QED-vacuum\,\footnote{Here and henceforth it is implied that the z-axis of the axial magnetic configuration passes through the center of the Coulomb source.}. The main difference between the induced charge density effects and magnetic vacuum polarization of such type is that the latter does not require the spontaneous emission of positrons to appear in full splendor, it is able to develop without any connection with the lepton number conservation.

The effect of spontaneous generation of an axial vacuum current and corresponding dipole-like magnetic field under the influence of an axially symmetric supercritical Coulomb source has been recently explored in a 2+1 D graphene-like QED-system~\cite{Grashin2020a,*Grashin2020b}. It is shown that this effect comes about for $Z \geqslant Z^{\ast}$ with $Z^{\ast}>Z_{cr,1}$ being a peculiar QED-analogue of the Curie point in ferromagnetics and is above all caused by the discrete levels diving into the lower continuum. The properties of such vacuum ferromagnetic state are studied in detail. It is also shown that the thus arising magnetic component of supercritical vacuum polarization leads to a significant decrease of the total VP-energy of the system.

In this paper we explore the spontaneous generation of the axial vacuum current and the corresponding magnetic field in the 3+1 D supercritical QED system under the influence of a spherically symmetric Coulomb source. It is significantly more complicated compared to the 2+1 D case, since the development in the multipole form of the axial vacuum current occurs, whereas in the two-dimensional case the latter is restricted by the dipole-like configuration. At the same time, indeed such multipole configuration of the axial current leads to a substantial decrease of the VP-energy into the negative range, which successfully cancels the electrostatic repulsion between the components of the external Coulomb source and so could provide the formation of a stable superheavy nuclear cluster. The main difficulties in solving such problems, are caused by absence of analytic or even semi-analytic solutions of the Dirac-Maxwell (DM) equations for the crossed radial Coulomb-like forces and axial magnetic fields. The latter circumstance makes it impossible to calculate the VP-energy via separate contributions from the continua and the discrete levels, which is shown to be very powerful in purely Coulomb problems~\cite{Davydov2017,Sveshnikov2017,Grashin2022a,*Krasnov2022a}. However, the elaborated recently in Refs.~\cite{Voronina2019c, *Voronina2019d} method of vacuum energy evaluation via Wichmann-Kroll contour integration of $\ln [|\mathrm{Wronskian}|]$ solves these problems quite effectively. Moreover, while the most promising range of external Coulomb charges for spontaneous positron emission (if allowed by lepton number) is $300 < Z < 600$, the corresponding $Z$ for such magnetic VP-effects turns out to be even larger, namely, $600 < Z < 1000$\,\footnote{Such  values of $Z$  should not be misleading. Charge configurations of this type are unlikely to be realized in accelerator-based experiments. However, they may well be implemented using other experimental techniques. See below for details.}. We will consider those VP-effects, which do not conflict with the lepton number.

Such a study is performed within the DM problem  with external quasi-static spherically-symmetric Coulomb potential, created by a uniformly charged spherical shell with external radius $R_1$ and internal  $R_2$. The shell radii are chosen in such a way that imitates a system of bare uranium (U) nuclei, whose centers are located at an equal distance $a$ from each other at the vertices of a polyhedron inscribed in a sphere of radius $R$. This means
\beq
\label{1.0}
R_1=R+ R_U \ , \quad R_2=R- R_U \ ,
\eeq
where
\beq
\label{1.0a}
R_U = 1.23\, (238)^{1/3} \simeq  7.62\, \hbox{fm} \ ,
\eeq
and so the width of the shell equals to
\beq
\label{1.0b}
\D R =2 R_U \simeq 15.24\, \hbox{fm} \ .
\eeq
The total number of nuclei, the geometry of the polyhedron and the relation between the distance $a$ between its vertices and the radius $R$ will be specified later. The corresponding Coulomb potential is defined by the following expression
\begin{multline}
\label{1.1}
V(r)=- Z\a\,\[ \tt(R_2-r){3 (R_1 + R_2) \over 2 (R_1^2 + R_1 R_2 + R_2^2)} + \right. \\ \left. + \tt(R_2 \leq r \leq R_1) {3 R_1^2/2 - r^2/2 - R_2^3/r \over R_1^3 - R_2^3} + \right. \\ \left. +  \tt(r-R_1) {1 \over r} \] \ .
\end{multline}
The distance $a$ between centers of the neighboring nuclei and so the middle radius $R$ of the shell vary in the range from
\beq
\label{1.8}
a_{min}= 2 R_U + \D a \ \ ,
\eeq
where $\D a \simeq 0.6-0.7 $ fm, altering up to certain $a_{max}$ corresponding to $R_{max}$ of order of one electron Compton length, where the VP-effects are already small and show up as  $O(1/R)$-corrections. The particular choice of $a$ is to be presented in detail in the subsequent sections. The calculation of $\E_{VP}(Z,R)$ as a function of $R$ in the range $R_{min} \leq R \leq R_{max}$ for fixed $Z$ is the main goal of the present paper.

As in basic works on this topic ~\cite{Wichmann1956,Gyulassy1975, McLerran1975a,*McLerran1975b,*McLerran1975c,Greiner1985a,Plunien1986,Greiner2012,Ruffini2010, Rafelski2016}, radiative corrections from virtual photons are neglected. Henceforth, if it is not stipulated separately, relativistic units $\hbar=m_e=c=1$ and the standard representation of  Dirac matrices are used. Concrete calculations, illustrating the general picture, are performed for $\a=1/137.036$ by means of Computer Algebra Systems (such as Maple 21) to facilitate the analytic calculations  and GNU Octave code to increase efficiency in the numerical work.

This paper is generally organized as follows. In Section \ref{Sec_GenProblem} the problem statement is formulated in terms of DM equation and vacuum current density, particular prerequisites of the methodology to attain  pursued goal are clarified. The Dirac equation transformations in Section \ref{Sec_WK} leads to infinite system of differential equations for Dirac spinor components. Also the chosen magnetic potential is introduced. Onwards, in Section \ref{Sec_Var} we move towards the variational approach to VP-effects evaluation that results in exposure of searching for energy minimum techniques. In Section \ref{Sec_CoulombSource} the applications of taken Coulomb source are discussed in detail. In addition, potential's magnitudes are shown to be connected to the criteria of non-collapse of the taken nuclear system. Eventually, Section \ref{Sec_Results} is dedicated to the numerical calculation results for cluster with concrete $Z$ that matches the range of perceptibility to the magnetized field phase, also providing the reader with dependency of total system energy on its effective size. In Section \ref{Sec_Conclusion} the main conclusions and prospects of the current study are outlined.

\section{General problem statement}\label{Sec_GenProblem}

The most efficient approach to study such essentially non-perturbative VP-effects is based on the  Wichmann and Kroll (WK) framework \  ~\cite{Wichmann1956,Gyulassy1975,Mohr1998}, and was performed similarly in ~\cite{Grashin2020a}. The starting point is the VP-current density $j^{\mu}_{VP}(\vec{r})$
\begin{multline}\label{3.1}
 j^{\mu}_{VP}(\vec{r})=-\frac{|e|}{2}\(\sum\limits_{\e_{n}<\e_{F}} \bp_{n}(\vec{r})\g^{\mu}\p_{n}(\vec{r}) - \right. \\ \left. -  \sum\limits_{\e_{n}\geqslant \e_{F}} \bp_{n}(\vec{r})\g^{\mu}\p_{n}(\vec{r}) \) \ .
\end{multline}
 In (\ref{3.1}) $\e_F=-1$ is the Fermi level, which in such problems with strong external Coulomb fields is chosen at the lower threshold, while $\e_{n}$ and $\p_n(\vec{r})$ are the eigenvalues and properly normalized set of  eigenfunctions of the corresponding Dirac equation (DE). The expression (\ref{3.1}) for the  VP-charge density is a direct consequence of the general Schwinger prescription for the  fermionic current in terms of the fermionic fields commutators
\beq
\label{3.1a}
 j^{\mu}(\v r, t)=-{|e| \over 2}\, \[ \bp(\v r, t)\,, \g^{\mu} \p(\v r, t)\] \ .
 \eeq

An important point here is that in the case under study $j^{\mu}_{VP}(\v r)$ turns out to be not an average, but an eigenvalue of the fermionic charge density operator $j^{\mu} (\v r, t)$ acting on the vacuum state. The last statement follows immediately from the expansion of the fermionic fields in terms of creation-annihilation operators in the Fourier picture
 \beq
\label{3.1b}
\p(\v r, t)=\sum\limits_{\e_{n}\geqslant \e_{F}}  b_{n}\, \p_{n}(\vec{r})\,\mathrm{e}^{-i\e_{n}t}  + \sum\limits_{\e_{n} < \e_{F}}  d_{n}^{\dagger}\, \p_{n}(\vec{r})\,\mathrm{e}^{-i\e_{n}t}  \ ,
 \eeq
where
\beq
\label{3.1c}
 \{ b_n\,, b_{n'}^{\dagger} \} = \d_{n n'} \ , \quad \{ d_n\,, d_{n'}^{\dagger} \}=\d_{n n'} \ ,
 \eeq
while all the other anticommutators vanish, and the complete  set  $\{\p_{n}(\v r)\}$ is chosen in accordance with aforementioned definitions. Then the current density operator can be equivalently represented as the sum of its normal-ordered form and the VP-density introduced in (\ref{3.1})
\beq
\label{3.1d}
 j^{\mu}(\v r, t)=: j^{\mu}(\v r, t): + j^{\mu}_{VP}(\v r)  \ ,
 \eeq
and since in the Furry picture the vacuum state $|vac\>$ is subject of relations
\beq
\label{3.1e}
  b_n\, |vac\> = d_n\,|vac\> = 0 \ ,
 \eeq
one obtains
\beq
\label{3.1f}
 j^{\mu}(\v r, t)\, |vac\> =j^{\mu}_{VP}(\v r)\, |vac\>  \ .
 \eeq
Therefore $j^{\mu}_{VP}(\v r)$ turns out to be the true c-number component of the current density.

In the next step, from  the  Maxwell equations applied to the vacuum state, one finds that $F^{\m\n}_{VP}(\v r)$ and so  $A^\m_{VP}(\v r)$ turn out also to be the true c-numbered functioms of $\v r$. Here, it should be noted that the natural gauge condition for the problem under study, which includes the external quasi-static EM-source, is the Coulomb one
\beq
\label{3.11d}
\v \nabla \v A=0  \ .
 \eeq
In this gauge the scalar potential $A^0(\v r, t)$ is determined via Poisson equation without retardation
 \beq
\label{3.11e}
\D A^0(\v r, t)=-4 \pi j^0(\v r, t)  \ ,
 \eeq
and so the induced scalar VP-component $A_{VP}^0(\v r)$ is defined directly by VP-charge density
 \beq
\label{3.11f}
A_{VP}^0(\v r)=\int \! d\vec{r}\, '\, {j_{VP}^0(\vec{r}\, ') \over |\v r - \vec{r}\, '|} \ .
 \eeq
However, without spontaneous positrons the induced charge density  $j^0_{VP}(\v r)$ reduces  to a small perturbation of the Uehling type~\cite{Greiner2012,Grashin2023a}.  Therefore we can freely ignore in  DE an addition $\D V(\v r)= e A_{VP}^0 (\v r)$ to  the external  potential $V(r)$.

 Proceeding further, let us consider the matrix element of the type $\< n\hbox{-th one-fermion state}| \dots |vac\>$  of the operator-valued DE in the external scalar potential $V(r)$
\beq
\label{3.11a}
i \pd_t \p(\v r, t) = \[\v \a \(\v p - e \v A(\v r, t)\)  + V(r)  \] \p(\v r, t) \ .
 \eeq
Due to vanishing  commutator
\beq
\label{3.11b}
\[\v A(\v r, t),\p(\v r', t) \]=0  \ ,
 \eeq
one finds
\beq
\label{3.11c}
\e_n \p_n (\v r) = \[\v \a \(\v p - e \v A_{VP}(\v r)\) +  V(r) \] \p_n(\v r) \ ,
 \eeq
which is written completely in true c-numbers.

For quadratic in fermion fields theories like QED and other gauge field models the same statement holds also for VP-energy. In QED it happens because the Schwinger prescription for the current (\ref{3.1a}) dictates the following form of the Dirac Hamiltonian in the external EM-field (for details see, e.g., Refs.~\cite{Sveshnikov2017,Plunien1986})
\begin{multline}
\label{3.1g}
\ham_D (\v r, t) = \\ = {1\over 4i}\,\Big\{ \[ \p(\v r, t)^{\dagger}\,, \v\a\,  \v \nabla \p(\v r, t)\] + \[ \p(\v r, t)\,,  \v \nabla \p(\v r, t)^{\dagger}\, \v\a\] \Big\} \ + \\ +  {1 \over 2} \[ \p(\v r, t)^{\dagger}\,, \b\,  \p(\v r, t)\] + j_{\mu}(\v r, t) A^{\mu}(\v r, t) \ ,
 \end{multline}
whence by the same arguments as for the current one obtains
\beq
\label{3.1h}
\ham_D (\v r, t)\, |vac\> = h_{D,VP}(\v r)\, |vac\>  \ ,
 \eeq
where
\begin{multline}
\label{3.1i}
h_{D,VP}(\vec{r})=\frac{1}{2}\(\sum\limits_{\e_{n}<\e_{F}} \e_n\,\p_{n}(\vec{r})^{\dagger}\p_{n}(\vec{r}) \ - \right. \\ \left. - \ \sum\limits_{\e_{n}\geqslant \e_{F}} \e_n\,\p_{n}(\vec{r})^{\dagger}\p_{n}(\vec{r}) \) \ .
\end{multline}
Upon integrating eq.(\ref{3.1i}) over the whole space we derive the expression for the VP-energy from the polarized Dirac sea
\beq
\label{3.1ii}
\E_{D,VP}=\frac{1}{2}\(\sum\limits_{\e_{n}<\e_{F}} \e_n  -   \sum\limits_{\e_{n}\geqslant \e_{F}} \e_n \) \ ,
\eeq
which also appears to be the eigenvalue of the total Dirac Hamiltonian acting on the vacuum state.

As a matter of fact, $\E_{D,VP}$ provides the main contribution to the total VP-energy of the system. However, within the DM problem there appears an additional contribution to the total VP-energy, reproducing the interaction between VP-currents. It takes its origin in the structure of the total QED-Hamiltonian with external Coulomb field $A^0_{ext}$ in the Coulomb gauge (see,e.g., Refs.~\cite{Bjorken1965,Jentschura2022})
\begin{multline}
\label{3.68}
\ham_{QED}  =  {1\over 4}\,\Big\{ \[ \p^{\dagger}\,, \v\a\,  (\v p - e \v A) \p\] + \[ \p\,,   (\v p - e \v A)\p^{\dagger}\, \v\a\] \Big\}  + \\ +  {1 \over 2} \[ \p^{\dagger}\,, \b\,  \p\] + j_0 A^0_{ext} + {1\over 8 \pi}\, \[ \cvE^2+ \cvH^2 \] \ ,
\end{multline}
where
\beq\begin{gathered}
\label{3.69}
\cvE(\v r,t)=-\v \nabla A^0(\v r,t) - {\pd \over \pd t} \v A (\v r,t)= \cvE_{||}(\v r,t) + \cvE_{\perp}(\v r,t) \ , \\ \cvH(\v r,t)=\v \nabla \times \v A(\v r,t)  \ .
\end{gathered}\eeq
For the proper contribution from the EM-field to the total $H_{QED}$ one finds
\beq
\label{3.71}
H_{EM} = {1\over 8 \pi}\,\int \! d{\v r}\,  \cvE_{||}^2 + {1\over 8 \pi}\,\int \! d{\v r}\, \[  \cvE_{\perp}^2 + \cvH^2 \]\ .
\eeq
The first term in $H_{EM}$ pronto transforms into the operator-valued Coulomb interaction~\cite{Bjorken1965,Jentschura2022}. Being  combined with the relations (\ref{3.1f},\ref{3.11f}), the latter yields the Coulomb interaction term in the total VP-energy
\beq
\label{3.72a}
\E_{C,VP}=\frac{1}{2} \int \! d{\v r}\,d{\vec{r}\, '}\, {j^0_{VP}(\v r) \ j^0_{VP}({\vec{r}\, '}) \over |\v r - {\vec{r}\, '}|} \ .
\eeq

However, in the supercritical region the main contribution to VP-charge density $j^0_{VP}(\v r)$ is produced via vacuum shells formation, which change the total VP-charge $Q_{VP}$. At the same time, vacuum shells appear as empty vacancies after levels diving into the lower continuum and become negatively charged only after the resonance decay, accompanied with positron emission~\cite{Greiner1985a,Plunien1986,Greiner2012}. Otherwise they remain empty and so do not contribute to $j^0_{VP}(\v r)$. Hence, the induced charge remains a small perturbative effect not significantly affecting the whole picture~\cite{Grashin2023a}.

In the adiabatic picture we consider, the term with $\cvE_{\perp}^2$ in (\ref{3.71}) can be omitted, while the last one yields the magnetic vacuum  self-interaction. In contrast to the Coulomb term (\ref{3.72a}), the magnetic one cannot be neglected at once. So the ultimate expression for the total VP-energy  reads
\beq
\label{3.1iv}
\E_{VP}= \E_{D,VP} +   {1\over 8 \pi}\int \! d{\v r} \ \cvH_{VP}^2  \  .
\eeq
An important point here is that with respect to $\E_{D,VP}$ the vacuum vector potential $\v A_{VP}(\v r)$ serves as an external one regarding the equation (\ref{3.11c}).

The purpose of this work is to deal with the expression (\ref{3.1iv}) in order to obtain  the correct  behavior of the VP-energy in the supercritical region. The following moments should be outlined here. First, if the vacuum shells get charged via positron emission, the properly renormalized $\E_{C,VP}^{ren}$ turns out to be negative~\cite{Grashin2023a}. This effect is similar to the perturbative Uehling one, where the spatial distribution of induced by the point Coulomb source VP-density is opposite to the classical picture ~\cite{Greiner1985a,Plunien1986,Greiner2012}. And although $\E_{C,VP}^{ren}$ turns out to be much smaller than the Dirac sea contribution~\cite{Grashin2023a}, the negative sign of  $\E_{C,VP}^{ren}$  indicates that the total correctly renormalized $\E_{VP}^{ren}$ should correspond to the picture, where the interaction of the vacuum magnetic currents is also opposite to the classical one. Speaking otherwise, there should take place a kind of anti-pinch effect with attraction of opposing currents and repulsion of parallel ones. In turn, since the evolution of magnetic VP-currents  proceeds without positron emission and so doesn't conflict with lepton number, such anti-pinch VP-effect can reach significant magnitudes and lead to a substantial diving of the total VP-energy into the negative range.   It should be specially noted that this effect critically depends on the parameters of the external supercritical Coulomb source, since the axial quasi-static magnetic field itself cannot lower the levels up to the lower continuum, without which the vacuum magnetic polarization cannot take place. In any case, however, at first stage it would be instructive to consider the evaluation of the Dirac sea contribution to the total VP-energy (\ref{3.1ii}) within WK-techniques via contour integration of logarithm of the Wronskian modulus.

\section{VP-energy in WK-contour integration}\label{Sec_WK}

\begin{figure}
\center
\includegraphics[scale=0.20]{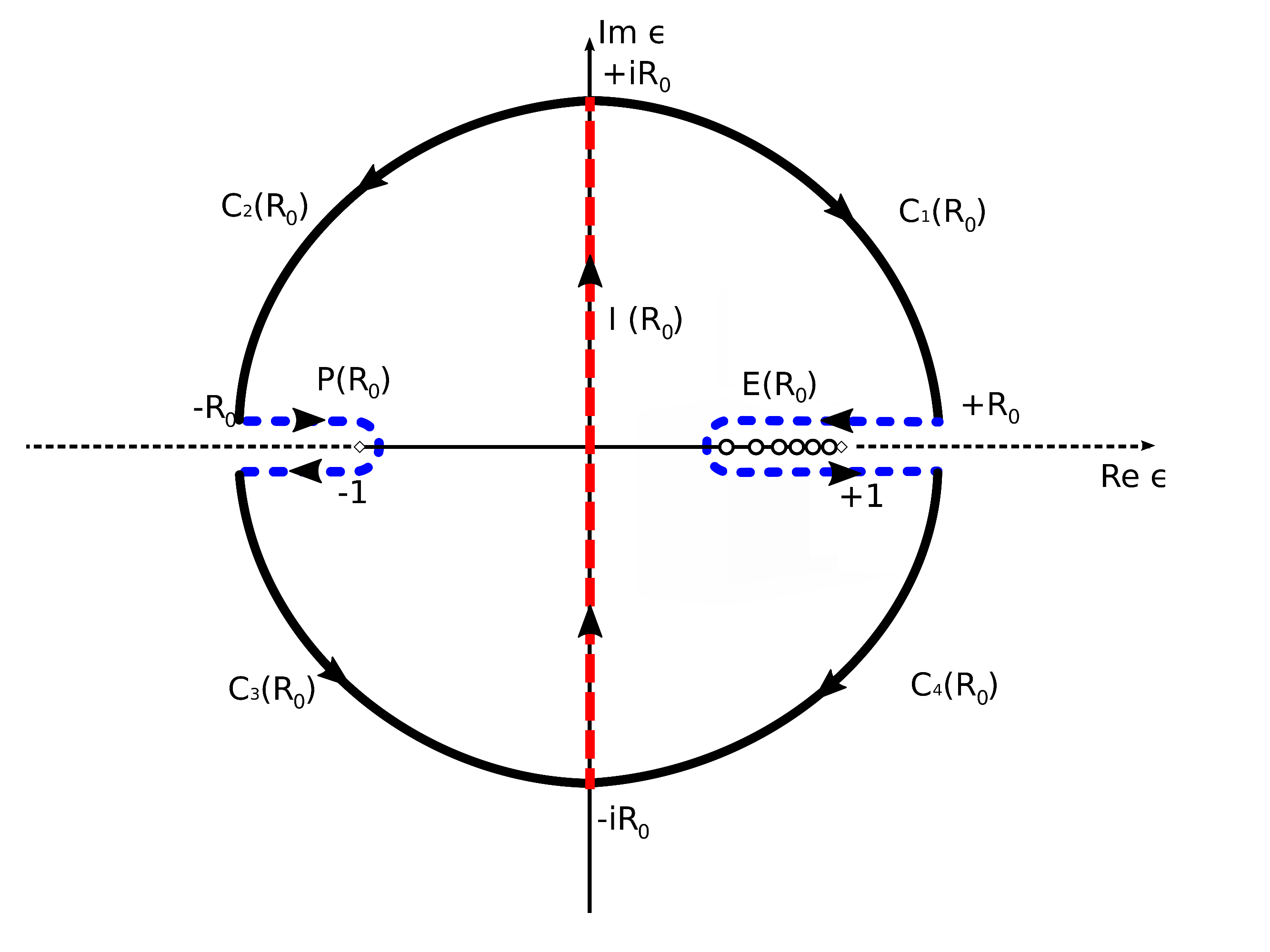} \\
\caption{\small (Color online) WK-contours in the complex energy plane, used for representation of the VP-quantities via contour integrals. The direction of contour integration is chosen in correspondence with (\ref{3.3}).}
\label{WK}
\end{figure}

The essence of the WK-techniques is the representation of the current density (\ref{3.1}) in terms of contour integrals on the first sheet of the Riemann energy plane, containing the trace of the Green function of the corresponding DM problem. In our case, the Green function is defined via equation\,\footnote{Here and henceforth we'll use for brevity the denotation $\v A (\v r)$ without VP-label for the vacuum magnetic fields. The reason is that in what follows the most part of calculations will be implemented indeed in terms of the induced vector potentials rather than of the VP-currents.}
\beq
\label{3.2}
\[\v{\a}\,\(\v p - e \v A (\v r)\) +\b + V(r) -\e \]G(\vec{r},\vec{r}\,' ;\e)  =\d(\vec{r}-\vec{r}\,' ) \ .
\eeq
The formal solution of (\ref{3.2}) reads
\beq
\label{3.3}
G(\vec{r},\vec{r}\, ';\e)=\sum\limits_{n}\frac{\p_{n}(\vec{r})\p_{n}(\vec{r}\, ')^{\dagger}}{\e_{n}-\e} \ .
\eeq
Following Ref.~\cite{Wichmann1956}, the scalar density $j^0_{VP}(\v r )$ is expressed via integrals along the contours $P(R_0)$ and $E(R_0)$ on the first sheet of the complex energy surface  (Fig.\ref{WK})
\begin{multline}
\label{3.4}
j^0_{VP}(\vec{r}) = \\ = -\frac{|e|}{2} \lim_{R_0\rightarrow \infty}\( \frac{1}{2\pi i}\int\limits_{P(R_0)} \! d\e\, \mathrm{Tr}G(\vec{r},\vec{r}\, ';\e)|_{\vec{r}\, '\rightarrow \v r} \ + \right. \\ \left.+ \ \frac{1}{2\pi i}\int\limits_{E(R_0)} \! d\e\, \mathrm{Tr}G(\vec{r},\vec{r}\, ';\e)|_{\vec{r}\, '\rightarrow \v r} \) \ .
\end{multline}
Note that the Green function in this relation must be properly regularized to insure that the limit $\vec{r}\, ' \to \v r$ exists and that the integrals over $d\e$ converge. This regularization is discussed below in current section. On this stage, though, all expressions are to be understood to involve only regulated Green functions. One of the main consequences of the last convention is the uniform asymptotics of the integrands in (\ref{3.4}) on the large circle $|\e| \to \inf$ at least as $O(1/\e)$, which allows for deforming the contours  $P(R_0)$ and $E(R_0)$ to the imaginary axis segment $I(R_0)$ and taking the limit $R_0 \to \inf$, that gives
\begin{multline}
\label{3.4a}
j^0_{VP}(\vec{r}) =|e| \[ \sum \limits_{-1 \leqslant \e_n < 0} \p_{n}(\vec{r})^{\dagger}\p_{n}(\vec{r}) + \right. \\ \left. + {1 \over 2\pi}\,  \int\limits_{-\inf}^{\inf} \! dy\, \mathrm{Tr}G(\vec{r},\vec{r}\, ';iy)|_{\vec{r}\, '\rightarrow \v r}\]  \ ,
\end{multline}
where $\{\p_n(\v r)\}$ are the normalized eigenfunctions of negative discrete levels with $-1 \leqslant \e_n <0$.

Applying the same techniques to the vector components of $j^\m_{VP}$ one finds
\begin{multline}
\label{3.4b}
\v{j}_{VP}(\vec{r}) =|e| \[ \sum \limits_{-1 \leqslant \e_n < 0} \p_{n}(\vec{r})^{\dagger}\,\v{\a}\,\p_{n}(\vec{r}) + \right. \\ \left. + {1 \over 2\pi}\,  \int\limits_{-\inf}^{\inf} \! dy\, \mathrm{Tr}\[G(\vec{r},\vec{r}\, ';iy)\,\v{\a}\]|_{\vec{r}\, '\rightarrow \v r}\]  \ .
\end{multline}

For the axial  vacuum current and corresponding magnetic field in eqs.(\ref{3.11c},\ref{3.2}) the total angular momentum $\vec j$ is not conserved, there remains only its projection $m_j$. So it would be pertinent to represent the Dirac bispinor with fixed $m_j$ in the form
\beq\label{3.5}
\p_{m_j}(\vec{r})= \begin{pmatrix}
 \vf_{m_j}(\v r)\\
-i \c_{m_j}(\v r)
\end{pmatrix} \ ,
\eeq
 where the spinors $\varphi_{m_j}, \chi_{m_j}$  are defined as the partial series over integer orbital momentum $l$
\beq\begin{gathered}
\label{3.6}
\vf_{m_j}(\v r)= \sum\limits_{l=|m_j|-1/2}^{\inf} \  \( u_{l m_j}(r)\, \W_{l m_j}^{(+)} (\v n) +  v_{l m_j}(r)\, \W_{l+1, m_j}^{(-)} (\v n)\) \ , \\
\c_{m_j}(\v r)= \sum\limits_{l=|m_j|-1/2}^{\inf} \ \( p_{l m_j}(r)\, \W_{l m_j}^{(+)} (\v n) + q_{l m_j}(r)\, \W_{l+1, m_j}^{(-)} (\v n)\) \ ,
\end{gathered}\eeq
with $\v n=\v r /r$ and   $\W_{l m_j}^{(\pm)}(\v n)$ being the spherical spinors with total momentum $j=l\pm 1/2$ and fixed $j_z=m_j$. Each term in  parentheses in series (\ref{3.6})
correspond to $j=l+1/2$, hence $l \geq |m_j|-1/2$, while from the structure of DE there follows that the radial functions $u_{l m_j}(r), v_{l m_j}(r), p_{l m_j}(r), q_{l m_j}(r)$  can be always chosen real.

The spherical spinors are defined as follows
\begin{multline}
\label{3.7}
\W_{l m_j}^{(\pm)} (\v n)=  \begin{pmatrix} \sqrt{{l \pm m_j+1/2 \over 2l+1}}\, Y_{l, m_j-1/2} \\ \pm \sqrt{{l \mp m_j+1/2 \over 2l+1}}\, Y_{l, m_j+1/2} \ \end{pmatrix} \ ,
\end{multline}
where the phase of spherical functions is chosen in a standard way, providing
\beq\begin{gathered}
\label{3.7a}
(\v \s \v n )\,\W_{l m_j}^{(+)} (\v n)=\W_{l+1, m_j}^{(-)} (\v n) \ , \\ l_{\pm}\,Y_{lm}=\sqrt{(l\mp m)(l\pm m +1)}\,Y_{l, m \pm 1} \ ,
\end{gathered}\eeq
and $Y_{l, -|m|}=(-1)^{|m|}\,Y_{l, |m|}^{\ast}$.

The  spontaneously induced axial vector potential
\beq\label{3.8c}
    \v {A} ({\v r})=A(r, \vt)\,\v {\bf e}_{\vf}
\eeq
is specified by the following multipole expansion. Here and henceforth the odd labels $n\,,n_1\,,n_2$ correspond always to the magnetic multipole components.
\beq\label{3.8}
      A(r, \vt) =2\,e \sum_{n \in \mathrm{odd}}\, {1 \over \sqrt{n(n+1)}}\,A_n(r)\, P_n^1(\cos\vt) \ ,
\eeq
where $A_n(r)$ are the radial magnetic multipole components, which magnitudes are controlled by the common multiplier $A$\, that in a more general approach can be separate for each multipole.
\beq\label{3.8b}
    A_n(r)=A\,f_n(r) \ .
\eeq
So for the fixed geometry of the Coulomb source (\ref{1.0}-\ref{1.0b}) and fixed profiles $f_n(r)$ of magnetic multipoles, both $\E_{D,VP}$ and the total VP-energy reduce  actually to  functions of two input parameters $Z$ and $A$.

Within  representations (\ref{3.6},\ref{3.8}) the initial DE (\ref{3.11c}) for any $m_j$ transforms into the (infinite) system
\begin{widetext}
\begin{multline}
\label{3.9}
 \\ {du_{l m_j}(r) \over dr}-{l \over r}\, u_{ l m_j}(r) - (\e-V(r)+1)\,q_{l m_j}(r)=
 \\- 2\,\a\sum_{n,s}^{\inf}\,\[ W_+^+(m_j,n,l+1,s)\,u_{s m_j}(r)+ W_-^+(m_j,n,l+1,s+1)v_{sm_j}(r) \]\,A_n(r) \  , \\
 {dq_{lm_j}(r) \over dr}+{l+2 \over r}\,q_{l m_j}(r) + (\e-V(r)-1)u_{l m_j}(r) = 
 \\2\,\a\sum_{n,s}^{\inf}\[ W_{+}^{-}(m_j,n,l,s)\,p_{s m_j}(r) + W_{-}^{-}(m_j,n,l,s+1)\,q_{s m_j}(r) \]\, A_n(r) \ , \\
 {dv_{l m_j}(r) \over dr}+{l+2 \over r}\,v_{l m_j}(r) - (\e-V(r)+1)\,p_{l m_j}(r) =
 \\2\,\a\sum_{n,s}^{\inf}\, \[W_{+}^{-}(m_j,n,l,s)\,u_{s m_j}(r)+W_{-}^{-}(m_j,n,l,s+1)\,v_{s m_j}(r)\]\,A_n(r) \ , \\
{dp_{l m_j}(r) \over dr}-{l \over r}\,p_{l m_j}(r) + (\e-V(r)-1)\,v_{l m_j}(r) =
\\- 2\,\a\sum_{n,s}^{\inf}\,\[W_{+}^{+}(m_j,n,l+1,s)\,p_{s m_j}(r)+W_{-}^{+}(m_j,n,l+1,s+1)\,q_{s m_j}(r)\]\,A_n(r) \ , \\
\end{multline}
where $l\,,s \geq |m_j|-1/2$, while the effective coupling constants between radial spinor and magnetic multipole components are given by
\beq\begin{gathered}
\label{3.10}
W^+_+(m,n,l,s)=\sqrt{(l+m+1/2)(s+m+1/2)}\,\w_{n,m}(l,s) + \sqrt{(l-m+1/2)(s-m+1/2)}\,\w_{n,m}(s,l) \ , \\
W^+_-(m,n,l,s)=\sqrt{(l+m+1/2)(s-m+1/2)}\,\w_{n,m}(l,s) - \sqrt{(l-m+1/2)(s+m+1/2)}\,\w_{n,m}(s,l) \ , \\
W^-_+(m,n,l,s)=\sqrt{(l-m+1/2)(s+m+1/2)}\,\w_{n,m}(l,s)-\sqrt{(l+m+1/2)(s-m+1/2)}\,\w_{n,m}(s,l) \ , \\
W^-_-(m,n,l,s)=\sqrt{(l-m+1/2)(s-m+1/2)}\,\w_{n,m}(l,s) + \sqrt{(l+m+1/2)(s+m+1/2)}\,\w_{n,m}(s,l) \ ,
\end{gathered}\eeq
and
\begin{multline}
\label{3.11}
\w_{n,m}(l,s)= \sqrt{{4\,\pi \over (2n+1)\,(2l+1)\,(2s+1)}}\,\int \! d\W\, Y_{l,m+1/2}^*\,Y_{n,1}\,Y_{s,m-1/2}= \\ = (-1)^{m+1/2}\begin{pmatrix}l & n & s\\ -(m+1/2) & 1 & m-1/2\end{pmatrix} \begin{pmatrix}l &n &s\\0&0&0\end{pmatrix} \ .
\end{multline}\end{widetext}
It can be easily verified that
\beq\begin{gathered}\label{3.12}
    W^{+}_{+}(m,n,l,s) = W^{-}_{-}(m,n,s,l) \ , \\
     W^{+(-)}_{-(+)}(m,n,l,s)= - W^{+(-)}_{-(+)}(m,n,s,l) \ .
\end{gathered}\eeq
Proceeding further, one finds that
\beq\label{3.12a}
\w_{n,-m}(l, s) =-\w_{n,m}(s, l) \ ,
\eeq
 whence there follows
\beq\begin{gathered}\label{3.12b}
    W^{+(-)}_{+(-)}(-m,n,l,s) = -W^{+(-)}_{+(-)}(m,n,l,s) \ ,
    \\
    W^{+(-)}_{-(+)}(-m,n,l,s)=  W^{+(-)}_{-(+)}(m,n,l,s) \  .
\end{gathered}\eeq

By means of these symmetry relations it is easy to verify that the system (\ref{3.9}) is invariant under following transformation
\begin{multline}\label{3.13c}
V \to -V \ , \quad  \v {A} \to - \v {A} \ , \quad m_j \to -m_j \ , \quad \e \to -\e \ , \\
u_{l m_j}(r), q_{l m_j}(r), p_{l m_j}(r), v_{l m_j}(r) \to \\ \to p_{l m_j}(r), v_{l m_j}(r), -u_{l m_j}(r), -q_{l m_j}(r) \ .
\end{multline}
In the next step from the Maxwell eqs. for the vector VP-components
\beq\label{3.13b}
   \D \v A_{VP}(\v r)=-4 \pi \v j_{VP}(\v r)
\eeq
one finds the relation between components of magnetic vector potential $A_n(r)$ and radial spinor components:

\begin{widetext}
\begin{multline}
\label{3.13a}
A_n(r)=\({1\over r}\, \int\limits_0^r \! \({r' \over r} \)^n +  \int\limits_r^\inf \! {1\over r'}\,\({r \over r'} \)^n \)\, {r'}^2\, dr' \sum_{m_j}\sum_{l,s}\,\Big[W_+^-(m_j,n,l,s)\,p_{l m_j}(r')\, u_{s m_j}(r') + \\ W_-^-(m_j,n,l,s+1)\, p_{l m_j}(r')\, v_{s m_j}(r')- W_+^+(m_j,n,l+1,s)\,q_{l m_j}(r')\, u_{s m_j}(r')- W_-^+(m_j,n,l+1,s+1)\, q_{l m_j}(r')\, v_{s m_j}(r')\Big]  \ .
\end{multline}

Proceeding further, the DE (\ref{3.9}) should be splitted by parity into two subsystems. The first one (called even) corresponds to the even parity of the spinors $\vf_{m_j} (\v r), \chi_{m_j} (\v r) $, while the second one --- to the odd. The explicit form of partial series (\ref{3.6}) should be replaced by

\beq\begin{gathered}
\label{3.14}
\vf_{m_j}(\v r)= \sum\limits_{k=N_1(m_j)}^{\inf} \   u_k(r)\, \W_{2k , m_j}^{(+)} (\v n) + \sum\limits_{k=N_2(m_j)}^{\inf} \ v_k(r)\, \W_{2k+2 ,  m_j}^{(-)} (\v n)  \ , \\
\c_{m_j}(\v r)= \sum\limits_{k=N_2(m_j)}^{\inf} \  p_k(r)\, \W_{2k+1 ,  m_j}^{(+)} (\v n) + \sum\limits_{k=N_1(m_j)}^{\inf} \ q_k(r)\, \W_{2k+1 ,  m_j}^{(-)} (\v n) \ ,
\end{gathered}\eeq
in the even case, while in the odd one
\beq\begin{gathered}
\label{3.15}
\vf_{m_j}(\v r)= \sum\limits_{k=N_2(m_j)}^{\inf} \  u_k(r)\, \W_{2k+1 , m_j}^{(+)} (\v n) +  \sum\limits_{k=N_1(m_j)}^{\inf} \  v_k(r)\, \W_{2k+1 ,  m_j}^{(-)} (\v n) \ , \\
\c_{m_j}(\v r)= \sum\limits_{k=N_1(m_j)}^{\inf} \  p_k(r)\, \W_{2k ,  m_j}^{(+)} (\v n) +  \sum\limits_{k=N_2(m_j)}^{\inf} \ q_k(r)\, \W_{2k+2 , m_j}^{(-)} (\v n) \ ,
\end{gathered}\eeq
where the counters $N_i(m)$ are defined as follows

\beq\begin{gathered}
\label{4.70}
N_1(m)=
\left\{\bal
&0, |m_j|=1/2 \ ,  \\
&\frac{2|m_j|-1}{4}, \frac{2|m_j|-1}{4} \in \mathbb{Z} \ , \\
&\frac{2|m_j|-3}{4} + 1, \frac{2|m_j|-3}{4} \in \mathbb{Z} \ ,    
\eal \right. , \quad
N_2(m)=
\left\{\bal
&\frac{2|m_j|-1}{4}, \frac{2|m_j|-1}{4} \in \mathbb{Z} \ , \\
&\frac{2|m_j|-3}{4}, \frac{2|m_j|-3}{4} \in \mathbb{Z} \   .
\eal \right. .
\end{gathered}
\eeq
These counters restrict the allowed values of the integer $k$ in expressions (\ref{3.14},\ref{3.15}) from below, since for the given $m_j$ the total angular moment $j$ cannot be less than $|m_j|$. In particular, $N_1(\pm 1/2)=N_2(\pm 1/2)=0$, since  for such $m_j$ there are allowed all the values of $j \geq 1/2$ and so of the integer counter $k$. However, $N_1(\pm 3/2)=1\,, N_2(\pm 3/2)=0$, since in the even case the lowest permitted $j=3/2$ cannot be achieved for $u_0(r)$ and $q_0(r)$, while in the odd one the same restriction holds for $v_0(r)$ and $p_0(r)$. At the same time, $N_1(\pm 5/2)=1\,, N_2(\pm 5/2)=1$, since now the lowest permitted $j=5/2$ forbids $k=0$ both in even and in odd cases.

The subsystems of each  parity take the form
\beq\begin{gathered}
\label{3.16a}
 k\geq N_1(m_j) \ , \quad {du_{k m_j}(r) \over dr}-{2 k \over r}\, u_{ k m_j}(r)=(\e-V(r)+1)\,q_{k m_j}(r) \ - \\
 - \ 2\,\a \sum_{n} \[\sum_{s=N_1(m_j)}^{\inf} W_+^+(m_j,n,2k+1,2s)\,u_{s m_j}(r) \ +  \sum_{s=N_2(m_j)}^{\inf} W_-^+(m_j,n,2k+1,2s+2)v_{s m_j}(r) \] A_n(r) \  , \\
 k\geq N_1(m_j) \ , \quad {dq_{k m_j}(r) \over dr}+{2k +2 \over r}\,q_{k m_j}(r)=-(\e-V(r)-1)u_{k m_j}(r) \ +
    \\
 + \ 2\,\a \sum_{n} \[\sum_{s=N_1(m_j)}^{\inf} W_-^-(m_j,n,2k,2s+1)\,q_{s m_j}(r) \ +  \sum_{s=N_2(m_j)}^{\inf} W_+^-(m_j,n,2k,2s+1)\,p_{s m_j}(r) \] A_n(r) \ , \\
 k\geq N_2(m_j) \ , \quad  {dv_{k m_j}(r) \over dr}+{2k+3 \over r}\,v_{k m_j}(r)=(\e-V(r)+1)\,p_{k m_j}(r) \ +
     \\ + \ 2\,\a \sum_{n} \[\sum_{s=N_2(m_j)}^{\inf} W_-^-(m_j,n,2k+1,2s+2)\,v_{s m_j}(r) \ + \ \sum_{s=N_1(m_j)}^{\inf} W_{+}^{-}(m_j,n,2k+1,2s)\,u_{s m_j}(r)\] A_n(r) \ , \\
 k\geq N_2(m_j) \ , \quad {dp_{k m_j}(r) \over dr}-{2k+1 \over r}\,p_{k m_j}(r)=-(\e-V(r)-1)\,v_{k m_j}(r) \ -
    \\ -  \ 2\,\a \sum_{n}\[\sum_{s=N_2(m_j)}^{\inf} W_+^+(m_j,n,2k+2,2s+1)\,p_{s m_j}(r) \ + \sum_{s=N_1(m_j)}^{\inf} W_-^+(m_j,n,2k+2,2s+1)\,q_{s m_j}(r)\] A_n(r) \ ,
\end{gathered}\eeq
in the even case, while in the odd one
\beq\begin{gathered}
\label{3.16b}
k\geq N_2(m_j) \ , \quad {du_{k m_j}(r) \over dr}-{2 k+1 \over r}\, u_{ k m_j}(r)=(\e-V(r)+1)\,q_{k m_j}(r) \ - \\
 - \  2\,\a \sum_{n} \[\sum_{s=N_2(m_j)}^{\inf}\, W_+^+(m_j,n,2k+2,2s+1)\,u_{s m_j}(r) \ + \ \sum_{s=N_1(m_j)}^{\inf}\, W_-^+(m_j,n,2k+2,2s+1)v_{s m_j}(r) \] A_n(r) \  , \\
k\geq N_2(m_j) \ , \quad {dq_{k m_j}(r) \over dr}+{2k +3 \over r}\,q_{k m_j}(r)=-(\e-V(r)-1)u_{k m_j}(r) \ +
    \\
 + \ 2\,\a \sum_{n}\[\sum_{s=N_2(m_j)}^{\inf} W_-^-(m_j,n,2k+1,2s+2)\,q_{s m_j}(r) \ + \ \sum_{s=N_1(m_j)}^{\inf} W_+^-(m_j,n,2k+1,2s)\,p_{s m_j}(r)  \] A_n(r) \ , \\
k\geq N_1(m_j) \ , \quad  {dv_{k m_j}(r) \over dr}+{2k+2 \over r}\,v_{k m_j}(r)=(\e-V(r)+1)\,p_{k m_j}(r) \ +
     \\ + \ 2\,\a \sum_{n}\[\sum_{s=N_1(m_j)}^{\inf}\,W_-^-(m_j,n,2k,2s+1)\,v_{s m_j}(r) \ + \ \sum_{s=N_2(m_j)}^{\inf}\, W_{+}^{-}(m_j,n,2k,2s+1)\,u_{s m_j}(r) \] A_n(r) \ , \\
k\geq N_1(m_j) \ , \quad {dp_{k m_j}(r) \over dr}-{2k \over r}\,p_{k m_j}(r)=-(\e-V(r)-1)\,v_{k m_j}(r) \ -
    \\ - \ 2\,\a \sum_{n} \[\sum_{s=N_1(m_j)}^{\inf}\,W_+^+(m_j,n,2k+1,2s)\,p_{s m_j}(r) \ + \ \sum_{s=N_2(m_j)}^{\inf}\,W_-^+(m_j,n,2k+1,2s+2)\,q_{s m_j}(r)\] A_n(r) \ .
\end{gathered}\eeq
\end{widetext}

In terms of DE (\ref{3.9}-\ref{3.16a},\ref{3.16b}) the logarithm of the Wronskian modulus approach to the Dirac sea term in the total VP-energy proceeds as follows. In the first step, $\E_{D,VP}$ is represented as a partial series over $m_j$ and parity $(\pm)$
\beq
\label{ED}
\E_{D,VP}=\sum\limits_\pm \sum_{m_j=\pm 1/2,\, \pm 3/2,...}\E_{D,VP,m_j}^\pm \ .
\eeq
Next, it is easy to see that the functions
\beq
F_{m_j}^\pm (\e)=\e\,{ dJ_{m_j}^{\pm}(\e)/d \e \over J_{m_j}^\pm(\e)} \ ,
\label{F}
\eeq
with $J_{m_j}^\pm(\e)$ being the Wronskians of  spectral problems (\ref{3.16a},\ref{3.16b}), reveal all the pole properties including the  degeneracy factor $g_{\n,m_j}^\pm$ of eigenvalues $\e_{\n,m_j}^\pm$, required for WK-techniques. It should be specially remarked here that in the purely Coulomb case $g_{\n,m_j}^\pm$ are trivial and so can be included as common factors for each partial term in $\E_{D,VP}$. However, in the case of crossed radial Coulomb and axial magnetic fields the levels with the same $m_j$ and parity can intersect and so $g_{\n,m_j}^\pm$ depend in a highly nontrivial way on concrete configuration of fields. The particular example illustrating this effect is submitted in the Appendix.

Therefore, quite similar to VP-currents, each partial $\E_{D,VP,m_j}^\pm $  can be represented via WK-contour integral with $F_{m_j}^\pm(\e)$ as the integrand. Being transformed  to imaginary axis, the latter yields the following expression
\begin{multline}
\label{3.13}
 \E_{D,VP,m_j}^\pm= {1 \over 2\pi i}\,  \int\limits_{-i\inf}^{i\inf} \! d\e\, \[\( F_{m_j}^\pm(\e)\)_{Z,A} - \(F_{m_j}^\pm(\e)\)_0\] - \\ -  \sum \limits_{-1 \leqslant \e_{\n,m_j}^\pm < 0} g_{\n,m_j}^\pm\,\e_{\n,m_j}^\pm\   \ ,
\end{multline}
where the label $Z,A$ denotes the  non-vanishing external Coulomb and the induced magnetic fields, while the label $0$ corresponds to the free case.  Defined in such a way, the total VP-energy (\ref{3.1iv}) vanishes by turning off the external source, while by turning on it contains only the interaction effects. Moreover,  the subtraction of the free field inside the integration over $d \e$  reduces the asymptotics of the integrand on the large circle to $O(1/\e)$ and so provides the deformation of the contours $P(R_0)$ and $E(R_0)$ into the imaginary axis segment $I(R_0)$ and taking the limit $R_0 \to \inf$. However, in the integration over imaginary axis the logarithmic divergence remains and must be removed via corresponding renormalization procedure.

In the purely Coulomb case the general result, obtained in Ref.~\cite{Gyulassy1975} within the expansion of $j^0_{VP}(r)$ in powers of  $Z\a$  with fixed $R$, is that all the divergences of $j^0_{VP}(r)$ originate from the fermionic loop with two external photon lines and free electron propagator, whereas  the next-to-leading orders of the expansion in $Z\a$ are already free from divergences (see also Ref.~\cite{Mohr1998} and refs. therein).

In the present case, likewise one- and two-dimensional models of super-criticality in QED-systems~\cite{Davydov2017,*Sveshnikov2017,Davydov2018a,*Davydov2018b,Sveshnikov2019a,*Sveshnikov2019b,Voronina2019a,*Voronina2019b,Grashin2020a, *Grashin2020b} and three-dimensional spherically-symmetric Coulomb case~\cite{Grashin2022a}, the renormalization of the Dirac sea term in VP-energy proceeds also via fermionic loop with the free electron propagator and two external photon lines. An essential point here is that,  likewise the Coulomb potential $V(r)$, the magnetic multipole components enter the DE (\ref{3.16a},\ref{3.16b}) as the external ones. Moreover, the transversal (gauge-invariant) part of the fermionic loop with two legs diverges indeed logarithmically~\cite{Bjorken1965,landau2012qed,Jentschura2022},  while the fourth-order terms in powers of external EM-fields are already finite. The last circumstance follows from the well-known properties of the fermionic loop with four legs, which can be described in terms of the corresponding  tensor $M_{\l \m \n \r}(k_1\,,k_2\,,k_3\,, k_4)$ (see, e.g., Ref.~\cite{landau2012qed,Jentschura2022}). In terms of the latter the gauge invariance simply reads
\beq
\label{3.63}
k_1^\l\,M_{\l \m \n \r}= k_2^\m\,M_{\l \m \n \r}= ... =0 \ ,
\eeq
and reduces to the condition $M_{\l \m \n \r}(0\,,0\,,0\,, 0)=0$, while the logarithmically divergent part of the tensor can be represented by the following integral
\beq
\label{3.64}
 \int \mathrm{Tr}\[\g_\l\,(\g q)\, \g_\m\,(\g q)\, \g_\r\,(\g q)\, \g_\n\,(\g q) \]\, { d^4 q \over \( q^2 \)^2 } \ ,
\eeq
which upon averaging over all directions of the 4-vector $q$ reduces to
\beq
\label{3.65}
\( g_{\l\m}\,g_{\n\r}+g_{\l\n}\,g_{\m\r} - 2\,g_{\l\r}\,g_{\m\n}\)\, \int  { d^4 q \over \( q^2 \)^2 } \ .
\eeq

It is well-known~\cite{landau2012qed,Jentschura2022}, that the combination of indices in (\ref{3.65}) vanishes upon symmetrization. At the same time, such symmetrization is equivalent to summing up all the diagrams of such type, which is indeed the case of VP-energy calculation. Furthermore, the  calculation of VP-energy via $\ln [|\mathrm{Wronskian}|]$ does not require for a special account of gauge condition (\ref{3.63}), since it deals at any stage only with gauge-invariant quantities  without any application to Green functions. So the renormalization via fermionic loop turns out to be a universal tool, that removes the divergence both in the purely perturbative and in the essentially non-perturbative vacuum polarization regimes by the external EM-field.

Thus, in the complete analogy with the renormalization of VP-energy considered  in~\cite{Davydov2017,*Sveshnikov2017,Davydov2018a,*Davydov2018b,Sveshnikov2019a,*Sveshnikov2019b,Voronina2019a,*Voronina2019b,Grashin2020a, *Grashin2020b,Grashin2022a}, the general approach to renormalization of the Dirac sea is reduced to separate replacement of quadratic Coulomb and magnetic field, henceforth Born, components in $\E_{D,VP}$ by corresponding perturbative counterterms under condition that the geometry of the external Coulomb source  and the profiles of magnetic multipoles $f_n(r)$ remain unchanged. That means
\beq
\label{4.59}
\E^{ren}_{D,VP}(Z,A)=\E_{D,VP}+ \z_{C}Z^2 +\z_{A}A^2\ ,
\eeq
where the renormalization coefficients $\z_{C}$ and $\z_{A}$ are defined in the following way
\beq
\label{4.60}
\z_{C} = \lim\limits_{Z_0 \to 0}  \[{\E_{D,VP}^{PT}(Z_0,0)-\E_{D,VP}(Z_0,0) \over Z_0^2}\]_{(R, f_n(r))} \ , \eeq
\beq
\label{4.61}
\z_{A} = \lim\limits_{A_0 \to 0}  \[{\E_{D,VP}^{PT}(0,A_0)-\E_{D,VP}(0,A_0) \over A_0^2}\]_{(R,f_n(r))} \ .
\eeq

Let us mention that the renormalization coefficients are calculated separately for Coulomb and magnetic components according to divergences in fermionic loop containing the free electron propagator. The need in such renormalization via fermionic loop follows also from the analysis of the properties of $j^0_{VP}$, which shows that without such UV-renormalization the integral VP-charge cannot acquire the expected integer value in units of $(-2|e|)$~\cite{Grashin2022a,Krasnov2022,*Krasnov2022a,Grashin2023a}. In fact, the properties of $j^0_{VP}$ play here the role of a controller, which provides the implementation of the required physical conditions for a correct description of VP-effects beyond the scope of PT, since the latters cannot be tracked via direct evaluation of $\E_{D,VP}$ by means of the initial relations (\ref{3.1ii},\ref{3.13}). Moreover, being combined with minimal subtraction prescription, it eliminates the inevitable arbitrariness in isolating the divergences in the initial expressions for VP-currents and VP-energy.

That procedure removes the remaining logarithmic divergences in $\E_{D,VP}$ and so allows for the following representation of Dirac sea contribution
\begin{multline}
\label{4.62}
\E^{ren}_{D,VP}(Z,A)=  \E_{D,VP}(Z,A)-\E_{D,VP}(Z,0)+ \z_{A}A^2 \\ + \E_{D,VP}(Z,0)-\E_{D,VP}(0,0)+ \z_{C}Z^2 = \\ = \E_1(Z,A)+\E_2(Z) \ .
\end{multline}

In the next step, the $\ln [|\mathrm{Wronskian}|]$ techniques is applied to the first term in (\ref{4.62})
\beq\label{4.63}
\E_1(Z,A)=  \E_{D,VP}(Z,A)-\E_{D,VP}(Z,0)+ \z_{A}A^2
\eeq
and yields
\begin{widetext}\begin{multline}
\label{4.64}
\E_1(Z,A)= \sum\limits_{m_j}\[ \int\limits_{-\inf}^{\inf} \! dy\,\(-{1 \over 2\pi }\,\sum\limits_{\pm} \(\ln\[J_{m_j}^\pm(iy)\]_{Z,A} - \ln\[J_{m_j}^\pm(iy)\]_{Z,0}\) -  B(m_j,y)\, A^2\) \ - \right. \\ \left.   - \  \(\sum \limits_{-1 \leqslant \e_{\n,m_j}^\pm < 0} g_{\n,m_j}^\pm\,\e_{\n,m_j}^\pm\)_{Z,A} \ + \ \(\sum \limits_{-1 \leqslant \e_{\n,m_j}^\pm < 0} \e_{\n,m_j}^\pm\)_{Z,0} \] +\E_{D,VP}^{PT}(0,A) \ .
\end{multline}\end{widetext}

The integration by parts in the integral over $dy$ proceeds now without any extra terms, since the logarithmically divergent $O(1/|y|)$ asymptotics for large $|y|$ of $\ln \[J_{m_j}^\pm(iy)\]_{Z,A}$ contains separate quadratic contributions from $Z$ and  $A$, which are canceled by corresponding terms in the asymptotics of $\ln\[J_{m_j}^\pm(iy)\]_{Z,0}$ and of the magnetic Born term $B(m_j,y) \, A^2$. Note also that in the purely Coulomb case the degeneracy factors of negative discrete levels are trivial $g_{\n,m_j}^\pm=1$.

Proceeding further, from the global structure of DE (\ref{3.9}-\ref{3.16a},\ref{3.16b}) there follows
\beq
\label{4.65}
 J_{m_j}^\pm(iy)^\ast=J_{m_j}^\pm(-iy) \ ,
\eeq
whence one obtains the final form of the integral over $dy$ in the expression (\ref{3.64}), which is the essence of the $\ln [|\mathrm{Wronskian}|]$ techniques
\begin{widetext}
\begin{multline}
\label{4.66}
   \int\limits_{-\inf}^{\inf} \! dy\, \(\ln\[J_{m_j}^\pm(iy)\]_{Z,A} - \ln\[J_{m_j}^\pm(iy)\]_{Z,0}\) \to   \\ \to   \int\limits_{-\inf}^{\inf} \! dy\, \(\ln\[|J_{m_j}^\pm(iy)|\]_{Z,A} - \ln\[|J_{m_j}^\pm(iy)|\]_{Z,0}\) =  2\, \int\limits_{0}^{\inf} \! dy\, \(\ln\[|J_{m_j}^\pm(iy)|\]_{Z,A} - \ln\[|J_{m_j}^\pm(iy)|\]_{Z,0}\)   \ .
\end{multline}
For the magnetic Born term $B(m_j,y)$, which is defined via
 \beq\begin{gathered}
\label{4.67}
  \lim\limits_{A_0 \to 0} \[{\E_{D,VP}(0,A_0) \over A_0^2}\]_{(R,f_n(r))} = \sum\limits_{m_j} \int\limits_{-\inf}^{\inf} \! dy\, B(m_j,y) \ ,
\end{gathered}\eeq
  by means of a straightforward, but some cumbersome calculations one obtains
\begin{multline}\label{4.68}
B(m,y)=\(-{8 \a^2 \over
\pi}\)\g(y)^2\sum\limits_{n_1,n_2} \\ \Bigg(  \sum\limits_{k=N_1(m)}\sum\limits_{s=N_1(m)}\,\Big[W^+_+(m,n_1,2k+1,2s)\,
Y_1(n_1,n_2,k,s,y)\,W^+_+(m,n_2,2s+1,2k) \ + \\ + \ W^+_+(m,n_1,2k+1,2s)\,
Y_2(n_1,n_2,k,s,y)\,W^-_-(m,n_2,2s,2k+1)\Big] \ + \\ + \ \sum\limits_{k=N_2(m)}\sum\limits_{s=N_2(m)}\,\Big[W^+_+(m,n_1,2k+2,2s+1)\,
Y_1(n_1,n_2,k+1/2,s+1/2,y)\,W^+_+(m,n_2,2s+2,2k+1) \ + \\ + \ W^+_+(m,n_1,2k+2,2s+1)\,
Y_2(n_1,n_2,k+1/2,s+1/2,y)\,W^-_-(m,n_2,2s+1,2k+2)\Big] \ - \\ - \ \sum\limits_{k=N_2(m)}\sum\limits_{s=N_1(m)}\,\Big[W^-_+(m,n_1,2k+1,2s)\,
Y_3(n_1,n_2,k,s,y)\,W^+_-(m,n_2,2s+1,2k+2) \ + \\ + \ W^-_+(m,n_1,2k+1,2s)\,
Y_2(n_1,n_2,k,s,y)\,W^-_+(m,n_2,2s,2k+1) \ + \\ + \ W^+_-(m,n_1,2k+2,2s+1)\,
Y_2(n_1,n_2,k+1/2,s+1/2,y)\,W^+_-(m,n_2,2s+1,2k+2) \ + \\ + \ W^+_-(m,n_1,2k+2,2s+1)\,
Y_4(n_1,n_2,k,s,y)\,W^-_+(m,n_2,2s,2k+1)\Big]
\Bigg) \ ,
 \end{multline}
where $\g(y)=\sqrt{1+y^2}$,
 \begin{multline}\label{4.69}
Y_1(n_1,n_2,k,s,y)=\int\limits_0^\inf \! r\,dr\,f_{n_1}(r)\,I_{2k+3/2}\(\g(y)\,r\)\,I_{2s+1/2}\(\g(y)\,r\)\,\int\limits_r^\inf \! r'\,dr'\, f_{n_2}(r')\, K_{2k+1/2}\(\g(y)\,r'\)\,K_{2s+3/2}\(\g(y)\,r'\)\ , \\
Y_2(n_1,n_2,k,s,y)=\int\limits_0^\inf \! r\,dr\,f_{n_1}(r)\,I_{2k+3/2}\(\g(y)\,r\)\,I_{2s+1/2}\(\g(y)\,r\)\,\int\limits_r^\inf \! r'\,dr'\, f_{n_2}(r')\, K_{2k+3/2}\(\g(y)\,r'\)\,K_{2s+1/2}\(\g(y)\,r'\)\ , \\
Y_3(n_1,n_2,k,s,y)=\int\limits_0^\inf \! r\,dr\,f_{n_1}(r)\,I_{2k+3/2}\(\g(y)\,r\)\,I_{2s+1/2}\(\g(y)\,r\)\,\int\limits_r^\inf \! r'\,dr'\, f_{n_2}(r')\, K_{2k+5/2}\(\g(y)\,r'\)\,K_{2s+3/2}\(\g(y)\,r'\)\ , \\
Y_4(n_1,n_2,k,s,y)=\int\limits_0^\inf \! r\,dr\,f_{n_1}(r)\,I_{2k+5/2}\(\g(y)\,r\)\,I_{2s+3/2}\(\g(y)\,r\)\,\int\limits_r^\inf \! r'\,dr'\, f_{n_2}(r')\, K_{2k+3/2}\(\g(y)\,r'\)\,K_{2s+1/2}\(\g(y)\,r'\) \ ,
\end{multline}
with $I_\n(z)\,, K_\n(z)$ being the Infeld and McDonald functions. It is easy to verify that
\beq
\label{4.69a}
 B(m,y)=B(-m,y) \ , \quad B(m,y)=B(m,-y) \ .
\eeq
Furthermore, for $|y| \to \infty$ all the integrals (\ref{4.69}) expose the same asymptotics
\beq
\label{4.69b}
Y_i(n_1,n_2,k,s,y) \to {1 \over (2\g(y))^3}\, \int\limits_0^\inf  \!dr\, f_{n_1}(r)\,f_{n_2}(r)  \ ,
\eeq
which provides the required $O(1/|y|)$ asymptotics of the Born term in the expression (\ref{4.64}). The performed calculations reveal excellent coincidence between the coefficients in the $O(1/|y|)$ asymptotics  of the difference of  Wronskians in the expression (\ref{4.66}) and that of the magnetic Born term. The rest converges as $O(1/|y|^3)$, just as the phase integral in the pure Coulomb case~\cite{Grashin2022a,Krasnov2022,*Krasnov2022a}.

The perturbative counterterm for magnetic component takes the form
\beq
\label{4.71}
\E_{D,VP}^{PT}(0,A) =\(-{\a \over \pi^2}\)\, \sum\limits_{n} {4 \pi \over 2 n+1}\, \int\limits_0^{\inf} \! d q \  q^4\, \Pi_R(-q^2)\,   \( \int\limits_0^{\inf} \! r^2\, d r\, j_n(q r)\, A_n (r) \)^2  \  ,
\eeq
where $j_n(z)$ are the spherical Bessel functions, while the polarization function  $\Pi_R(q^2)$ is defined via general relation $\Pi_R^{\m\n}(q)=\(q^\m q^\n - g^{\m\n}q^2\)\Pi_R(q^2)$ and so is dimensionless. In the adiabatic case under consideration  $q^0=0$ and $\Pi_R(-{\v q}^2)$ takes the  form
\beq
\label{4.72}
\Pi_R(-{\v q}^2) =  {2 \a \over \pi}\, \int \limits_0^1 \! d\b\,\b(1-\b)\,\ln\[1+\b(1-\b)\,{{\v q}^2 \over m^2-i\e}\] = {\a \over \pi}\, S\(|\v q|/m \) \ ,
\eeq
where
\beq\label{4.73}
S(x)= -5/9 + 4/3 x^2 + (x^2- 2)\, \sqrt{x^2+4}\, \ln \[ \(\sqrt{x^2+4}+x\) \Big/ \(\sqrt{x^2+4}-x\) \]/3x^3  \ ,
\eeq
\end{widetext}

The remaining part of $\E^{ren}_{D,VP}(Z,A)$
\beq\label{4.74}
\E_2(Z)=\E_{D,VP}(Z,0)-\E_{D,VP}(0,0)+ \z_{C}Z^2
\eeq
is nothing else but the correctly renormalized VP-energy $\E_{C,VP}^{ren}(Z)$ of the Dirac sea for the pure Coulomb case with the external source of the form (\ref{1.1}). It can be found either via the same $\ln [|\mathrm{Wronskian}|]$ techniques, presented above, or via partial phase analysis of Refs.~\cite{Grashin2022a,Krasnov2022,*Krasnov2022a}, which can be applied to any spherical source via slicing method described in~\cite{Grashin2022a}. However, the Coulomb Born term turns out to be not less cumbersome than the magnetic one, and so here we restrict the calculation to the slicing method, since it is already well elaborated and the source configuration is smooth enough to be effective.

So the ultimate expression for the renormalized $\E^{ren}_{D,VP}(Z,A)$ reads
\begin{widetext}\beq \label{4.75}
\E^{ren}_{D,VP}(Z,A)= 2\, \int\limits_{0}^\inf \! dy\, h(Z,A,y) + \D \E_{neglev}(Z,A)  + \E_{D,VP}^{PT}(0,A)  + E_{C,VP}^{ren}(Z) \ ,
\eeq
where
\beq
\label{4.76}
h(Z,A,y)= -{1 \over 2\pi }\,\sum\limits_{\pm}\[\sum\limits_{m_j}\(\ln\[|J_{m_j}^\pm(iy)|\]_{Z,A} - \ln\[|J_{m_j}^\pm(iy)|\]_{Z,0}\)\] \ - \ 2\,A^2 \sum\limits_{m_j \geqslant 1/2}B(m_j,y) \ ,
\eeq
and
\beq \label{4.77}
 \D \E_{neglev}(Z,A)=-  \(\sum \limits_{-1 \leqslant \e_{\n,m_j}^\pm < 0} g_{\n,m_j}^\pm\,\e_{\n,m_j}^\pm\)_{Z,A}  +  \(\sum \limits_{-1 \leqslant \e_{\n,m_j}^\pm < 0} \e_{\n,m_j}^\pm\)_{Z,0} \ .
 \eeq

The renormalization of the  remaining magnetic self-interaction term in the total VP-energy (\ref{3.1iv}) turns out to be very simple, since it is quadratic in $A$ and so should be replaced by the corresponding perturbative expression. That means
 \beq\label{4.78}
 {1\over 8 \pi}\int \! d{\v r} \ \cvH_{VP}^2 \to \E_{M,VP}^{ren}(A) \ ,
 \eeq
 where
\beq
\label{4.79}
\E_{M,VP}^{ren}(A) ={\a^3 \over \pi^4}\, \sum\limits_{n} {4 \pi \over 2 n+1}\, \int\limits_0^{\inf} \! d q \  q^4\, S^2(q)\,   \( \int\limits_0^{\inf} \! r^2\, d r\, j_n(q r)\, A_n (r) \)^2  \  ,
\eeq
and so the ultimate expression for the total renormalized VP-energy reads
\beq \label{4.80}
\E^{ren}_{VP}(Z,A)= \E_{D,VP}^{ren}(Z,A)  + \E_{M,VP}^{ren}(A) \ .
\eeq
 \end{widetext}

 \section{Variational approach and concomitant circumstances}\label{Sec_Var}

The most efficient approach to essentially  non-perturbative evaluation of such magnetic VP-effects is the variational one. One of the main arguments in favor for such a choice is that the vacuum currents and magnetic fields arise from the very beginning as the true c-numbered functions. Therefore, they turn out to be the natural functional arguments when performing the variational estimates for the effect of spontaneous generation of magnetic component of the vacuum state.

Let us note that the self-consistency equation (\ref{3.13b}) can be easily obtained from the initial expression for the VP-energy (\ref{3.1iv}) via variational approach by means of the Schwinger relation
\beq \label{5.1}
\d\E_{D,VP}= \int \! d \v r\, j^\m_{VP}(\v r)\, \d A_\m (\v r) \ .
\eeq
However, for our purposes to access the reliable configuration of spontaneously generated vacuum magnetic field the direct way is not the best one, since the latter must  minimize the total renormalized VP-energy (\ref{4.80}). Due to complexity of the  functional (\ref{4.80}) the direct minimization of the latter poses serious problems. Moreover, there appear some additional aspects of the problem, which should be taken into account right away.

The first one is that in systems with Dirac fermions the variational approach, unlike the Schroedinger case, does not necessarily lead to the  energy minimum. A pertinent  example of such kind is given by the hydrogen-like atom. In the non-relativistic case the $1s$-ground state provides the absolute minimum of the Shroedinger energy functional. However, in the relativistic case, described by the corresponding DE, the situation is different. The correct $1s_{1/2}$-ground state eigenfunction with spin projection $+1/2$ takes the form
\begin{multline}
\label{5.2}
\p_{1s_{1/2}} (\v r)= N\,(2\,Z \a r)^{\g-1}\,\mathrm{e}^{-Z\a r}\, \begin{pmatrix} 1 \\ 0 \\ i\,{(1-\g) \over Z\a}\,\cos \vt \\ i\,{(1-\g) \over Z\a}\,\mathrm{e}^{i\vf}\,\sin \vt \end{pmatrix} \ ,
\end{multline}
where $\g=\sqrt{1-(Z\,\a)^2}$, while the normalization factor $N$ is given by
\beq \label{5.3}
N^2= {(2\,Z\a)^3 \over 4 \pi}\,{1+\g \over 2\, \G(1+2\g) } \ .
\eeq
The corresponding eigenvalue of the Dirac-Coulomb problem equals to
\beq \label{5.4}
\e_{1s}=\g=\sqrt{1-(Z\,\a)^2} .
\eeq
Now let us consider the modified $1s_{1/2}$-wavefunction $\p_{1s_{1/2}}^\s$ of the form
\begin{multline}
\label{5.5}
\p_{1s_{1/2}}^\s (\v r)= \\ = N(\s)\,(2\,Z \a r)^{\g-1}\,\mathrm{e}^{-Z\a \s r}\, \begin{pmatrix} 1 \\ 0 \\ i\,{(1-\g) \over Z\a}\,\cos \vt \\ i\,{(1-\g) \over Z\a}\,\mathrm{e}^{i\vf}\,\sin \vt \end{pmatrix} \ ,
\end{multline}
where the parameter $\s>0$ in the exponent looks like, but is not identical to the effective screening parameter of the Coulomb source, while the normalization factor is replaced by
\beq \label{5.6}
N^2(\s)= {(2\,Z\a)^3 \over 4 \pi}\,{1+\g \over 2\, \G(1+2\g) }\,\s^{2\g+1} \ .
\eeq
The most important feature of the modified wavefunction $\p_{1s_{1/2}}^\s (\v r)$ is that it yields the same result for the average energy as the true one (\ref{5.2})
\beq \label{5.7}
\e_{1s_{1/2}}(\s)=\<\p_{1s_{1/2}}^\s | H_D|\p_{1s_{1/2}}^\s \>=\g=\sqrt{1-(Z\,\a)^2} \ ,
\eeq
regardless of the specific value of the parameter $\s>0$. Therefore the true $1s_{1/2}$-ground state (\ref{5.2}) turns out to be neither minimum nor maximum of the Dirac-Coulomb energy functional, rather it corresponds to the inflection point of the latter. This result should not be misleading, since in the Dirac-Coulomb problem  the spectrum is not bounded from below. In the application to the problem under study, this circumstance manifests itself in the fact that the VP-energy of the Dirac sea turns out to be unstable with respect to contraction of the Coulomb source and collapsing of the magnetic field configuration with growing magnitude. In particular, for fixed $Z>250$ and decreasing $R$ the Coulomb polarization VP-energy decays into the negative range $\sim - C_1/R$ ~\cite{Krasnov2022b,Grashin2023a}, whereas the calculations performed below show that the magnetic polarization energy decreases one order faster $\sim - C_2/R^2$, while the size of the Coulomb source is limited from below by the minimal radius of superheavy nucleus
\beq
\label{5.8}
R_{min}(Z) \simeq 1.2\, (2.5\,Z)^{1/3}\, \ \hbox{fm} \ ,
\eeq
the magnetic vacuum field configuration is formally free from any restrictions and in principle can collapse down up to infinitesimal scales. In turn, at such scales the VP-energy decays into the negative infinity and the self-consistent solution is reached precisely in this limit.

However, this result holds only in the pure QED. At the same time, the problem under study is more diverse, including the nuclear structure of the Coulomb source and properties of the low-energy hadronic matter. The last two lead to some extra restrictions, which make this limit non-physical. First, there should exist an absolute limit on the size of magnetic field from below, provided  by low-energy hadronic scale of several pion Compton lengths (actually not less than 3-5), since at such scales a sufficiently strong  magnetic field  should cause an essentially nonlinear excitation of quark-pion and further of gluon condensates, which presumably are the main components of low-energy hadronic matter (see, e.g., Refs.~\cite{Hosaka2001,Weigel2007}). Since both condensates are bosonic, their excitation must be accompanied by a significant increase in energy. And although a fully consistent model of low-energy hadronic matter including confinement problems is still far from being completed, it would be natural to assume that the associated energy scales should be incomparably larger than those of QED.

Second, the spontaneously generated vacuum magnetic field  acts on the charged nuclear component of the Coulomb source as the external one, which causes the corresponding Lorentz force and hence, the rotation of the Coulomb source as a whole. For a spherically-symmetric charge configuration the axial structure of the vector-potential  (\ref{3.8c}) implies that the generated this way nuclear current must be of the same form
\beq\label{5.9}
    \v j_{nucl} ({\v r})=j_{nucl}(r, \vt)\,\v {\bf e}_{\vf} \ ,
\eeq
being generated by the z-component of the induced magnetic field
\beq\label{5.10}
    {\cal H}_z (r, \vt)={ \cos \vt \over r\, \sin \vt}\, \pd_\vt\(\sin\vt\, A(r,\vt)\)+ {\sin\vt \over r}\pd_r\(r A(r,\vt)\) \ .
\eeq
In terms of magnetic multipoles expansion (\ref{3.8}) ${\cal H}_z ({r, \vt})$ takes the form
 \begin{widetext}\begin{multline}\label{5.11}
      {\cal H}_z (r, \vt)={2\,e \over r\,\sin\vt}\,\sum\limits_n {1 \over \sqrt{n\,(n+1)}}\,  \[A_n(r) \(P_{n}^1 (x)+(x^2-1)\,x\,\pd_x P_n^1(x)\) \ +  \ r\pd_rA_n(r)\,(1-x^2)\,P_n^1(x)\] = \\ = {2\,e \over r\,\sin\vt}\,\sum\limits_n {P_n^1(x) \over n\,(n+1)}\, \Bigg[\sqrt{n\,(n+1)}\,A_n(r) \ +  \\  + \ \sum_{m}{n+1/2 \over \sqrt{m\,(m+1)}}\,\(A_m(r)\,\int_{-1}^1 \! dx'\,P_n^1(x')\,(x'^2-1)\,x'\,\pd_{x'} P^1_m (x') \ +  \ r\pd_r A_m(r)\,\int_{-1}^1 \! dx'\, P_n^1(x')\,(1-x'^2)\,P_m^1(x') \)\Bigg] \ ,
\end{multline} \end{widetext}
where $x=\cos \vt$.

Now let us consider for clarity a charged sphere with radius $R$ instead of the shell (\ref{1.0}). The Lorentz force acting at each proton located on the sphere with the polar angle $\vt$ is generated via axial rotation of the sphere as a whole and equals to
\begin{multline}\label{5.12}
    \v F_L(R, \vt)=|e|\,v(R,\vt)\,{\cal H}_z (R, \vt)\,\[\v {\bf e}_{\vf} \times \v {\bf e}_{z}\] \ = \\ = \ |e|\,v(R,\vt)\,{\cal H}_z ({R, \vt})\,\v {\bf e}_{\vr} \ ,
\end{multline}
where $v(R,\vt)=\W (R)\, R\sin\vt$ is the axial velocity of any  point located on the circle $0\leqslant \vf \leqslant 2\pi$ with coordinates $R\,, \vt$, while $\v {\bf e}_{\vr}$ is the unit vector in  the direction of the planar radius.

The most important point here is that to provide the rotation of the sphere as a whole, the angular velocity of axial rotation $\W (R)$ cannot depend on the polar angle $\vt$. The latter is the direct consequence of assumption that the nuclear forces should be incomparably larger than those of QED. Otherwise, any non-uniform rotation of spherical sectors corresponding to different polar angles may cause the destruction  of the Coulomb source. This possibility cannot be excluded at once and such a process lies far beyond the scope of the present paper.

At the same time, the axial rotation of the sphere is generated by the centripetal force acting at each elementary cell of the sphere, which consists of one proton plus $\simeq 1.5$ neutrons. To provide the required rotation of the sphere as a rigid object, such force acting on the elementary cell located at the  same point with coordinates $R\,, \vt$, should be equal to
\beq\label{5.13}
    \v F_C (R, \vt)=- M_N^{\ast}\,\W^2(R)\,R\,\sin\vt\, {\bf e}_{\vr} \ ,
\eeq
where $M_N^{\ast} \simeq 2.5\,M_N$ with $M_N$ being the average mass of any nucleon for heavy ions. Since $ \v F_L(R, \vt)$ and $ \v F_C (R, \vt)$ should coincide, from (\ref{5.12},\ref{5.13}) one obtains
\beq\label{5.14}
    \W(R) = e\,{{\cal H}_z ({R, \vt})  \over M_N^{\ast}} \ ,
\eeq
which is nothing else but the cyclotron frequency, but not the Larmor one, since the latter describes the precession of the magnetic moment. But in the present case the magnetic moment contains only the z-component and so doesn't rotate.

However, ${\cal H}_z ({r, \vt})$ in the r.h.s. of the expression  (\ref{5.14}) depends on $\vt$. So we are led to an additional condition imposed on the structure of ${\cal H}_z ({r, \vt})$, which implies that in the expression (\ref{5.11}) all the multipole components with $n \geqslant 3$ should vanish. It happens because only the component, corresponding to the first multipole $P_1^1(\cos \vt)=\sin \vt$,  leads to a constant $\W (R)$, while all the others multipoles $P_n^1(\cos \vt)$ with $n \geqslant 3$, being orthogonal to $P_1^1(\cos \vt)$,  contain more complicated  dependence on $\vt$ up to $\sin (n\vt)$. Therefore the radial  multipole components $f_n(r)$, introduced via relation (\ref{3.8b}), should be subject of the following set of relations
\beq\label{5.15}
    K_{n}(R)=0 \ , \quad n \geqslant 3 \ ,
\eeq
where
\beq\label{5.16}
    K_{n}(r)=S_{nm}\,f_m(r) + T_{nm}\,r\,\pd_r f_m(r) \ ,
\eeq
and
\begin{multline}\label{5.17}
    S_{nm}=\sqrt{n\,(n+1)}\,\d_{nm} \ + \\ + {n+1/2 \over \sqrt{m\,(m+1)}}\,\int_{-1}^1 \! dx\,P_n^1(x)\,(x^2-1)\,x\,\pd_{x} P^1_m (x) \ ,
\end{multline}
\beq\label{5.18}
    T_{nm}={n+1/2 \over \sqrt{m\,(m+1)}}\,\int_{-1}^1 \! dx\,P_n^1(x)\,(1-x^2)\, P^1_m (x)  \ .
\eeq
The resulting angular velocity equals to
\beq\label{5.19}
    \W_{rot,sphere}(R) ={2\,\a\,A \over 5\,M_N R}\, K_{1}(R) \ ,
\eeq
and yields the following rotational energy of the sphere
\beq\label{5.20}
    \E_{rot,sphere}(R) ={2\, Z\,\a^2\,A^2 \over 15\,M_N }\, K^2_{1}(R) \ .
\eeq

The generalization to the spherical shell (\ref{1.0}) is quite straightforward. The main difference is in the area of integration, that now is $R_2 \leqslant r \leqslant R_1 , \ 0\leqslant \vf \leqslant 2\pi\ $ with polar angle in the interval $\(\vt\, ,\vt + d\vt\)$ and an integration parameter $d\vt$. Upon imposing the condition that the angular velocity $\W_{rot,shell}$ of axial rotation  of all such cropped cones must be independent of $\vt$, by means of the same arguments as above one obtains the following set of conditions
\beq\label{5.21}
    \int_{R_2}^{R_1}\! dr\, r^3 K_{n}(r) =0 \ , \quad n \geqslant 3 \ .
\eeq
The angular velocity equals to
\beq\label{5.22}
    \W_{rot,shell} ={2\,\a\,A \over M_N\, (R_1^5-R_2^5) }\,\int_{R_2}^{R_1}\! dr\, r^3 K_{1}(r) \ ,
\eeq
while the corresponding rotational energy is given by the following expression
\begin{multline}\label{5.23}
    \E_{rot,shell} ={2\,Z\,\a^2\,A^2 \over M_N\,(R_1^3-R_2^3)\, (R_1^5-R_2^5) } \times \\ \times \(\int_{R_2}^{R_1}\! dr\, r^3 K_{1}(r)\)^2 \ .
\end{multline}
It is easy to verify that by taking the limit $R_2 \to R_1$ the expressions (\ref{5.22},\ref{5.23}) transform into the  corresponding answers for the sphere (\ref{5.19},\ref{5.20}). Moreover, $\E_{rot,shell}$ should  be necessarily included into the final energy balance of the system at the same footing with the total $\E_{VP}^{ren}$  and electrostatic self-energy of the Coulomb source.

It should be noted also that in addition ${\cal H}_z ({r, \vt})$ gives rise to the axial rotation of each proton over its own z-axis. This rotation, however, is completely compensated inside the shell, there remains only the surface axial current, which direction on the external and internal surfaces of the shell is opposite and so the resulting effect is very small compared to the rotation of the shell as a whole.

 Further steps in searching for the magnetic configuration, which would satisfy all the conditions formulated above and simultaneously provide such a decrease in the total $\E_{VP}^{ren}$ that would reliably guarantee the spontaneous generation of such VP-magnetic effects, proceed as follows. Since there is no possibility to implement  the direct minimization of the total VP-energy (\ref{4.80}), it would be pertinent to  simulate the procedure of spontaneous symmetry breaking, starting with the axial vacuum current fluctuations. For these purposes we consider
\beq\label{5.24}
    \v j_{fluct} ({\v r})=j_{fluct}(r, \vt)\,\v {\bf e}_{\vf} \ ,
\eeq
where
\beq\label{5.25}
      j_{fluct}=2\,e \sum_{n}\, {2 n+1 \over 4\pi\,\sqrt{n(n+1)}}\,j_n(r)\, P_n^1(\cos\vt) \ ,
\eeq
while $j_n(r)$ are the radial multipole components of  current fluctuations. In turn, $\v j_{fluct} ({\v r})$ gives rise to $\v A_{fluct} ({\v r})$ of the same axial form as in (\ref{3.8}) with magnetic multipoles $a_n(r)$, which are related to $j_n(r)$ in the next way
\beq\label{5.26}
      \(-\D_r+{n(n+1) \over r^2}\)\,a_n(r)=\(2n+1\)\,j_n(r) \ .
\eeq
or, equivalently,
\begin{multline}
\label{5.27}
a_n(r)=\({1\over r}\, \int\limits_0^r \! \({r' \over r} \)^n +  \int\limits_r^\inf \! {1\over r'}\,\({r \over r'} \)^n \)\, {r'}^2\, dr' j_n(r')  \ .
\end{multline}
Simulation of current fluctuations proceeds further via specifying  $j_n(r)$ as a set of $\d$-functions
\beq
\label{5.28}
j_n(r)=\sum\limits_m \! \g_{nm}\,\d(r-\x_{nm})  \ .
\eeq
The pertinent choice of coefficients $\g_{nm}$ and location radii $\x_{nm}$ is subject of a numerical experiment, which includes:
\\

\noindent (a) solving the spectral DE (\ref{3.11c}) with $\v A_{fluct} ({\v r})$ as the vector-potential and taking special account for the set of discrete levels, which dive into the lower continuum;
\\

\noindent
(b) following the approach, considered in Refs.~\cite{Grashin2020a,*Grashin2020b}, the wavefunctions of these levels at the threshold of the lower continuum are used further for restoration of the induced this way axial vacuum current $\v j_{VP,seed}(\v r)$, which in turn  gives rise to the  corresponding magnetic field $\v A_{VP,seed}(\v r)$ through the self-consistency equation (\ref{3.13b});
\\

\noindent
(c)
 achieved this way $A_{VP,seed}(\v r)$ determines the  seed magnetic multipoles $f_n(r)$ for VP-energy calculation, described in the preceding Section. The latter includes the minimization of $\E_{VP}^{ren}$ with respect to uniform scaling of multipoles $f_n(r) \to f_n(r/\l)$ under conditions (\ref{5.21}), and common amplitude factor $A$.
\\

In essence, we are playing out here a kind of the trial function method  with a special selection of both its initial profile and variation parameters, likewise E.A.Hylleraas variational approach to the $H^-$ problem~\cite{Hylleraas2000} (see also Ref.~\cite{Bethe1977}). The Hylleraas trial function and the set of variational parameters are quite complicated, however, such approach provides a  precise estimate of the $H^-$ bound state energy and its main properties. Our scenario of spontaneous breakdown in the magnetic VP-component also turns out to be complicated, but leads to quite significant results for VP-energy decrease and so for the very possibility of such spontaneous magnetic VP-effects.

\section{The Coulomb source and magnetic field configurations}\label{Coulomb_source}\label{Sec_CoulombSource}

 \begin{figure}
 \center
\includegraphics[scale=0.53]{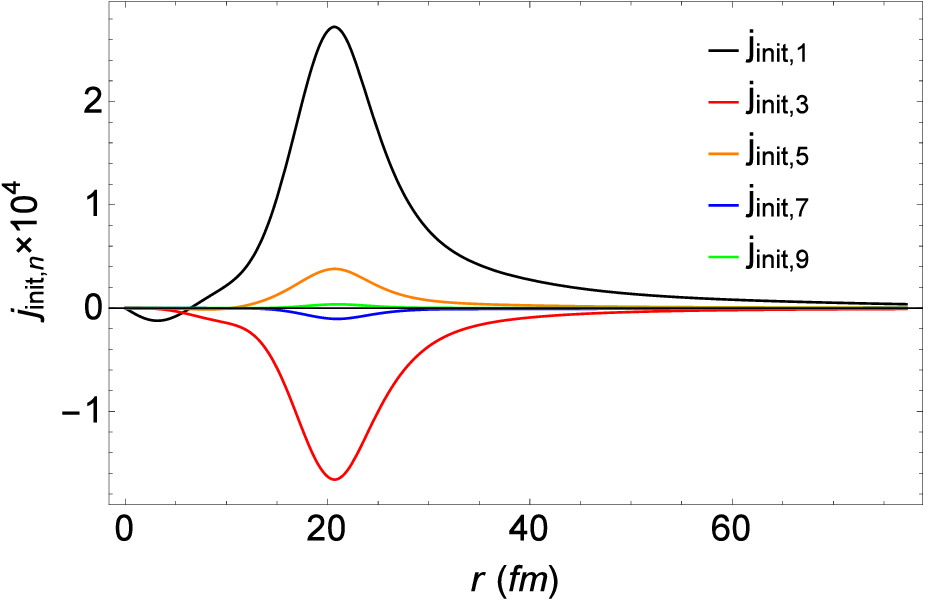} \\
\caption{(Color online) First five multipole components of the initial VP-current.}
\label{j_init-13579}
\end{figure}

\begin{figure*}[ht!]
\subfigure[]{
		\includegraphics[width=1\columnwidth]{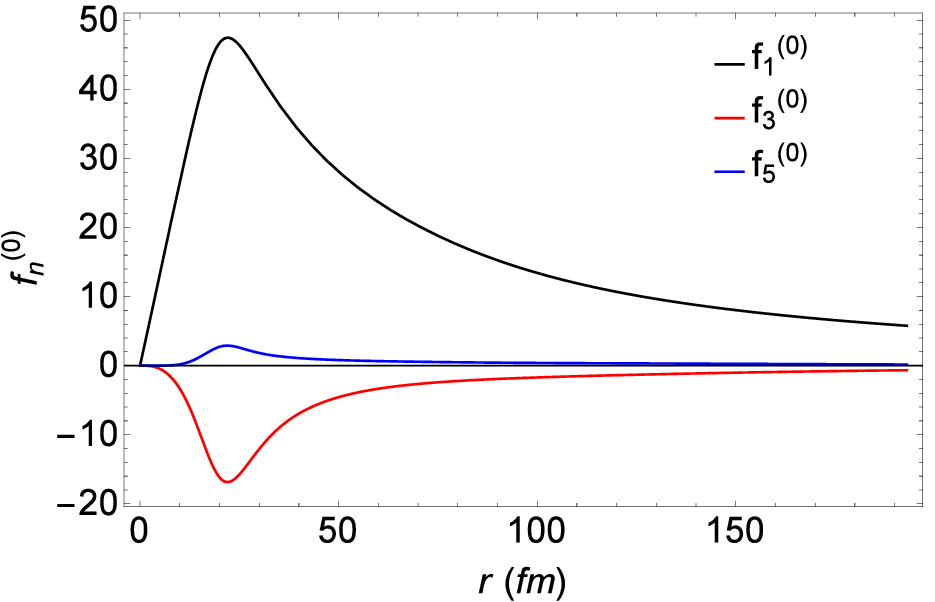}
}
\hfill
\subfigure[]{
		\includegraphics[width=1\columnwidth]{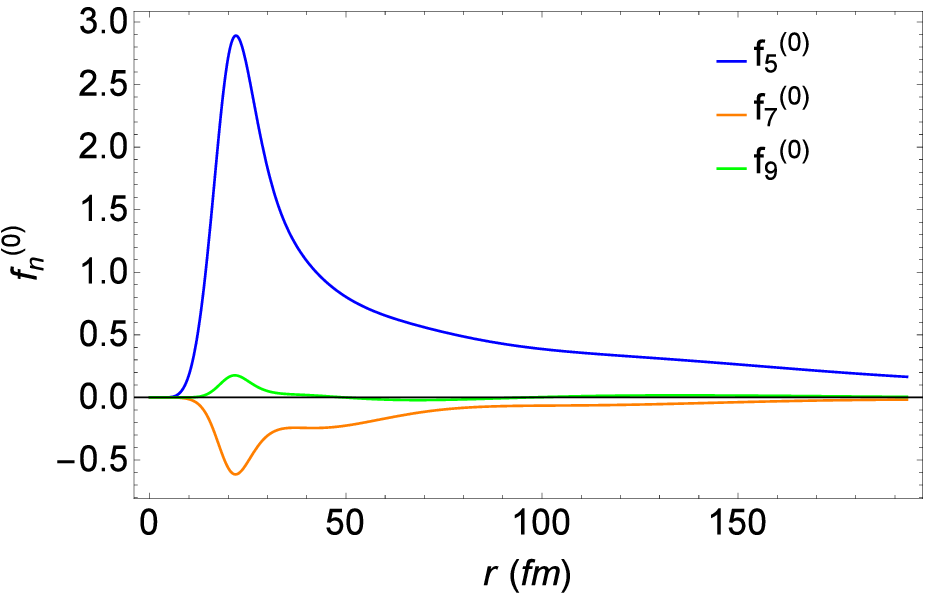}
}
\caption{(Color online) First five multipole components of  the initial VP-magnetic field $\v A_{init}(\v r)$ (a):  $n=1\,,3\,,5$ and  (b):  to $n=5\,,7\,,9$. }
	\label{f_init-135-579}	
\end{figure*}

Now - having dealt with the general disposition  in the problem of how to explore the spontaneous generation of  magnetic VP-effects this way - let us start with the more detailed description of the Coulomb source configuration (\ref{1.0}-\ref{1.8}). It represents a uniformly charged spherical shell, which is chosen in such a way that imitates a system of bare U nuclei, whose centers are located at an equal distance $a$ from each other at the vertices of a polyhedron inscribed in a sphere of radius $R$. The concrete number of vertices must be 9, hence, the total charge of the shell is $Z=9\times 92=828$~\footnote{It should be repeatedly reminded that charge configurations of such type stay out of the area of atomic scale systems permitted to exist naturally, so it is worth to highlight the nuclear scale questions in order to convince if the regarded cluster is experimentally conceivable not delving deeply into the implementation details. Any other hypotheses on this subject are omitted for now.}. Such a polyhedron with 9 vertices inscribed in a sphere has been considered first by K.Schuette and B.L. van der Waerden~\cite{vanderWaerden1951}. The more comprehensive review is given in Ref.~\cite{Toth2013}. Due to the fact that an electromagnetic interaction is determined by electric charge, which is identical for all taken ions, through the process of their relatively slow rapprochement the saturation and the curve of electromagnetic field lines are reasonable to consider to be maintained symmetrical. This immediately leads to the problem solved in another useful research \cite{melnykextr1977} (and refs. therein). In short, the polyhedron appears to be a product of dihedral group $D_{3h}$, having a tricapped prismatic geometry. It immediately leads to the solved (in accordance with given refs.) problem of minimizing the function of the distance between vertices distributed on a sphere. 

The studied polyhedron faces contain 8 regular triangles and 3 parallelograms. All the edges of the polyhedron are equal to $a$, which is related to the radius $R$ of the circumscribed sphere via relation
\beq
\label{6.1}
R={\sqrt{3} \over 2}\,a  \ .
\eeq
Such a choice is motivated by the desire to explicitly achieve the range of $600 \leqslant Z \leqslant 1000$ (mentioned in Section \ref{Sec_Intro}). The minimal distance $a_{min}$ between centers of neighboring nuclei is
\beq
\label{6.2}
a_{min}= 2 R_U + \D a \ ,
\eeq
where $\D a = 0.64 $  fm, what gives
\begin{equation}
\label{6.3}
a_{min} \simeq 15.88 \ \hbox{fm} \ , \quad  R_{min} \simeq 13.76 \ \hbox{fm} \ .
\end{equation}

The presence of $\D a$ is due to the location of Reed's potential, by means of which internuclear effects are accounted. A more expanded analysis of such a choice of $a_{min}$ and $\D a$ is given below in Section \ref{Sec_Results}. Here, it should be noted that for the van der Waerden polyhedron under study the starting charge configuration of minimal size are taken according to eqs.(\ref{6.3}) and the main goal of the present paper is calculation of $\E_{VP}(R)$ as a function of $R$ in the range $R_{min} \leq R \leq R_{max}$ for fixed $Z = 828$.

Proceeding further, we first define the initial (seed) set of magnetic multipoles $f_n^{(0)}(r)$. Hereinafter this set is obtained for the starting charge configuration with $a_{min}$ and $R_{min}$ and it does not change for any arbitrary $R > R_{min}$. It is build in the next way. First, it should be clear that the most productive vacuum current fluctuations must take place in the region of maximal strength of the external Coulomb field. According to items (a)-(c), described at the end of Section \ref{Sec_Var}, it is the external radius of the shell $R_{1,min}=R_{min}+R_U \simeq 21.38$ fm that appears as the most preferable. Its amplitudes are to satisfy relation
\beq
\label{6.4}
\g_n R_{1,min}^n=(-1)^{(n+1)/2-1}\g_1 R_{1,min}  \ .
\eeq
 The magnetic field fluctuations take the form
\beq
\label{6.5}
a_n(r)=\a_n r^n\, \tt(R_{1,min}-r) + {\m_n \over r^{n+1}}\,\tt(r-R_{1,min}) \ ,
\eeq
where
\beq
\label{6.6}
\a_n=\g_n \ , \quad \m_n=\g_n\,R_{1,min}^{2n+1} \ .
\eeq
Note that the relations (\ref{6.4}-\ref{6.6}) imply that the  maximal amplitudes of the magnetic fluctuations are equal, but alternating in sign.

The total number of multipole channels with current fluctuations turns out to be 3, including the first three $n=1\,,3\,,5$. Enlarging the number of multipole channels doesn't lead to significant changes in results~\footnote{This statement concerns only the charge configurations of the considered type build on the base of the van der Waerden polyhedron with 9 vertices and/or uniform pentagonal prism with 10 vertices and with total $Z$ in the range $820 \leqslant Z \leqslant 830$. For different kind configurations with another geometry and total charge, the structure of current fluctuations might be significantly different}. The generated according to items (a)-(b) of Section \ref{Sec_Var} five initial (seed) multipole components of the axial VP-current $\v j_{init}(\v r)$ are shown in Figs.~\ref{j_init-13579}. Because in what follows the most part of calculations will be implemented in terms of the induced vector potentials rather than of the VP-currents, in  Figs.~\ref{f_init-135-579} the  initial VP-magnetic multipoles are shown with in detail. The presentation of multipoles is restricted to first five with $n=1\,,3\,,5\,,7\,,9$. Actually, if this number of multipoles leads to the most significant decrease of the VP-energy or not --- is a subject for future investigations. So far, it is confirmed via numerical calculations, that 3-multipole ($n=1\,,3\,,5$) and 4-multipole ($n=1\,,3\,,5\,,7$) configurations are insufficient regarding the VP-effects under study. As for 5-multipole one, as expected, its components are alternating in sign, while their magnitudes depend essentially non-linearly on $\g_1$,  which value is selected to ensure the best result for VP-energy drop with such numbers of multipoles via numerical experiment. Note also that all the curves of the seed currents and magnetic multipoles exceed their maximal/minimal values exactly at $R_{1,min}$.

 At the next stage, for any $R$  in the range $R_{min} \leqslant R \leqslant R_{max}$ the magnetic seed configuration  must be scaled under  conditions (\ref{5.21}), which transforms  the initial  multipoles $f_n^{(0)}(r)$ into the current ones
\beq
\label{6.7}
f_n(\l,r)=c_n(\l)\, f_n^{(0)}(r/\l) \ ,
\eeq
where the coefficients $c_n(\l)$ are obtained from conditions (\ref{5.21}) in the following form
 \beq
\label{6.8}
c_1=1\,, \quad  c_n(\l)=-\sum\limits_{m=3}\[L^{-1}(\l)\]_{nm}\,L_{m1}(\l) \ ,  \quad n \geqslant 3  \ ,
\eeq
where
\begin{multline}\label{6.9}
 L_{nm}(\l)=\int\limits_{R_2/\l}^{R_1/\l} \! dr\, r^3\, \(S_{nm}\,f^{(0)}_m(r) + T_{nm}\,r\,\pd_r f^{(0)}_m (r)\)  \ , \\ n\,,m \geqslant 3 \ .
\end{multline}
In this case the expressions (\ref{5.22},\ref{5.23}) for the angular velocity and corresponding energy of the axial shell rotation should be modified by means of the following replacement
\begin{multline}\label{6.9a}
K_1(r) \to    K_{\l,1}(r)=S_{1m}\,f_m(\l,r) + T_{1m}\,r\,\pd_r f_m(\l,r)= \\
=\(6 \sqrt{2}/5\)\,f_1(\l,r)-\(16 \sqrt{3}/35\)\,f_3(\l,r) + \\ +  \(4 \sqrt{2}/5\)\,r\,\pd_r f_1(\l,r)-\(4 \sqrt{3}/35\)\,r\,\pd_r f_3(\l,r) \ ,
\end{multline}
with $f_m(\l,r)$ being defined in (\ref{6.7}).

 It is easy to see that the behavior of perturbative counterterms (\ref{4.71}) and (\ref{4.79}) under scaling poses no problems. However, the behavior of the magnetic Born term $B(m,y)$,  defined via expressions (\ref{4.68}) and (\ref{4.69}), requires some additional considerations. This is because the Born term actually depends on $\g(y)=\sqrt{1+y^2}$, rather than directly on $y$. Scaling with $\l \geqslant 1$ poses no problems, since in this case the replacement of arguments $r \to \l\,r$ in integrals (\ref{4.69}) leads to $\g(y) \to \l\,\g(y)$ and additional multiplier $\l^2$. So for the scaled Born term one obtains a very simple answer
 \beq
\label{6.9b}
B(m,y) \to \l^2\,B(m,\sqrt{\l^2 y^2 + \l^2 -1}) \ ,
\eeq
with the only replacement $f_n(r) \to c_n f_n^{(0)}(r)$ in the integrals (\ref{4.69}). Hence, in this case the calculation of the Born term for any $R$  and $\l$ requires actually only the knowledge of the coefficients $c_n(\l)$ and integrals (\ref{4.69}), found for the initial multipoles $f_n^{(0)}(r)$.

However, for the collapsing process with $\l <1$  there appears  the region $0 \leqslant y < \sqrt{1-\l^2}/\l$, where $\l\,\g(y)<1$ and so the substitution (\ref{6.9b}) fails. In this region the Born term must be calculated separately via initial formulae (\ref{4.68}) and (\ref{4.69}) with the same replacement of arguments $r \to \l\,r$ in integrals (\ref{4.69}), while the final answer for the Born term in the entire region $0 \leqslant y < \inf$ contains two parts, automatically sewn together by continuity at $y=\sqrt{1-\l^2}/\l $, where $\l \g(y)=1$. The first one corresponds to $0 \leqslant y < \sqrt{1-\l^2}/\l$,  while the second for $ \sqrt{1-\l^2}/\l \leqslant y < \inf$ is achieved via substitution (\ref{6.9b}).

\begin{figure}
\center
\includegraphics[scale=0.5]{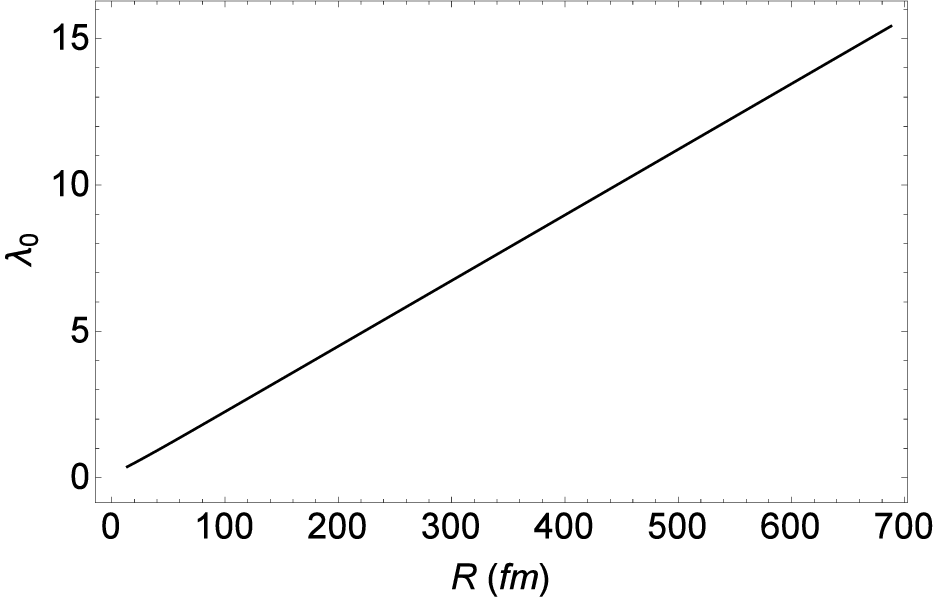} \\
\caption{\small (Color online) The behavior of the stopping points $\l_0(R)$ of the magnetic collapsing process.}
\label{Lambda0(R)}
\end{figure}

 At it was already stated above, the spontaneously generated vacuum magnetic field configuration reveals a clear tendency to collapse down up to infinitesimal scales, where the magnetic VP-energy enters the negative infinity, which must be prevented from general grounds. The properties of the initial magnetic configuration, considered above,  turn out to be such that they themselves define almost unambiguously the value of the scaling parameter $\l(R)$ for each $R$, which provides the restrictions imposed on spherical configuration collapse (see below). For each $R$ the process of reducing $\l$, which  imitates the process of nuclei oncoming to each other, there is a stopping point $\l_0(R)$, at which
 \beq
\label{6.10}
\det L(\l_0) =0 \ .
\eeq
The relation (\ref{6.10}) implies that at this point the condition (\ref{5.21}) cannot be solved in terms of $c_n$. This means that any non-uniform rotation of spherical sectors corresponding to different polar angles occurs resulting in the destruction of the Coulomb source. Furthermore, by approaching the point $\l_0(R)$ from above all the coefficients $c_n(\l)\,, \ n \geqslant 3$ increase infinitely in modulus, which leads to very fast infinite grow of VP-energy  at least as $O(1/(\l-\l_0)^4)$. So the VP-energy itself creates at this point an impenetrable barrier for further collapse of magnetic configuration. It should be specially noted that the dominating role in this effect  belongs to   by $\E^{ren}_{D,VP}$. Moreover, it turns out that for $R \geqslant R_{min}$ the stopping points $\l_0(R)$ satisfy all the conditions mentioned above. The behavior of $\l_0(R)$ as a function of $R$ in the range $R_{min} \leqslant R \leqslant 600$ fm is shown in Fig.\ref{Lambda0(R)}.
In particular, the lowest stopping point equals to $\l_0(R_{min})=0.3728$, which means that the region of maximal strength of magnetic multipoles is shifted  from $R_{1,min} \simeq 21.38$ fm up to $\simeq 7.97\,\hbox{fm} \sim 5.7 \l_{\pi}$, which is the lowest allowed value for the size of magnetic field from below, provided  by low-energy hadronic scales.

At the same time, before starting the infinite grow just before the stopping point, the VP-energy, being considered as a function of $\l$, reveals a smooth parabolic minimum, located in the region $\simeq 10^{-3}$  to the right of the stopping point. Therefore, the choice of the most pertinent current scaling parameter $\l(R)$ for any $R$ doesn't pose any problems. In particular, $\l(R_{min})=0.373$.

\section{Numerical calculation results}\label{Sec_Results}

To reveal the general pattern of $\E^{ren}_{VP}(Z,A)$ behavior in an absence of possibility to obtain the analyticcal solution for (\ref{3.16a}),( \ref{3.16b}), we fix $Z = 828$ now and so forth and solve their analogues built the next way.

It is a natural assumption\footnote{By complete analogy with, for instance, Hartree-Fock approach to the calculation of an infinite crystal structure energy or any of scattering problems.} that only a certain finite amount of energy levels diving into the lower continuum at every concrete $A$ for the specified Coulomb source contributes to the non-linear effects of vacuum polarization under study in a substantial way. Thus we introduce the cutoff parameter $k_{max}\equiv Num$ for orbital momentum $l$ and narrow the infinite subsystems of equations for corresponding partial channels (\ref{3.16a}),( \ref{3.16b}) down to finite ones containing $4(Num+1)$ unknown functions for every given $Num$. The spinor components are thus defined by
\begin{widetext}
\begin{equation}
    \begin{gathered}
	\label{3.14cl}
	\tilde{\vf}_{m_j}(\v r)= \sum\limits_{k=N_1(m_j)}^{Num} \   \tilde{u}_{km_j}(r)\, \W_{2k , m_j}^{(+)} (\v n) +  \sum\limits_{k=N_2(m_j)}^{Num} \ \tilde{v}_{km_j}(r)\, \W_{2k+2 ,  m_j}^{(-)} (\v n)  \ , \\
	\tilde{\c}_{m_j}(\v r)= \sum\limits_{k=N_2(m_j)}^{Num} \  \tilde{p}_{km_j}(r)\, \W_{2k+1 ,  m_j}^{(+)} (\v n) + \sum\limits_{k=N_1(m_j)}^{Num} \ \tilde{q}_{km_j}(r)\, \W_{2k+1 ,  m_j}^{(-)} (\v n) \ 
    \end{gathered}
\end{equation}
for the even case and
\begin{equation}
    \begin{gathered}
	\label{3.15cl}
	\tilde{\vf}_{m_j}(\v r)= \sum\limits_{k=N_2(m_j)}^{Num} \  \tilde{u}_{km_j}(r)\, \W_{2k+1 , m_j}^{(+)} (\v n) +  \sum\limits_{k=N_1(m_j)}^{Num} \  \tilde{v}_{km_j}(r)\, \W_{2k+1 ,  m_j}^{(-)} (\v n) \ , \\
	\tilde{\c}_{m_j}(\v r)= \sum\limits_{k=N_1(m_j)}^{Num} \  \tilde{p}_{km_j}(r)\, \W_{2k ,  m_j}^{(+)} (\v n) +   \sum\limits_{k=N_2(m_j)}^{Num} \ \tilde{q}_{km_j}(r)\, \W_{2k+2 , m_j}^{(-)} (\v n) \ 
    \end{gathered}
\end{equation}
for the odd one.

Here $\tilde{u}_{km_j}(r),\tilde{p}_{km_j}(r),\tilde{v}_{km_j}(r),\tilde{q}_{km_j}(r)$ - the sought solutions of the corresponding cut subsystems with $0\leq k \leq Num$. \footnote{For the $\ln [|\mathrm{Wronskian}|]$ method the same replacement is performed without breaking the convergence in $Num$.} By means of $Num$ increase, the approximate $\E^{ren}_{VP}(Z,A,Num)$ tends to the exact $\E^{ren}_{VP}(Z,A)$, which is true for arbitrary fixed $Z$.
\end{widetext}

\begin{figure*}[ht!]
	\subfigure[]{
		\includegraphics[width=1\columnwidth]{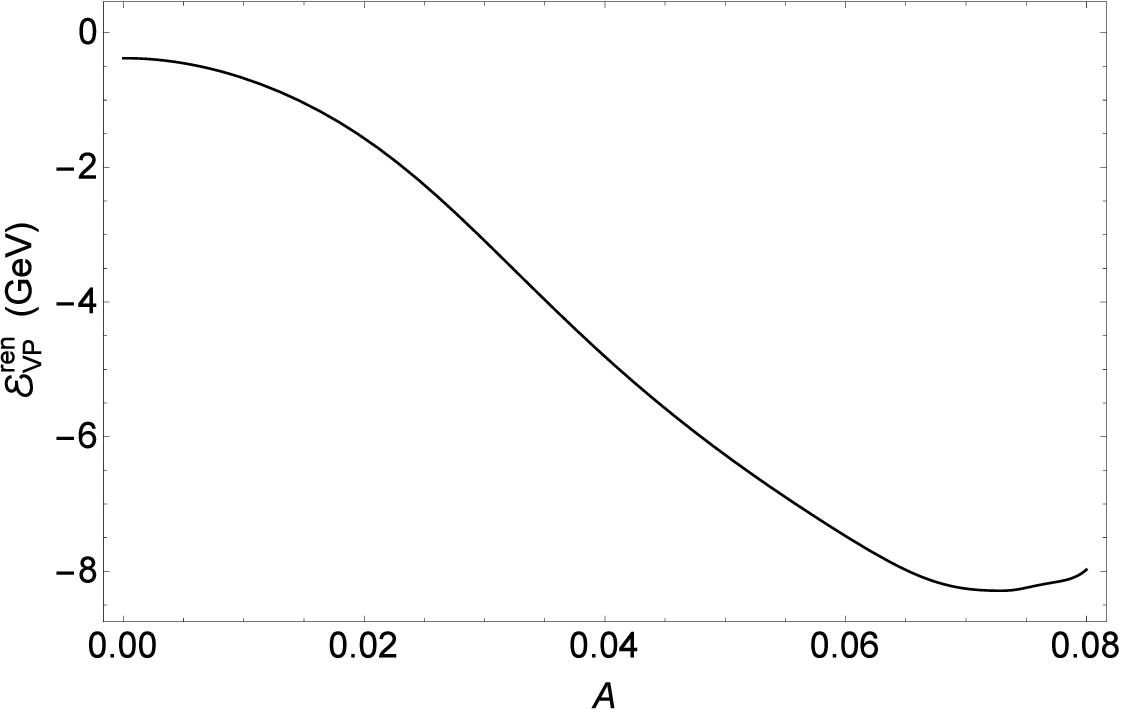}
	}
	\hfill
	\subfigure[]{
		\includegraphics[width=1\columnwidth]{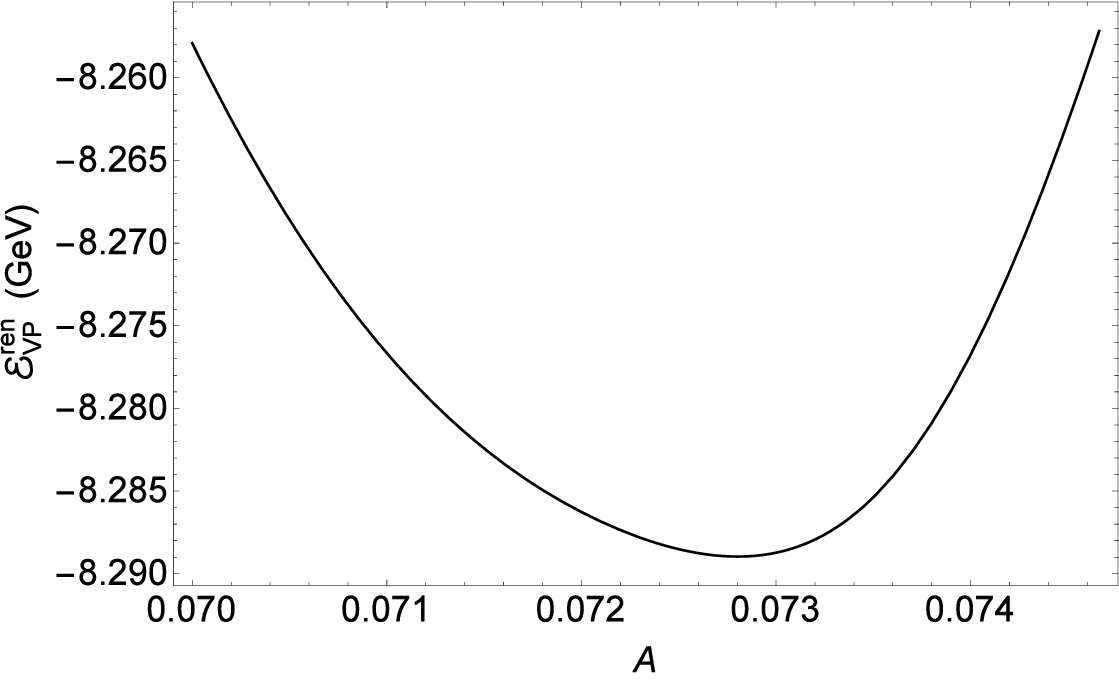}
	}
	\caption{\small (Color online) The dependence of VP-energy for the source with total charge $Z = 828$ and radius $R\simeq82.54$ fm and $Num=11$ on common multiplier $A$ of magnetic field for: (a) $0\leq A\leq0.08$  (b) $0.07\leq A\leq0.0747$. \small }
	\label{Evac_Atot}	
\end{figure*}

In Figs. \ref{Evac_Atot} the dependence of VP-energy on magnetic amplitude factor for pointed configuration parameters is presented. Such a well-expressed minimum revealing the DE spectrum reconstruction with respect to $A$ occurs for the next reason. With growing field strength, the structure of the lowest discrete levels wavefunctions changes in such a way that at first stage the levels with corresponding sign of $m_j$ fall down faster and faster, but afterwards their wavefunctions deform in such a way that the components with large orbital moments begin to produce an increasing positive contribution due to  centrifugal energy. As a consequence, the reverse process of levels rising takes place. Meanwhile, it is indeed the discrete levels diving into the lower continuum, which affects in the most significant manner nonlinear effects of vacuum polarization under study. Those terms responsible for magnetic self-interaction (\ref{4.79}) and for negative discrete energy levels $\D \E_{neglev}(A)$ appear to make a minor contribution to the picture.

\begin{figure*}[ht!]
	\subfigure[]{
		\includegraphics[width=1\columnwidth]{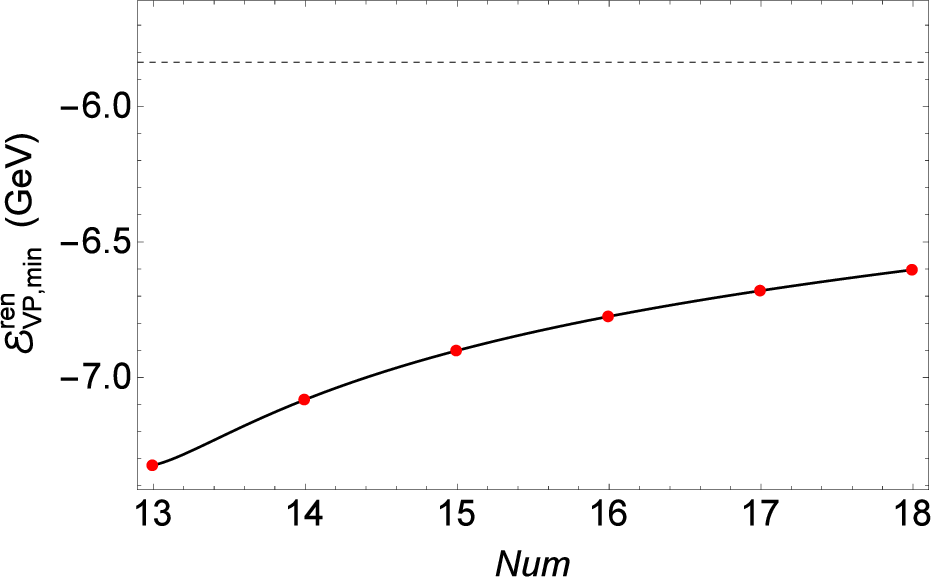}
	}
	\hfill
	\subfigure[]{
		\includegraphics[width=1\columnwidth]{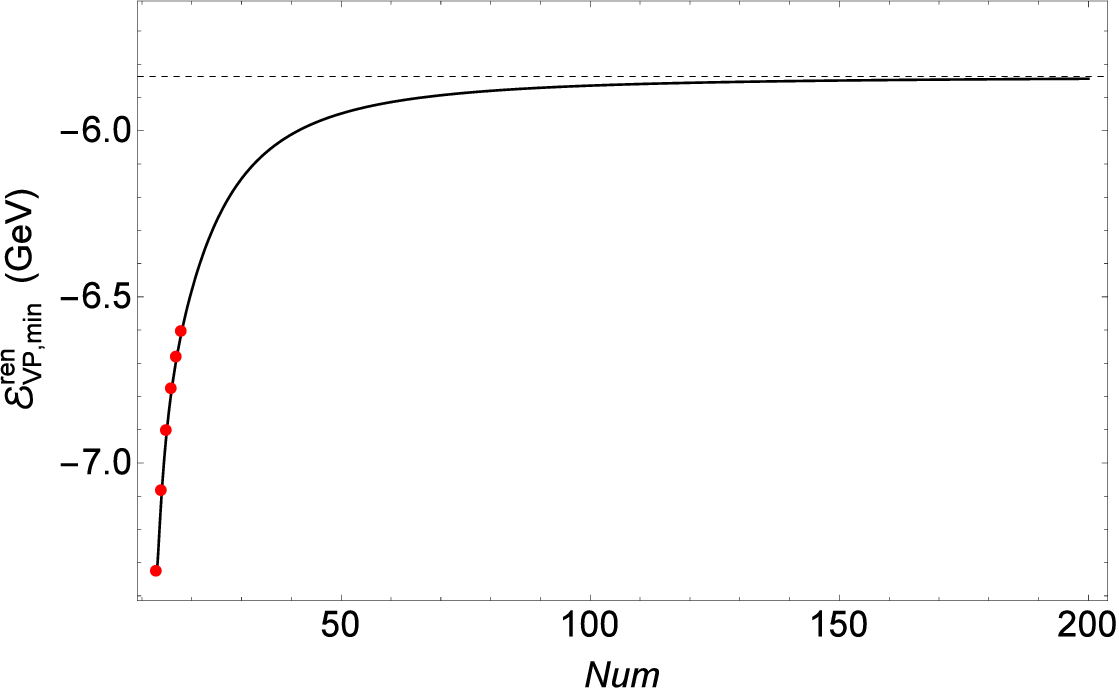}
	}
	\caption{\small (Color online) The dependence of VP-energy in the minimum for the source with total charge $Z = 828$ and radius $R = 82.54$ fm on the cutoff parameter $Num$ for: (a) $13\leq Num\leq18$  (b) $13\leq Num\leq200$. The red dots and black line correspond to calculated values of VP-energy and extrapolation function respectively. \small. }
	\label{Evac_Num}	
\end{figure*}

However, by varying the $Num$ one ascertains that its increase impacts on the curve minimum, gradually shifting its position towards smaller values of $A$, while the corresponding absolute value of VP-energy decreases. Such a tendency is transparently expected: the higher levels are located above the negative range boundary, so the more terms in (\ref{3.14cl}),(\ref{3.15cl}) are calculated, the more negative energy contribution is countervailed. To insure the energy minimum doesn't evaporate with $Num\longrightarrow \infty$ an approximation for $\E^{ren,min}_{VP}$ is performed via application of the Aitken $\D^2$-process (see \cite{Sidi2023}), the results of which are shown in the Fig. \ref{Evac_Num}. It can be seen that the sequence $\{\E^{ren}_{VP,min}(Num)\}$ indeed converges to the concrete value $\E^{ren}_{VP,min}$, although quite slowly: the ratio $\big( \E^{ren}_{VP,min}  \big) / \big( \E^{ren}_{VP,min}(Num) \big)$ achieves 99$\%$ at $Num\sim 100$.

Proceeding further, to obtain a reliable comparative estimation of the ultimate  energy of highly magnetized 9 U configuration, its total energy is assembled as
\begin{equation}\label{4.80num}
    \E_{tot} = \E^{ren}_{VP,min} + \E_{coul} + \E_{nucl} \,
\end{equation}
where $Z = 828$,  $\a=1/137.036$. $\E_{coul}$ is the Coulomb repulsion energy exactly determined by
\begin{equation}
    \E_{coul} = 3Z^2\a \frac{2R_1^5-5R_1^2R_2^3+3R_2^5}{10\(R_1^3-R_2^3\)^2} \,
\end{equation}
while the appropriate construction of the nuclear interaction term $\E_{nucl}$ is a point to dwell on as follows.

It is now to be renoted that the internuclear forces regulate the position of a starting point $a_{min}$ (\ref{6.2}), (\ref{6.3}). The essential point is about ensuring whether their contribution remains negligible throughout the entire range of $R$. From one side, it should not wreck the prospect of constructing the taken cluster, from another --- prevent from dealing with an accurate calculation of the surface distribution of charge density. As a matter of fact, taking into account nuclear deformations will only reduce the nuclear interaction energy contribution, but in case of a rather small values of $\E_{nucl}$ would not affect the $\E_{tot}$ profile, permitting to avoid this non-trivial task at all. Thereafter a rigid upper bound for nuclear forces estimation consents ions to have a spherical form.
\begin{figure}[h!]
    \centering
    \includegraphics[width=0.75\columnwidth]{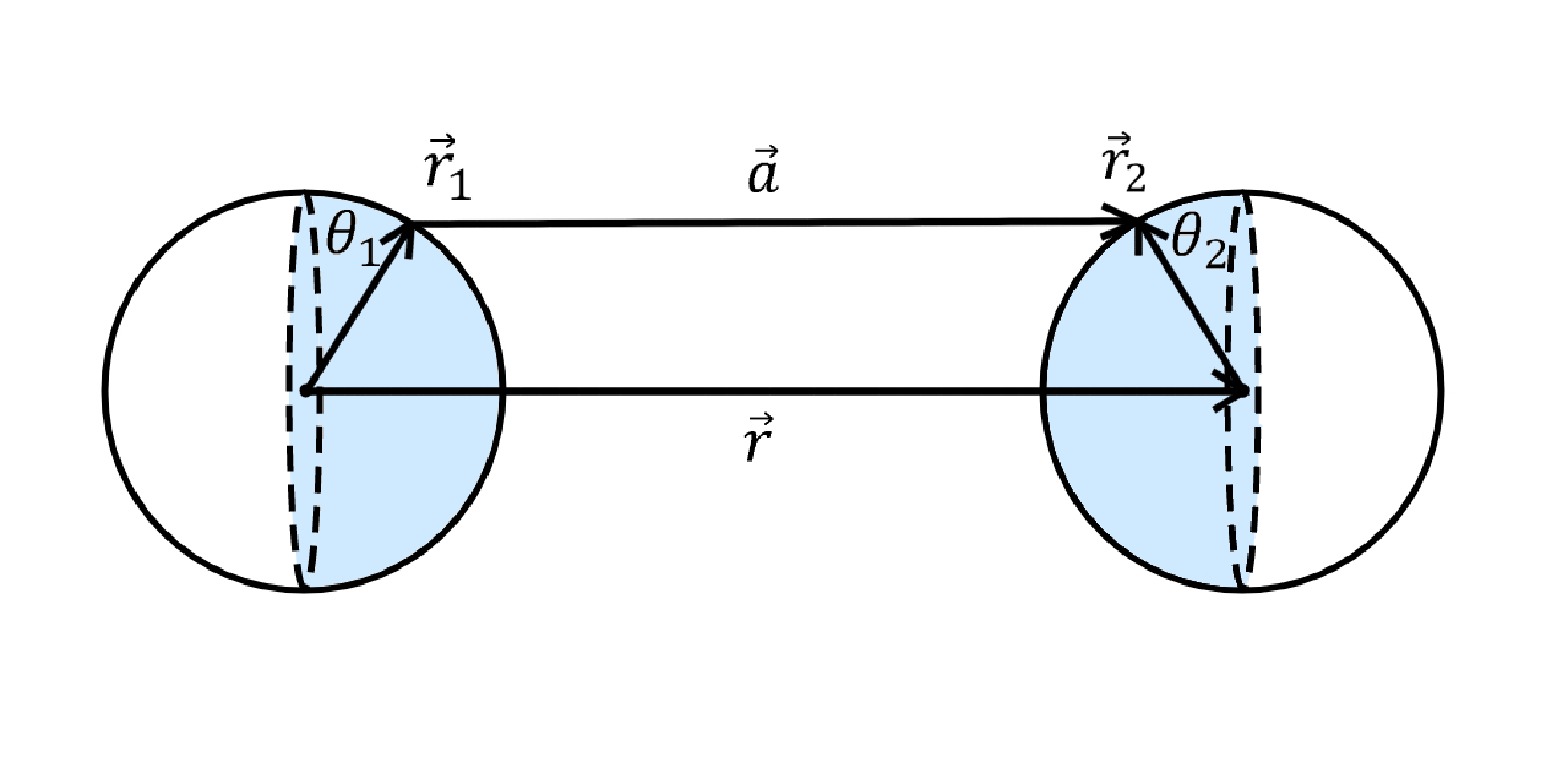}
    \caption{(Color online) The area of integration (colored). The coordinate system is joined in such a way that the axis $Oy$ connects the centers of two nuclei and the axis $Oz$ lies in the scheme plane.}
    \label{nuclear_interaction}
\end{figure}

In itself $\E_{nucl}$ is short-range, so it permits to effectively eliminate ion bulks and leave only the facing parts of each pair of interacting ions to integrate over (see Fig. \ref{nuclear_interaction}). The latters, in turn, can be treated as a summation of outward parts of nucleons as a product of the "tight-packaging" model. Albeit, instead of their actual form, the cross-sectional area of the nucleon is accepted as the determinant factor. Introducing $\sigma = 1/(\pi R_{nucleon}^2) = 1/(S_{nucleon}^{eff})$ --- a surface nucleon density that rapidly decreases with package inflation and the nuclear potential smeared over the nucleon interaction area, one obtains (for more details see \cite{katoot1989microscopic}-\cite{zagrebaev2006low}) 
\begin{equation}
    \E_{nucl}^{pair} = R_{U}^4 \sigma^2   \int\limits_{0}^{\pi} d\theta_1 \int\limits_{0}^{\pi} d\varphi_1 \int\limits_{0}^{\pi} d\theta_2 \int\limits_{\pi}^{2\pi} d \varphi_2 sin\theta_1 \thinspace sin\theta_2 \thinspace V(|\vec{a}|),
\end{equation}
where $\vec{a} = \vec{r} + \vec{r_2} - \vec{r_1}$ --- the distance between the points placed on the opposite surfaces of two neighboring nuclei.

Eventually, regarding the choice of $V(|\vec{a}|)$, the minimal $\vec{a}$ corresponds to the zero of the Reid potential in the $s$-channel \cite{reid1968local}, \cite{Blocki1977}, \cite{Brown1976}:
\begin{equation}
    V(r)=-\frac{h e^{-x}}{x}-1650.6 \thinspace \frac{e^{-4x}}{x}+6484.3 \thinspace \frac{e^{-7x}}{x},
\end{equation}
where $h=10.463$ MeV, $x=\mu r$, $\mu=0.7$ fm$^{-1}$.

\begin{figure*}[ht!]
	\subfigure[]{
		\includegraphics[width=1\columnwidth]{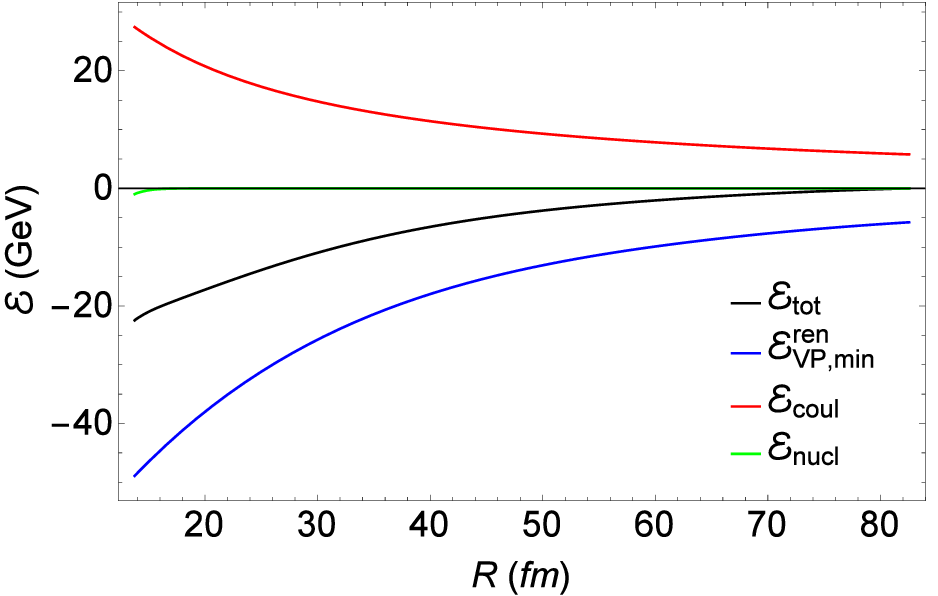}
	}
        \hfill
        \subfigure[]{
		\includegraphics[width=1\columnwidth]{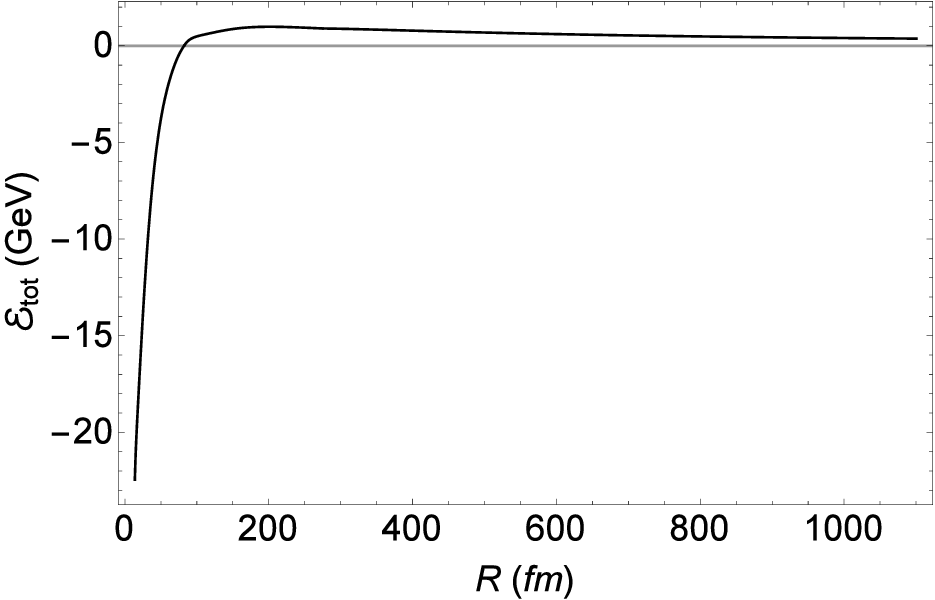}
	}
	\caption{\small (Color online) The dependence of the total energy of nuclear cluster with total charge $Z = 828$ on its radius $R$ for: (a) $13.76$ fm $\leq R\leq 82.54$ fm (b) $13.76$ fm $\leq R\leq1100$ fm. \small}	\label{Etot}	
\end{figure*}

This form of $V(r)$ dictates $a_{min}$ to be shifted from the limiting value, at which the space distribution of protons collapses. Otherwise, polyhedron symmetry (as well as the taken spherical shell model) loses its implication, since accounting nuclear surface currents ceases to be avoidable. Being generally acknowledged this approach, based on nuclear proximity forces, appears to be the most attractive for establishing the choice of $a_{min}$ (\ref{6.2}), (\ref{6.3}), also leading  to  $\D a \simeq -1.43  $ fm, whence there follows $a_{min}\simeq 13.81$ fm  and $R_{min} \simeq 11.96$ fm, while the total internuclear interaction energy after performing the summation of $\E_{nucl}^{pair}$ over all possible pairs of ions reaches $\simeq -1770$ MeV. At larger distances it anticipatedly vastly converges to zero (in Fig. \ref{Etot} (a) it is shown in comparison with other contributions).

Having all terms in (\ref{4.80num}) discussed we now pass to the total energy of the system $E_{tot}$ with fixed $Z=828$ dependence on the scaling parameter $R$ shown in Fig.\ref{Etot} with several points emphasized in Table~\ref{table_E_tot}). It must be noted that the initial stated problem is not evolutionary: technically, we are dealing with a consequent pointwise set of cluster states realized in such a way that vacuum structure has enough time to rearrange. This means the profile of $E_{tot}$ validly imitates an adiabatically slow nuclei moving towards each other\footnote{In complete analogy with Born-Oppenheimer approximation}. At distances $\simeq 1000$ fm we get a Coulomb repulsion domination, that is expected due to VP-effects being perceptible in a relatively nearby area of charged sources. To put it our way, ions should be closer to each other to initiate the vacuum reconstruction. Diminishing $R$, repulsion, in turn, gradually increases as $\sim (1/R)$, that gives rise for a slight rise right near $\simeq 200$ fm, generating a barrier to overcome. The latter requires a transfer channel, that could allow the cluster system to dive into the potential well and enter the bound state. Since the positron emission is an opened question, it makes sense to discuss if there are any alternative ways of energy release.

\begin{table*}
\caption{Indicative approximate values of dominating contributions selected from Fig. \ref{Etot} (a), (b).}
\begin{ruledtabular}
\begin{tabular}{cccc} \label{table_E_tot}
 $R$, $R_{min}$ & $\E^{ren}_{VP,min}$, MeV & $E_{coul}$, MeV & $E_{tot}$, MeV \\ \hline
 1 & -48900 & 27400 & -22400 \\
 2.15 &-26200 & 15000 & -11200 \\
 4 & -11300 & 8500 & -2800 \\
 5 &-7900 & 6900 & -1000 \\
 5.98 & -5800 & 5800 & 0 \\
 8 & -3800 & 4400 & 600 \\
 12 & -2000 & 2900 & 900\\
 14.63 & -1400 & 240 & 1000 \\
 22 & -700 & 1700 & 900 \\
 73 & -100 & 500 & 400 \\
 \end{tabular} 
\end{ruledtabular}
\end{table*}

The taken spherically symmetric heavy Coulomb source has every chance to be treated as an object of classical electrodynamics. Thus the first candidate for energy outcome is the radiation from rotating charged ball layer. Supposing it has the charge $Z$ and fixed radii $R_1$ (inner one) and $R_2$ (outer) and rotates with angular velocity $\omega$ (see \ref{5.12}) around an axis $z$, we get its charge density as
\begin{equation}
\rho = \frac{3Z|e|}{4\pi(R_2^3 - R_1^3)},
\end{equation}

Then the total magnetic moment is by definition equal to

\begin{equation}
\vec{M} = \frac{1}{2c}\int\limits_Vd\vec{r}[\vec{r},\rho [\vec{\omega}, \vec{r}]] = \frac{Z|e|\omega(R_2^5 - R_1^5)}{5c(R_2^3 - R_1^3)} \vec{e}_z .
\end{equation}

In full accordance with the definition of magnetic dipole radiation one can obtain, that in expansion process from some $r_1$ to some $r_2$ the emitted energy evaluates as
\begin{widetext}
\begin{equation}
E_{rad} = \int\limits_{t_1}^{t_2}dt I =  \int\limits_{t_1}^{t_2}dt \frac{2}{3c}|\ddot{\vec{M}}|^2 
= \frac{2}{3cm^{3/2}} \int\limits_{r_1}^{r_2}\frac{1}{\sqrt{2 (E-V(r))}}\left( -\frac{dV(r)}{dr} \frac{dM(r)}{dr}+ 2 \left(E-V(r)\right)\frac{d^2M(r)}{dr^2} \right)^2dr.
\end{equation}
\end{widetext}

Taking $r_1$ equal to $R_{min}$, $r_2 = 200.85$ fm corresponding to the potential maximum $E = \E_{tot}(r_2)$ required for above-barrier transit, we get $E_{rad} = 1.34 * 10^{-9} m_e c^2 = 6.83 * 10^{-4}$ eV. As seen, the classical component of the dipole magnetic radiation captivates a portion of the energy insufficient to transit the system from the scattering state to the bound state. At the same time, there is an alternative channel --- nuclear energy ejection --- to refer. Being licit to take the form of gamma quanta production, it takes away that particular $\E_{nucl} = 1770$ MeV permitting the bound state formation. 

Ultimately the spontaneous magnetic component activates enough to deepen the total system energy down into the negative range, which according to Table \ref{table_E_tot} takes its start at $5.98 R_{min} \approx 82.3$ fm. Onwards, the absolute $E_{tot}$ rapidly reaches GeVs, achieving $\simeq 22.4 GeV$ at $R_{min}$. The calculations has no right to be proceeded to lower values of $R$ due to the restrictions discussed in previous sections. At this point some crucial features must be outlined. This minimum value of total energy is undoubtedly the ground to claim that the non-perturbative VP-effects substantially affect vacuum structure enough to form an extra heavy nuclear cluster. However the given energy bottom value should be considered with prudence. At such a system charge $Z = 828$ and energies on the order of GeVs and dozens of GeV it is renormalization of coupling constant should be involved, requiring a separate fully-fledged research. Anyway, regarding the increase of $\alpha$ causes no injure to the VP-effects under study, on the contrary, it would only be enhanced by the corresponding running electron charge enlarging, so the presumed energy bottom would descend even lower in negative continuum.

\section{Concluding remarks}\label{Sec_Conclusion}

To finalize, it should be emphasized, that, despite the rough character of taken model, it still brings to light some new points to concern about. In accordance with the main goal of the study, we are provided with convergent essentially non-trivial solution, that exposes the inevitability of magnetic phase affecting the vacuum structure in a drastic way. Calculations are performed via the peculiar technique with the $\ln [|\mathrm{Wronskian}|]$, achieved by means of well-established structures and natural assumptions only, thence observed results reliably fit the QED-model. 

Needless to mention the present paper should be considered as a very first step to the area in all verity new, giving the significant boost to heavy ion physics applications as well as investigations of their intrinsic properties, some of which arise forthwith. Primary item supplementing the obtained result is to circumscribe the diapason of magnetic phase major influence with respect to the Coulomb charge $Z$ and the number of considered multipole components. Apparently this identification of the Curie point is enhanced by the problem discussed in Section \ref{Sec_CoulombSource}. In connection with experimental setting, it requires the revealment of a way to reduce the polyhedron vertices number maintaining the adequacy of source's spherical symmetry and uniform charge approximations. In addition, the $\alpha$-renormalization procedure should be performed to at least specify the current result.

One of the most intriguing factors providing the practical interest is the lepton number's nature that is currently a topic of considerable controversy. The substantial diminution of vacuum energy (likewise the one obtained above) should be accompanied by the positron production of the amount sufficient enough to become unambiguously discernible. Vacuum shells should gain the charge with concomitant intermediate state appearing with the first emitted positron. The number of emitted positrons is $(-1)\times$their total number. Even in view of the nuclear system might quickly acquire any electrons, which would partly compensate the bare proton charge, $Z = 828$ is a lot enough to be secure regarding the stated question.

The feasibility of investigating the area of charge overcriticality is also warranted by the real possibility of creation of extraordinary composite ion systems to shed light on new physics standing behind all this and discover any yet hidden effects. However, this preferable range strongly exceeds the highest achievable nowadays $Z=192$ for Cm+Cm, so to access the announced overcritical area serous additional implementation efforts are to undertake. In any case, looking for a preferable or at least possible way to reach QED at a supercritical Coulomb field is definitely of interest, but for now this challenge requires a separate discussion.

\section*{ACKNOWLEDGEMENTS}

The authors are very indebted to Dr. P.K.Silaev, Dr. A.V.Borisov, Dr. Yu.S.Voronina, Dr. O.V.Pavlovsky and A.A.Krasnov from MSU Department of Physics, Dr. A.S.Davydov from Kurchatov Institute and Dr. A.A. Roenko from JINR (Dubna) for interest and helpful discussions. This work has been supported in part by the RF Ministry of Sc. $\&$ Ed. Scientific Research Program, projects No. 01-2014-63889, A16- 116021760047-5, and by RFBR grant No. 14-02-01261. The research is carried out using the equipment of the shared research facilities of HPC computing resources at Moscow Lomonosov State University, project No. 2226. A large amount of numerical calculations (more than 50 $\%$) has been performed using computing resources of the federal collective usage center Complex for Simulation and Data Processing for Mega science Facilities at NRC Kurchatov Institute, research group No. g-0142.
\\

\section*{APPENDIX}\label{Appendix}

\begin{figure*}[ht!]
	\center
	\includegraphics[scale=1.2]{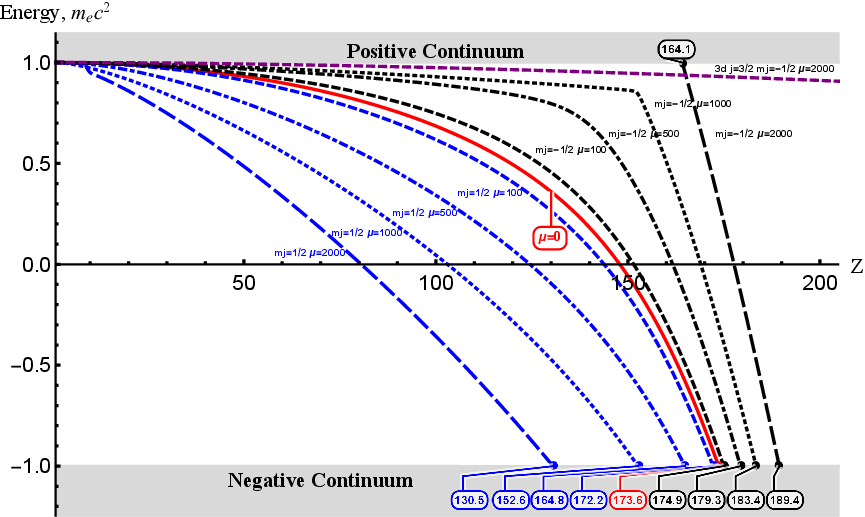} \\
	\caption{\small (Color online) The nonlinear Dirac-Zeeman splitting of $1s_{1/2}$-level in the charged sphere model with total charge $Z$ under influence of the magnetic dipole field, generated by the axial current $\v {j} ({\v r})=j(r, \vt)\,\v {\bf e}_{\vf}$, in dependence on the dipole magnetic moment $\m$. Level crossing takes place for $\m=2000$ (in units of $2\,\m_B=|e|\h/m_ec$) between $1s_{1/2}$- and $3d_{3/2}$-levels with $m_j=-1/2$ at $Z\simeq 165$.}
	\label{LevelCrossing_1s}
\end{figure*}

In Fig.\ref{LevelCrossing_1s} there is presented the nonlinear Dirac-Zeeman splitting of the lowest $1s_{1/2}$-level in the charged sphere model with total charge $Z$ under influence of the magnetic dipole field, generated by the axial current $\v {j} ({\v r})=j(r, \vt)\,\v {\bf e}_{\vf}$ with amplitude $j(r, \vt)=j_0\,r\,\mathrm{e}^{-\l \sqrt{r}}\,\sin \vt$. The corresponding magnetic moment equals to  $\m=9!\,(8\pi/3\,\l^{10})\,j_0$. The parameter $\l=2\,\sqrt{8\,Z\,\a}$ is chosen in accordance with the asymptotics of Coulomb levels at the threshold of the lower continuum, where in the current case  $Z=Z_{cr,1}=173.6$. The magnetic moment $\m$ is varied in a sufficiently wide range to provide a clear picture of level splitting and crossing. Level crossing takes place for  $\mu=2000$ (in units of $2\,\m_B=|e|\h/m_ec$). In this case,  the level $1s_{1/2}$ with $m_j=-1/2$ for small $Z$  is firstly pushed out off the discrete spectrum into the upper continuum, but for  $Z\simeq 164$ returns back and dives very quickly into the lower continuum, intersecting along the way with level $3d_{3/2}$ with the same $m_j=-1/2$,  which falls significantly slowly, since its critical charge exceeds 400.

\bibliography{VP3DC}

\begin{thebibliography}{54}%
\makeatletter
\providecommand \@ifxundefined [1]{%
 \@ifx{#1\undefined}
}%
\providecommand \@ifnum [1]{%
 \ifnum #1\expandafter \@firstoftwo
 \else \expandafter \@secondoftwo
 \fi
}%
\providecommand \@ifx [1]{%
 \ifx #1\expandafter \@firstoftwo
 \else \expandafter \@secondoftwo
 \fi
}%
\providecommand \natexlab [1]{#1}%
\providecommand \enquote  [1]{``#1''}%
\providecommand \bibnamefont  [1]{#1}%
\providecommand \bibfnamefont [1]{#1}%
\providecommand \citenamefont [1]{#1}%
\providecommand \href@noop [0]{\@secondoftwo}%
\providecommand \href [0]{\begingroup \@sanitize@url \@href}%
\providecommand \@href[1]{\@@startlink{#1}\@@href}%
\providecommand \@@href[1]{\endgroup#1\@@endlink}%
\providecommand \@sanitize@url [0]{\catcode `\\12\catcode `\$12\catcode
  `\&12\catcode `\#12\catcode `\^12\catcode `\_12\catcode `\%12\relax}%
\providecommand \@@startlink[1]{}%
\providecommand \@@endlink[0]{}%
\providecommand \url  [0]{\begingroup\@sanitize@url \@url }%
\providecommand \@url [1]{\endgroup\@href {#1}{\urlprefix }}%
\providecommand \urlprefix  [0]{URL }%
\providecommand \Eprint [0]{\href }%
\providecommand \doibase [0]{http://dx.doi.org/}%
\providecommand \selectlanguage [0]{\@gobble}%
\providecommand \bibinfo  [0]{\@secondoftwo}%
\providecommand \bibfield  [0]{\@secondoftwo}%
\providecommand \translation [1]{[#1]}%
\providecommand \BibitemOpen [0]{}%
\providecommand \bibitemStop [0]{}%
\providecommand \bibitemNoStop [0]{.\EOS\space}%
\providecommand \EOS [0]{\spacefactor3000\relax}%
\providecommand \BibitemShut  [1]{\csname bibitem#1\endcsname}%
\let\auto@bib@innerbib\@empty
\bibitem [{\citenamefont {Rafelski}\ \emph {et~al.}(2017)\citenamefont
  {Rafelski}, \citenamefont {Kirsch}, \citenamefont {M\"uller}, \citenamefont
  {Reinhardt},\ and\ \citenamefont {Greiner}}]{Rafelski2016}%
  \BibitemOpen
  \bibfield  {author} {\bibinfo {author} {\bibfnamefont {J.}~\bibnamefont
  {Rafelski}}, \bibinfo {author} {\bibfnamefont {J.}~\bibnamefont {Kirsch}},
  \bibinfo {author} {\bibfnamefont {B.}~\bibnamefont {M\"uller}}, \bibinfo
  {author} {\bibfnamefont {J.}~\bibnamefont {Reinhardt}}, \ and\ \bibinfo
  {author} {\bibfnamefont {W.}~\bibnamefont {Greiner}},\ }\enquote {\bibinfo
  {title} {Probing {QED} {Vacuum} with {Heavy} {Ions}},}\ in\ \href {\doibase
  10.1007/978-3-319-44165-8_17} {\emph {\bibinfo {booktitle} {New {Horizons} in
  {Fundamental} {Physics}}}},\ \bibinfo {series and number} {{FIAS}
  {Interdisciplinary} {Science} {Series}}\ (\bibinfo  {publisher} {Springer},\
  \bibinfo {year} {2017})\ pp.\ \bibinfo {pages} {211--251}\BibitemShut
  {NoStop}%
\bibitem [{\citenamefont {Davydov}\ \emph {et~al.}(2017)\citenamefont
  {Davydov}, \citenamefont {Sveshnikov},\ and\ \citenamefont
  {Voronina}}]{Davydov2017}%
  \BibitemOpen
  \bibfield  {author} {\bibinfo {author} {\bibfnamefont {A.}~\bibnamefont
  {Davydov}}, \bibinfo {author} {\bibfnamefont {K.}~\bibnamefont {Sveshnikov}},
  \ and\ \bibinfo {author} {\bibfnamefont {Y.}~\bibnamefont {Voronina}},\
  }\href {\doibase 10.1142/S0217751X17500543} {\bibfield  {journal} {\bibinfo
  {journal} {Int. J. Mod. Phys. A}\ }\textbf {\bibinfo {volume} {32}},\
  \bibinfo {pages} {1750054} (\bibinfo {year} {2017})}\BibitemShut {NoStop}%
\bibitem [{\citenamefont {Voronina}\ \emph {et~al.}(2017)\citenamefont
  {Voronina}, \citenamefont {Davydov},\ and\ \citenamefont
  {Sveshnikov}}]{Sveshnikov2017}%
  \BibitemOpen
  \bibfield  {author} {\bibinfo {author} {\bibfnamefont {Y.}~\bibnamefont
  {Voronina}}, \bibinfo {author} {\bibfnamefont {A.}~\bibnamefont {Davydov}}, \
  and\ \bibinfo {author} {\bibfnamefont {K.}~\bibnamefont {Sveshnikov}},\
  }\href {\doibase 10.1134/S004057791711006X} {\bibfield  {journal} {\bibinfo
  {journal} {Theor. Math. Phys.}\ }\textbf {\bibinfo {volume} {193}},\ \bibinfo
  {pages} {1647} (\bibinfo {year} {2017})}\BibitemShut {NoStop}%
\bibitem [{\citenamefont {Popov}\ \emph {et~al.}(2018)\citenamefont {Popov},
  \citenamefont {Bondarev}, \citenamefont {Kozhedub}, \citenamefont {Maltsev},
  \citenamefont {Shabaev}, \citenamefont {Tupitsyn}, \citenamefont {Ma},
  \citenamefont {Plunien},\ and\ \citenamefont {St{\"o}hlker}}]{Popov2018}%
  \BibitemOpen
  \bibfield  {author} {\bibinfo {author} {\bibfnamefont {R.}~\bibnamefont
  {Popov}}, \bibinfo {author} {\bibfnamefont {A.}~\bibnamefont {Bondarev}},
  \bibinfo {author} {\bibfnamefont {Y.}~\bibnamefont {Kozhedub}}, \bibinfo
  {author} {\bibfnamefont {I.}~\bibnamefont {Maltsev}}, \bibinfo {author}
  {\bibfnamefont {V.}~\bibnamefont {Shabaev}}, \bibinfo {author} {\bibfnamefont
  {I.}~\bibnamefont {Tupitsyn}}, \bibinfo {author} {\bibfnamefont
  {X.}~\bibnamefont {Ma}}, \bibinfo {author} {\bibfnamefont {G.}~\bibnamefont
  {Plunien}}, \ and\ \bibinfo {author} {\bibfnamefont {T.}~\bibnamefont
  {St{\"o}hlker}},\ }\href {\doibase 10.1140/epjd/e2018-90056-4} {\bibfield
  {journal} {\bibinfo  {journal} {Eur. Phys. J. D}\ }\textbf {\bibinfo {volume}
  {72}},\ \bibinfo {pages} {115} (\bibinfo {year} {2018})}\BibitemShut
  {NoStop}%
\bibitem [{\citenamefont {Novak}\ \emph {et~al.}(2018)\citenamefont {Novak},
  \citenamefont {Kholodov}, \citenamefont {Surzhykov}, \citenamefont
  {Artemyev},\ and\ \citenamefont {St{\"o}hlker}}]{Novak2018}%
  \BibitemOpen
  \bibfield  {author} {\bibinfo {author} {\bibfnamefont {O.}~\bibnamefont
  {Novak}}, \bibinfo {author} {\bibfnamefont {R.}~\bibnamefont {Kholodov}},
  \bibinfo {author} {\bibfnamefont {A.}~\bibnamefont {Surzhykov}}, \bibinfo
  {author} {\bibfnamefont {A.~N.}\ \bibnamefont {Artemyev}}, \ and\ \bibinfo
  {author} {\bibfnamefont {T.}~\bibnamefont {St{\"o}hlker}},\ }\href {\doibase
  https://doi.org/10.1103/PhysRevA.97.032518} {\bibfield  {journal} {\bibinfo
  {journal} {Phys. Rev. A}\ }\textbf {\bibinfo {volume} {97}},\ \bibinfo
  {pages} {032518} (\bibinfo {year} {2018})}\BibitemShut {NoStop}%
\bibitem [{\citenamefont {Maltsev}\ \emph {et~al.}(2018)\citenamefont
  {Maltsev}, \citenamefont {Shabaev}, \citenamefont {Popov}, \citenamefont
  {Kozhedub}, \citenamefont {Plunien}, \citenamefont {Ma},\ and\ \citenamefont
  {St{\"o}hlker}}]{Maltsev2018}%
  \BibitemOpen
  \bibfield  {author} {\bibinfo {author} {\bibfnamefont {I.~A.}\ \bibnamefont
  {Maltsev}}, \bibinfo {author} {\bibfnamefont {V.~M.}\ \bibnamefont
  {Shabaev}}, \bibinfo {author} {\bibfnamefont {R.~V.}\ \bibnamefont {Popov}},
  \bibinfo {author} {\bibfnamefont {Y.~S.}\ \bibnamefont {Kozhedub}}, \bibinfo
  {author} {\bibfnamefont {G.}~\bibnamefont {Plunien}}, \bibinfo {author}
  {\bibfnamefont {X.}~\bibnamefont {Ma}}, \ and\ \bibinfo {author}
  {\bibfnamefont {T.}~\bibnamefont {St{\"o}hlker}},\ }\href {\doibase
  10.1103/PhysRevA.98.062709} {\bibfield  {journal} {\bibinfo  {journal} {Phys.
  Rev. A}\ }\textbf {\bibinfo {volume} {98}},\ \bibinfo {pages} {062709}
  (\bibinfo {year} {2018})}\BibitemShut {NoStop}%
\bibitem [{\citenamefont {Roenko}\ and\ \citenamefont
  {Sveshnikov}(2018)}]{Roenko2018}%
  \BibitemOpen
  \bibfield  {author} {\bibinfo {author} {\bibfnamefont {A.}~\bibnamefont
  {Roenko}}\ and\ \bibinfo {author} {\bibfnamefont {K.}~\bibnamefont
  {Sveshnikov}},\ }\href {\doibase https://doi.org/10.1103/PhysRevA.97.012113}
  {\bibfield  {journal} {\bibinfo  {journal} {Phys. Rev. A}\ }\textbf {\bibinfo
  {volume} {97}},\ \bibinfo {pages} {012113} (\bibinfo {year}
  {2018})}\BibitemShut {NoStop}%
\bibitem [{\citenamefont {Maltsev}\ \emph {et~al.}(2019)\citenamefont
  {Maltsev}, \citenamefont {Shabaev}, \citenamefont {Popov}, \citenamefont
  {Kozhedub}, \citenamefont {Plunien}, \citenamefont {Ma}, \citenamefont
  {St\"ohlker},\ and\ \citenamefont {Tumakov}}]{Maltsev2019}%
  \BibitemOpen
  \bibfield  {author} {\bibinfo {author} {\bibfnamefont {I.~A.}\ \bibnamefont
  {Maltsev}}, \bibinfo {author} {\bibfnamefont {V.~M.}\ \bibnamefont
  {Shabaev}}, \bibinfo {author} {\bibfnamefont {R.~V.}\ \bibnamefont {Popov}},
  \bibinfo {author} {\bibfnamefont {Y.~S.}\ \bibnamefont {Kozhedub}}, \bibinfo
  {author} {\bibfnamefont {G.}~\bibnamefont {Plunien}}, \bibinfo {author}
  {\bibfnamefont {X.}~\bibnamefont {Ma}}, \bibinfo {author} {\bibfnamefont
  {T.}~\bibnamefont {St\"ohlker}}, \ and\ \bibinfo {author} {\bibfnamefont
  {D.~A.}\ \bibnamefont {Tumakov}},\ }\href {\doibase
  10.1103/PhysRevLett.123.113401} {\bibfield  {journal} {\bibinfo  {journal}
  {Phys. Rev. Lett.}\ }\textbf {\bibinfo {volume} {123}},\ \bibinfo {pages}
  {113401} (\bibinfo {year} {2019})}\BibitemShut {NoStop}%
\bibitem [{\citenamefont {Popov}\ \emph {et~al.}(2020)\citenamefont {Popov},
  \citenamefont {Shabaev}, \citenamefont {Telnov}, \citenamefont {Tupitsyn},
  \citenamefont {Maltsev}, \citenamefont {Kozhedub}, \citenamefont {Bondarev},
  \citenamefont {Kozin}, \citenamefont {Ma}, \citenamefont {Plunien},
  \citenamefont {St\"ohlker}, \citenamefont {Tumakov},\ and\ \citenamefont
  {Zaytsev}}]{Maltsev2020}%
  \BibitemOpen
  \bibfield  {author} {\bibinfo {author} {\bibfnamefont {R.~V.}\ \bibnamefont
  {Popov}}, \bibinfo {author} {\bibfnamefont {V.~M.}\ \bibnamefont {Shabaev}},
  \bibinfo {author} {\bibfnamefont {D.~A.}\ \bibnamefont {Telnov}}, \bibinfo
  {author} {\bibfnamefont {I.~I.}\ \bibnamefont {Tupitsyn}}, \bibinfo {author}
  {\bibfnamefont {I.~A.}\ \bibnamefont {Maltsev}}, \bibinfo {author}
  {\bibfnamefont {Y.~S.}\ \bibnamefont {Kozhedub}}, \bibinfo {author}
  {\bibfnamefont {A.~I.}\ \bibnamefont {Bondarev}}, \bibinfo {author}
  {\bibfnamefont {N.~V.}\ \bibnamefont {Kozin}}, \bibinfo {author}
  {\bibfnamefont {X.}~\bibnamefont {Ma}}, \bibinfo {author} {\bibfnamefont
  {G.}~\bibnamefont {Plunien}}, \bibinfo {author} {\bibfnamefont
  {T.}~\bibnamefont {St\"ohlker}}, \bibinfo {author} {\bibfnamefont {D.~A.}\
  \bibnamefont {Tumakov}}, \ and\ \bibinfo {author} {\bibfnamefont {V.~A.}\
  \bibnamefont {Zaytsev}},\ }\href {\doibase 10.1103/PhysRevD.102.076005}
  {\bibfield  {journal} {\bibinfo  {journal} {Phys. Rev. D}\ }\textbf {\bibinfo
  {volume} {102}},\ \bibinfo {pages} {076005} (\bibinfo {year}
  {2020})}\BibitemShut {NoStop}%
\bibitem [{\citenamefont {Grashin}\ and\ \citenamefont
  {Sveshnikov}(2022)}]{Grashin2022a}%
  \BibitemOpen
  \bibfield  {author} {\bibinfo {author} {\bibfnamefont {P.}~\bibnamefont
  {Grashin}}\ and\ \bibinfo {author} {\bibfnamefont {K.}~\bibnamefont
  {Sveshnikov}},\ }\href {\doibase 10.1103/PhysRevD.106.013003} {\bibfield
  {journal} {\bibinfo  {journal} {Phys. Rev. D}\ }\textbf {\bibinfo {volume}
  {106}},\ \bibinfo {pages} {013003} (\bibinfo {year} {2022})}\BibitemShut
  {NoStop}%
\bibitem [{\citenamefont {Krasnov}\ and\ \citenamefont
  {Sveshnikov}(2022{\natexlab{a}})}]{Krasnov2022}%
  \BibitemOpen
  \bibfield  {author} {\bibinfo {author} {\bibfnamefont {A.}~\bibnamefont
  {Krasnov}}\ and\ \bibinfo {author} {\bibfnamefont {K.}~\bibnamefont
  {Sveshnikov}},\ }\href {\doibase arXiv:2201.04829 [physics.atom-ph]}
  {\bibfield  {journal} {\bibinfo  {journal} {arXiv.org}\ } (\bibinfo {year}
  {2022}{\natexlab{a}}),\ arXiv:2201.04829 [physics.atom-ph]}\BibitemShut
  {NoStop}%
\bibitem [{\citenamefont {Krasnov}\ and\ \citenamefont
  {Sveshnikov}(2022{\natexlab{b}})}]{Krasnov2022a}%
  \BibitemOpen
  \bibfield  {author} {\bibinfo {author} {\bibfnamefont {A.}~\bibnamefont
  {Krasnov}}\ and\ \bibinfo {author} {\bibfnamefont {K.}~\bibnamefont
  {Sveshnikov}},\ }\href {\doibase 10.1142/S021773232250136X} {\bibfield
  {journal} {\bibinfo  {journal} {Mod. Phys.Lett. A}\ }\textbf {\bibinfo
  {volume} {37}},\ \bibinfo {pages} {2250136} (\bibinfo {year}
  {2022}{\natexlab{b}})}\BibitemShut {NoStop}%
\bibitem [{\citenamefont {Greiner}\ \emph {et~al.}(1985)\citenamefont
  {Greiner}, \citenamefont {M\"uller},\ and\ \citenamefont
  {Rafelski}}]{Greiner1985a}%
  \BibitemOpen
  \bibfield  {author} {\bibinfo {author} {\bibfnamefont {W.}~\bibnamefont
  {Greiner}}, \bibinfo {author} {\bibfnamefont {B.}~\bibnamefont {M\"uller}}, \
  and\ \bibinfo {author} {\bibfnamefont {J.}~\bibnamefont {Rafelski}},\ }\href
  {http://link.springer.com/book/10.1007/978-3-642-82272-8} {\emph {\bibinfo
  {title} {Quantum Electrodynamics of Strong Fields}}},\ \bibinfo {edition}
  {2nd}\ ed.\ (\bibinfo  {publisher} {Springer},\ \bibinfo {address} {Berlin},\
  \bibinfo {year} {1985})\BibitemShut {NoStop}%
\bibitem [{\citenamefont {Plunien}\ \emph {et~al.}(1986)\citenamefont
  {Plunien}, \citenamefont {M\"uller},\ and\ \citenamefont
  {Greiner}}]{Plunien1986}%
  \BibitemOpen
  \bibfield  {author} {\bibinfo {author} {\bibfnamefont {G.}~\bibnamefont
  {Plunien}}, \bibinfo {author} {\bibfnamefont {B.}~\bibnamefont {M\"uller}}, \
  and\ \bibinfo {author} {\bibfnamefont {W.}~\bibnamefont {Greiner}},\ }\href
  {\doibase 10.1016/0370-1573(86)90020-7} {\bibfield  {journal} {\bibinfo
  {journal} {Phys. Rep.}\ }\textbf {\bibinfo {volume} {134}},\ \bibinfo {pages}
  {87 } (\bibinfo {year} {1986})}\BibitemShut {NoStop}%
\bibitem [{\citenamefont {Greiner}\ and\ \citenamefont
  {Reinhardt}(2009)}]{Greiner2012}%
  \BibitemOpen
  \bibfield  {author} {\bibinfo {author} {\bibfnamefont {W.}~\bibnamefont
  {Greiner}}\ and\ \bibinfo {author} {\bibfnamefont {J.}~\bibnamefont
  {Reinhardt}},\ }\href {\doibase /10.1007/978-3-540-87561-1} {\emph {\bibinfo
  {title} {Quantum Electrodynamics}}},\ \bibinfo {edition} {4th}\ ed.\
  (\bibinfo  {publisher} {Springer-Verlag Berlin Heidelberg},\ \bibinfo {year}
  {2009})\BibitemShut {NoStop}%
\bibitem [{\citenamefont {Ruffini}\ \emph {et~al.}(2010)\citenamefont
  {Ruffini}, \citenamefont {Vereshchagin},\ and\ \citenamefont
  {Xue}}]{Ruffini2010}%
  \BibitemOpen
  \bibfield  {author} {\bibinfo {author} {\bibfnamefont {R.}~\bibnamefont
  {Ruffini}}, \bibinfo {author} {\bibfnamefont {G.}~\bibnamefont
  {Vereshchagin}}, \ and\ \bibinfo {author} {\bibfnamefont {S.-S.}\
  \bibnamefont {Xue}},\ }\href {\doibase 10.1016/j.physrep.2009.10.004}
  {\bibfield  {journal} {\bibinfo  {journal} {Phys. Rep.}\ }\textbf {\bibinfo
  {volume} {487}},\ \bibinfo {pages} {1 } (\bibinfo {year} {2010})}\BibitemShut
  {NoStop}%
\bibitem [{\citenamefont {M{\"u}ller-Nehler}\ and\ \citenamefont
  {Soff}(1994)}]{Mueller1994}%
  \BibitemOpen
  \bibfield  {author} {\bibinfo {author} {\bibfnamefont {U.}~\bibnamefont
  {M{\"u}ller-Nehler}}\ and\ \bibinfo {author} {\bibfnamefont {G.}~\bibnamefont
  {Soff}},\ }\href {\doibase 10.1016/0370-1573(94)90068-X} {\bibfield
  {journal} {\bibinfo  {journal} {Phys.Rep.}\ }\textbf {\bibinfo {volume}
  {246}},\ \bibinfo {pages} {101} (\bibinfo {year} {1994})}\BibitemShut
  {NoStop}%
\bibitem [{\citenamefont {Grashin}\ and\ \citenamefont
  {Sveshnikov}(2023)}]{Grashin2023a}%
  \BibitemOpen
  \bibfield  {author} {\bibinfo {author} {\bibfnamefont {P.}~\bibnamefont
  {Grashin}}\ and\ \bibinfo {author} {\bibfnamefont {K.}~\bibnamefont
  {Sveshnikov}},\ }\href {\doibase 10.1142/S0217751X23501257} {\bibfield
  {journal} {\bibinfo  {journal} {International Journal of Modern Physics A}\
  }\textbf {\bibinfo {volume} {38}},\ \bibinfo {pages} {2350125} (\bibinfo
  {year} {2023})},\ \Eprint
  {http://arxiv.org/abs/https://doi.org/10.1142/S0217751X23501257}
  {https://doi.org/10.1142/S0217751X23501257} \BibitemShut {NoStop}%
\bibitem [{\citenamefont {Gumberidze}\ \emph {et~al.}(2009)\citenamefont
  {Gumberidze}, \citenamefont {St{\"o}hlker}, \citenamefont {Beyer},
  \citenamefont {Bosch}, \citenamefont {Bräuning-Demian}, \citenamefont
  {Hagmann}, \citenamefont {Kozhuharov}, \citenamefont {K\"uhl}, \citenamefont
  {Mann}, \citenamefont {Indelicato}, \citenamefont {Quint}, \citenamefont
  {Schuch},\ and\ \citenamefont {Warczak}}]{FAIR2009}%
  \BibitemOpen
  \bibfield  {author} {\bibinfo {author} {\bibfnamefont {A.}~\bibnamefont
  {Gumberidze}}, \bibinfo {author} {\bibfnamefont {T.}~\bibnamefont
  {St{\"o}hlker}}, \bibinfo {author} {\bibfnamefont {H.~F.}\ \bibnamefont
  {Beyer}}, \bibinfo {author} {\bibfnamefont {F.}~\bibnamefont {Bosch}},
  \bibinfo {author} {\bibfnamefont {A.}~\bibnamefont {Bräuning-Demian}},
  \bibinfo {author} {\bibfnamefont {S.}~\bibnamefont {Hagmann}}, \bibinfo
  {author} {\bibfnamefont {C.}~\bibnamefont {Kozhuharov}}, \bibinfo {author}
  {\bibfnamefont {T.}~\bibnamefont {K\"uhl}}, \bibinfo {author} {\bibfnamefont
  {R.}~\bibnamefont {Mann}}, \bibinfo {author} {\bibfnamefont {P.}~\bibnamefont
  {Indelicato}}, \bibinfo {author} {\bibfnamefont {W.}~\bibnamefont {Quint}},
  \bibinfo {author} {\bibfnamefont {R.}~\bibnamefont {Schuch}}, \ and\ \bibinfo
  {author} {\bibfnamefont {A.}~\bibnamefont {Warczak}},\ }\href {\doibase
  https://doi.org/10.1016/j.nimb.2008.10.079} {\bibfield  {journal} {\bibinfo
  {journal} {Nucl. Instr. {\&} Meth. in Phys. Research B}\ }\textbf {\bibinfo
  {volume} {267}},\ \bibinfo {pages} {248} (\bibinfo {year}
  {2009})}\BibitemShut {NoStop}%
\bibitem [{\citenamefont {Ter-Akopian}\ \emph {et~al.}(2015)\citenamefont
  {Ter-Akopian}, \citenamefont {Greiner}, \citenamefont {Meshkov},
  \citenamefont {Oganessian}, \citenamefont {Reinhardt},\ and\ \citenamefont
  {Trubnikov}}]{Ter2015}%
  \BibitemOpen
  \bibfield  {author} {\bibinfo {author} {\bibfnamefont {G.~M.}\ \bibnamefont
  {Ter-Akopian}}, \bibinfo {author} {\bibfnamefont {W.}~\bibnamefont
  {Greiner}}, \bibinfo {author} {\bibfnamefont {I.}~\bibnamefont {Meshkov}},
  \bibinfo {author} {\bibfnamefont {Y.}~\bibnamefont {Oganessian}}, \bibinfo
  {author} {\bibfnamefont {J.}~\bibnamefont {Reinhardt}}, \ and\ \bibinfo
  {author} {\bibfnamefont {G.}~\bibnamefont {Trubnikov}},\ }\href {\doibase
  10.1142/S0218301315500160} {\bibfield  {journal} {\bibinfo  {journal} {Int.
  J. Mod. Phys. E}\ }\textbf {\bibinfo {volume} {24}},\ \bibinfo {pages}
  {1550016} (\bibinfo {year} {2015})}\BibitemShut {NoStop}%
\bibitem [{\citenamefont {Ma}\ \emph {et~al.}(2017)\citenamefont {Ma},
  \citenamefont {Wen}, \citenamefont {Zhang}, \citenamefont {Yu}, \citenamefont
  {Cheng}, \citenamefont {Yang}, \citenamefont {Huang}, \citenamefont {Wang},
  \citenamefont {Zhu}, \citenamefont {Cai}, \citenamefont {Zhao}, \citenamefont
  {Mao}, \citenamefont {Yang}, \citenamefont {Zhou}, \citenamefont {Xu},
  \citenamefont {Yuan}, \citenamefont {Xia}, \citenamefont {Zhao},
  \citenamefont {Xiao},\ and\ \citenamefont {Zhan}}]{MA2017169}%
  \BibitemOpen
  \bibfield  {author} {\bibinfo {author} {\bibfnamefont {X.}~\bibnamefont
  {Ma}}, \bibinfo {author} {\bibfnamefont {W.}~\bibnamefont {Wen}}, \bibinfo
  {author} {\bibfnamefont {S.}~\bibnamefont {Zhang}}, \bibinfo {author}
  {\bibfnamefont {D.}~\bibnamefont {Yu}}, \bibinfo {author} {\bibfnamefont
  {R.}~\bibnamefont {Cheng}}, \bibinfo {author} {\bibfnamefont
  {J.}~\bibnamefont {Yang}}, \bibinfo {author} {\bibfnamefont {Z.}~\bibnamefont
  {Huang}}, \bibinfo {author} {\bibfnamefont {H.}~\bibnamefont {Wang}},
  \bibinfo {author} {\bibfnamefont {X.}~\bibnamefont {Zhu}}, \bibinfo {author}
  {\bibfnamefont {X.}~\bibnamefont {Cai}}, \bibinfo {author} {\bibfnamefont
  {Y.}~\bibnamefont {Zhao}}, \bibinfo {author} {\bibfnamefont {L.}~\bibnamefont
  {Mao}}, \bibinfo {author} {\bibfnamefont {J.}~\bibnamefont {Yang}}, \bibinfo
  {author} {\bibfnamefont {X.}~\bibnamefont {Zhou}}, \bibinfo {author}
  {\bibfnamefont {H.}~\bibnamefont {Xu}}, \bibinfo {author} {\bibfnamefont
  {Y.}~\bibnamefont {Yuan}}, \bibinfo {author} {\bibfnamefont {J.}~\bibnamefont
  {Xia}}, \bibinfo {author} {\bibfnamefont {H.}~\bibnamefont {Zhao}}, \bibinfo
  {author} {\bibfnamefont {G.}~\bibnamefont {Xiao}}, \ and\ \bibinfo {author}
  {\bibfnamefont {W.}~\bibnamefont {Zhan}},\ }\href {\doibase
  https://doi.org/10.1016/j.nimb.2017.03.129} {\bibfield  {journal} {\bibinfo
  {journal} {Nucl. Instr. {\&} Meth. in Phys. Research B}\ }\textbf {\bibinfo
  {volume} {408}},\ \bibinfo {pages} {169} (\bibinfo {year}
  {2017})}\BibitemShut {NoStop}%
\bibitem [{\citenamefont {Grashin}\ and\ \citenamefont
  {Sveshnikov}(2020{\natexlab{a}})}]{Grashin2020a}%
  \BibitemOpen
  \bibfield  {author} {\bibinfo {author} {\bibfnamefont {P.}~\bibnamefont
  {Grashin}}\ and\ \bibinfo {author} {\bibfnamefont {K.}~\bibnamefont
  {Sveshnikov}},\ }\href {\doibase 10.1016/j.aop.2020.168094} {\bibfield
  {journal} {\bibinfo  {journal} {Ann. of Phys.}\ }\textbf {\bibinfo {volume}
  {532}},\ \bibinfo {pages} {168094} (\bibinfo {year}
  {2020}{\natexlab{a}})}\BibitemShut {NoStop}%
\bibitem [{\citenamefont {Grashin}\ and\ \citenamefont
  {Sveshnikov}(2020{\natexlab{b}})}]{Grashin2020b}%
  \BibitemOpen
  \bibfield  {author} {\bibinfo {author} {\bibfnamefont {P.}~\bibnamefont
  {Grashin}}\ and\ \bibinfo {author} {\bibfnamefont {K.}~\bibnamefont
  {Sveshnikov}},\ }\href {\doibase 10.1002/andp.201900351} {\bibfield
  {journal} {\bibinfo  {journal} {Ann. der Physik}\ }\textbf {\bibinfo {volume}
  {415}},\ \bibinfo {pages} {351} (\bibinfo {year}
  {2020}{\natexlab{b}})}\BibitemShut {NoStop}%
\bibitem [{\citenamefont {Voronina}\ \emph
  {et~al.}(2019{\natexlab{a}})\citenamefont {Voronina}, \citenamefont
  {Komissarov},\ and\ \citenamefont {Sveshnikov}}]{Voronina2019c}%
  \BibitemOpen
  \bibfield  {author} {\bibinfo {author} {\bibfnamefont {Y.}~\bibnamefont
  {Voronina}}, \bibinfo {author} {\bibfnamefont {I.}~\bibnamefont
  {Komissarov}}, \ and\ \bibinfo {author} {\bibfnamefont {K.}~\bibnamefont
  {Sveshnikov}},\ }\href {\doibase https://doi.org/10.1016/j.aop.2019.02.014}
  {\bibfield  {journal} {\bibinfo  {journal} {Ann. Phys.}\ }\textbf {\bibinfo
  {volume} {404}},\ \bibinfo {pages} {132 } (\bibinfo {year}
  {2019}{\natexlab{a}})}\BibitemShut {NoStop}%
\bibitem [{\citenamefont {Voronina}\ \emph
  {et~al.}(2019{\natexlab{b}})\citenamefont {Voronina}, \citenamefont
  {Komissarov},\ and\ \citenamefont {Sveshnikov}}]{Voronina2019d}%
  \BibitemOpen
  \bibfield  {author} {\bibinfo {author} {\bibfnamefont {Y.}~\bibnamefont
  {Voronina}}, \bibinfo {author} {\bibfnamefont {I.}~\bibnamefont
  {Komissarov}}, \ and\ \bibinfo {author} {\bibfnamefont {K.}~\bibnamefont
  {Sveshnikov}},\ }\href {\doibase 10.1103/PhysRevA.99.062504} {\bibfield
  {journal} {\bibinfo  {journal} {Phys. Rev. A}\ }\textbf {\bibinfo {volume}
  {99}},\ \bibinfo {pages} {062504} (\bibinfo {year}
  {2019}{\natexlab{b}})}\BibitemShut {NoStop}%
\bibitem [{\citenamefont {Wichmann}\ and\ \citenamefont
  {Kroll}(1956)}]{Wichmann1956}%
  \BibitemOpen
  \bibfield  {author} {\bibinfo {author} {\bibfnamefont {E.~H.}\ \bibnamefont
  {Wichmann}}\ and\ \bibinfo {author} {\bibfnamefont {N.~M.}\ \bibnamefont
  {Kroll}},\ }\href {\doibase 10.1103/PhysRev.101.843} {\bibfield  {journal}
  {\bibinfo  {journal} {Phys. Rev.}\ }\textbf {\bibinfo {volume} {101}},\
  \bibinfo {pages} {843} (\bibinfo {year} {1956})}\BibitemShut {NoStop}%
\bibitem [{\citenamefont {Gyulassy}(1975)}]{Gyulassy1975}%
  \BibitemOpen
  \bibfield  {author} {\bibinfo {author} {\bibfnamefont {M.}~\bibnamefont
  {Gyulassy}},\ }\href {\doibase 10.1016/0375-9474(75)90554-0} {\bibfield
  {journal} {\bibinfo  {journal} {Nucl. Phys. A}\ }\textbf {\bibinfo {volume}
  {244}},\ \bibinfo {pages} {497 } (\bibinfo {year} {1975})}\BibitemShut
  {NoStop}%
\bibitem [{\citenamefont {Brown}\ \emph
  {et~al.}(1975{\natexlab{a}})\citenamefont {Brown}, \citenamefont {Cahn},\
  and\ \citenamefont {McLerran}}]{McLerran1975a}%
  \BibitemOpen
  \bibfield  {author} {\bibinfo {author} {\bibfnamefont {L.}~\bibnamefont
  {Brown}}, \bibinfo {author} {\bibfnamefont {R.}~\bibnamefont {Cahn}}, \ and\
  \bibinfo {author} {\bibfnamefont {L.}~\bibnamefont {McLerran}},\ }\href
  {\doibase 10.1103/PhysRevD.12.581} {\bibfield  {journal} {\bibinfo  {journal}
  {Phys. Rev. D}\ }\textbf {\bibinfo {volume} {12}},\ \bibinfo {pages} {581}
  (\bibinfo {year} {1975}{\natexlab{a}})}\BibitemShut {NoStop}%
\bibitem [{\citenamefont {Brown}\ \emph
  {et~al.}(1975{\natexlab{b}})\citenamefont {Brown}, \citenamefont {Cahn},\
  and\ \citenamefont {McLerran}}]{McLerran1975b}%
  \BibitemOpen
  \bibfield  {author} {\bibinfo {author} {\bibfnamefont {L.}~\bibnamefont
  {Brown}}, \bibinfo {author} {\bibfnamefont {R.}~\bibnamefont {Cahn}}, \ and\
  \bibinfo {author} {\bibfnamefont {L.}~\bibnamefont {McLerran}},\ }\href
  {\doibase 10.1103/PhysRevD.12.596} {\bibfield  {journal} {\bibinfo  {journal}
  {Phys. Rev. D}\ }\textbf {\bibinfo {volume} {12}},\ \bibinfo {pages} {596}
  (\bibinfo {year} {1975}{\natexlab{b}})}\BibitemShut {NoStop}%
\bibitem [{\citenamefont {Brown}\ \emph
  {et~al.}(1975{\natexlab{c}})\citenamefont {Brown}, \citenamefont {Cahn},\
  and\ \citenamefont {McLerran}}]{McLerran1975c}%
  \BibitemOpen
  \bibfield  {author} {\bibinfo {author} {\bibfnamefont {L.}~\bibnamefont
  {Brown}}, \bibinfo {author} {\bibfnamefont {R.}~\bibnamefont {Cahn}}, \ and\
  \bibinfo {author} {\bibfnamefont {L.}~\bibnamefont {McLerran}},\ }\href
  {\doibase 10.1103/PhysRevD.12.609} {\bibfield  {journal} {\bibinfo  {journal}
  {Phys. Rev. D}\ }\textbf {\bibinfo {volume} {12}},\ \bibinfo {pages} {609}
  (\bibinfo {year} {1975}{\natexlab{c}})}\BibitemShut {NoStop}%
\bibitem [{\citenamefont {Mohr}\ \emph {et~al.}(1998)\citenamefont {Mohr},
  \citenamefont {Plunien},\ and\ \citenamefont {Soff}}]{Mohr1998}%
  \BibitemOpen
  \bibfield  {author} {\bibinfo {author} {\bibfnamefont {P.~J.}\ \bibnamefont
  {Mohr}}, \bibinfo {author} {\bibfnamefont {G.}~\bibnamefont {Plunien}}, \
  and\ \bibinfo {author} {\bibfnamefont {G.}~\bibnamefont {Soff}},\ }\href
  {\doibase 10.1016/S0370-1573(97)00046-X} {\bibfield  {journal} {\bibinfo
  {journal} {Phys. Rep.}\ }\textbf {\bibinfo {volume} {293}},\ \bibinfo {pages}
  {227 } (\bibinfo {year} {1998})}\BibitemShut {NoStop}%
\bibitem [{\citenamefont {Bjorken}\ and\ \citenamefont
  {Drell}(1965)}]{Bjorken1965}%
  \BibitemOpen
  \bibfield  {author} {\bibinfo {author} {\bibfnamefont {J.~D.}\ \bibnamefont
  {Bjorken}}\ and\ \bibinfo {author} {\bibfnamefont {S.~D.}\ \bibnamefont
  {Drell}},\ }\href@noop {} {\emph {\bibinfo {title} {Relativistic Quantum
  Fields}}}\ (\bibinfo  {publisher} {McGraw-Hill Book Co., New York, NY},\
  \bibinfo {year} {1965})\BibitemShut {NoStop}%
\bibitem [{\citenamefont {Jentschura}\ and\ \citenamefont
  {Adkins}(2022)}]{Jentschura2022}%
  \BibitemOpen
  \bibfield  {author} {\bibinfo {author} {\bibfnamefont {U.~D.}\ \bibnamefont
  {Jentschura}}\ and\ \bibinfo {author} {\bibfnamefont {G.~S.}\ \bibnamefont
  {Adkins}},\ }\href {\doibase doi.org/10.1142/12722} {\emph {\bibinfo {title}
  {Quantum Electrodynamics: Atoms, Lasers and Gravity}}}\ (\bibinfo
  {publisher} {World Scientific},\ \bibinfo {year} {2022})\ pp.\ \bibinfo
  {pages} {1--789}\BibitemShut {NoStop}%
\bibitem [{\citenamefont {Davydov}\ \emph
  {et~al.}(2018{\natexlab{a}})\citenamefont {Davydov}, \citenamefont
  {Sveshnikov},\ and\ \citenamefont {Voronina}}]{Davydov2018a}%
  \BibitemOpen
  \bibfield  {author} {\bibinfo {author} {\bibfnamefont {A.}~\bibnamefont
  {Davydov}}, \bibinfo {author} {\bibfnamefont {K.}~\bibnamefont {Sveshnikov}},
  \ and\ \bibinfo {author} {\bibfnamefont {Y.}~\bibnamefont {Voronina}},\
  }\href {\doibase 10.1142/S0217751X18500045} {\bibfield  {journal} {\bibinfo
  {journal} {Int. J. Mod. Phys. A}\ }\textbf {\bibinfo {volume} {33}},\
  \bibinfo {pages} {1850004} (\bibinfo {year}
  {2018}{\natexlab{a}})}\BibitemShut {NoStop}%
\bibitem [{\citenamefont {Davydov}\ \emph
  {et~al.}(2018{\natexlab{b}})\citenamefont {Davydov}, \citenamefont
  {Sveshnikov},\ and\ \citenamefont {Voronina}}]{Davydov2018b}%
  \BibitemOpen
  \bibfield  {author} {\bibinfo {author} {\bibfnamefont {A.}~\bibnamefont
  {Davydov}}, \bibinfo {author} {\bibfnamefont {K.}~\bibnamefont {Sveshnikov}},
  \ and\ \bibinfo {author} {\bibfnamefont {Y.}~\bibnamefont {Voronina}},\
  }\href {\doibase 10.1142/S0217751X18500057} {\bibfield  {journal} {\bibinfo
  {journal} {Int. J. Mod. Phys. A}\ }\textbf {\bibinfo {volume} {33}},\
  \bibinfo {pages} {1850005} (\bibinfo {year}
  {2018}{\natexlab{b}})}\BibitemShut {NoStop}%
\bibitem [{\citenamefont {Sveshnikov}\ \emph
  {et~al.}(2019{\natexlab{a}})\citenamefont {Sveshnikov}, \citenamefont
  {Voronina}, \citenamefont {Davydov},\ and\ \citenamefont
  {Grashin}}]{Sveshnikov2019a}%
  \BibitemOpen
  \bibfield  {author} {\bibinfo {author} {\bibfnamefont {K.}~\bibnamefont
  {Sveshnikov}}, \bibinfo {author} {\bibfnamefont {Y.}~\bibnamefont
  {Voronina}}, \bibinfo {author} {\bibfnamefont {A.}~\bibnamefont {Davydov}}, \
  and\ \bibinfo {author} {\bibfnamefont {P.}~\bibnamefont {Grashin}},\ }\href
  {\doibase doi.org/10.1134/S0040577919030024} {\bibfield  {journal} {\bibinfo
  {journal} {Theor. Math. Phys.}\ }\textbf {\bibinfo {volume} {198}},\ \bibinfo
  {pages} {331} (\bibinfo {year} {2019}{\natexlab{a}})}\BibitemShut {NoStop}%
\bibitem [{\citenamefont {Sveshnikov}\ \emph
  {et~al.}(2019{\natexlab{b}})\citenamefont {Sveshnikov}, \citenamefont
  {Voronina}, \citenamefont {Davydov},\ and\ \citenamefont
  {Grashin}}]{Sveshnikov2019b}%
  \BibitemOpen
  \bibfield  {author} {\bibinfo {author} {\bibfnamefont {K.}~\bibnamefont
  {Sveshnikov}}, \bibinfo {author} {\bibfnamefont {Y.}~\bibnamefont
  {Voronina}}, \bibinfo {author} {\bibfnamefont {A.}~\bibnamefont {Davydov}}, \
  and\ \bibinfo {author} {\bibfnamefont {P.}~\bibnamefont {Grashin}},\ }\href
  {\doibase doi.org/10.1134/S0040577919040056} {\bibfield  {journal} {\bibinfo
  {journal} {Theor. Math. Phys.}\ }\textbf {\bibinfo {volume} {199}},\ \bibinfo
  {pages} {533} (\bibinfo {year} {2019}{\natexlab{b}})}\BibitemShut {NoStop}%
\bibitem [{\citenamefont {Voronina}\ \emph
  {et~al.}(2019{\natexlab{c}})\citenamefont {Voronina}, \citenamefont
  {Sveshnikov}, \citenamefont {Grashin},\ and\ \citenamefont
  {Davydov}}]{Voronina2019a}%
  \BibitemOpen
  \bibfield  {author} {\bibinfo {author} {\bibfnamefont {Y.}~\bibnamefont
  {Voronina}}, \bibinfo {author} {\bibfnamefont {K.}~\bibnamefont
  {Sveshnikov}}, \bibinfo {author} {\bibfnamefont {P.}~\bibnamefont {Grashin}},
  \ and\ \bibinfo {author} {\bibfnamefont {A.}~\bibnamefont {Davydov}},\ }\href
  {\doibase https://doi.org/10.1016/j.physe.2018.08.013} {\bibfield  {journal}
  {\bibinfo  {journal} {Physica E}\ }\textbf {\bibinfo {volume} {106}},\
  \bibinfo {pages} {298 } (\bibinfo {year} {2019}{\natexlab{c}})}\BibitemShut
  {NoStop}%
\bibitem [{\citenamefont {Voronina}\ \emph
  {et~al.}(2019{\natexlab{d}})\citenamefont {Voronina}, \citenamefont
  {Sveshnikov}, \citenamefont {Grashin},\ and\ \citenamefont
  {Davydov}}]{Voronina2019b}%
  \BibitemOpen
  \bibfield  {author} {\bibinfo {author} {\bibfnamefont {Y.}~\bibnamefont
  {Voronina}}, \bibinfo {author} {\bibfnamefont {K.}~\bibnamefont
  {Sveshnikov}}, \bibinfo {author} {\bibfnamefont {P.}~\bibnamefont {Grashin}},
  \ and\ \bibinfo {author} {\bibfnamefont {A.}~\bibnamefont {Davydov}},\ }\href
  {\doibase https://doi.org/10.1016/j.physe.2018.09.026} {\bibfield  {journal}
  {\bibinfo  {journal} {Physica E}\ }\textbf {\bibinfo {volume} {109}},\
  \bibinfo {pages} {209 } (\bibinfo {year} {2019}{\natexlab{d}})}\BibitemShut
  {NoStop}%
\bibitem [{\citenamefont {Berestetskii}\ \emph {et~al.}(2012)\citenamefont
  {Berestetskii}, \citenamefont {Pitaevskii},\ and\ \citenamefont
  {Lifshitz}}]{landau2012qed}%
  \BibitemOpen
  \bibfield  {author} {\bibinfo {author} {\bibfnamefont {V.}~\bibnamefont
  {Berestetskii}}, \bibinfo {author} {\bibfnamefont {L.}~\bibnamefont
  {Pitaevskii}}, \ and\ \bibinfo {author} {\bibfnamefont {E.}~\bibnamefont
  {Lifshitz}},\ }\href@noop {} {\emph {\bibinfo {title} {Quantum
  Electrodynamics}}},\ Vol.~\bibinfo {volume} {4}\ (\bibinfo  {publisher}
  {Elsevier},\ \bibinfo {year} {2012})\BibitemShut {NoStop}%
\bibitem [{\citenamefont {Krasnov}\ and\ \citenamefont
  {Sveshnikov}(2022{\natexlab{c}})}]{Krasnov2022b}%
  \BibitemOpen
  \bibfield  {author} {\bibinfo {author} {\bibfnamefont {A.}~\bibnamefont
  {Krasnov}}\ and\ \bibinfo {author} {\bibfnamefont {K.}~\bibnamefont
  {Sveshnikov}},\ }\href {\doibase arXiv:2209.00125 [hep-ph]} {\bibfield
  {journal} {\bibinfo  {journal} {arXiv.org}\ } (\bibinfo {year}
  {2022}{\natexlab{c}}),\ arXiv:2209.00125 [hep-ph]}\BibitemShut {NoStop}%
\bibitem [{\citenamefont {Hosaka}\ and\ \citenamefont
  {Toki}(2001)}]{Hosaka2001}%
  \BibitemOpen
  \bibfield  {author} {\bibinfo {author} {\bibfnamefont {A.}~\bibnamefont
  {Hosaka}}\ and\ \bibinfo {author} {\bibfnamefont {H.}~\bibnamefont {Toki}},\
  }\href@noop {} {\emph {\bibinfo {title} {Quarks, Baryons and Chiral
  Symmetry}}}\ (\bibinfo  {publisher} {World Scientific},\ \bibinfo {year}
  {2001})\BibitemShut {NoStop}%
\bibitem [{\citenamefont {Weigel}(2007)}]{Weigel2007}%
  \BibitemOpen
  \bibfield  {author} {\bibinfo {author} {\bibfnamefont {H.}~\bibnamefont
  {Weigel}},\ }\href@noop {} {\emph {\bibinfo {title} {Chiral Soliton Models
  for Baryons}}},\ Vol.\ \bibinfo {volume} {743 of Lecture Notes in Physics}\
  (\bibinfo  {publisher} {Springer},\ \bibinfo {year} {2007})\BibitemShut
  {NoStop}%
\bibitem [{\citenamefont {Hylleraas}(2000)}]{Hylleraas2000}%
  \BibitemOpen
  \bibfield  {author} {\bibinfo {author} {\bibfnamefont {E.~A.}\ \bibnamefont
  {Hylleraas}},\ }in\ \href {\doibase 10.1142/9789812795762_0008} {\emph
  {\bibinfo {booktitle} {Quantum Chemistry}}},\ \bibinfo {series and number}
  {World Scientific Series in 20th Century Chemistry}\ (\bibinfo  {publisher}
  {World Scientific},\ \bibinfo {year} {2000})\ pp.\ \bibinfo {pages} {124 --
  139}\BibitemShut {NoStop}%
\bibitem [{\citenamefont {Bethe}\ and\ \citenamefont
  {Salpeter}(1977)}]{Bethe1977}%
  \BibitemOpen
  \bibfield  {author} {\bibinfo {author} {\bibfnamefont {H.~A.}\ \bibnamefont
  {Bethe}}\ and\ \bibinfo {author} {\bibfnamefont {E.~E.}\ \bibnamefont
  {Salpeter}},\ }\href {\doibase 10.1007/978-1-4613-4104-8} {\emph {\bibinfo
  {title} {Quantum Mechanics of One- and Two-Electron Atoms}}}\ (\bibinfo
  {publisher} {Springer New York, NY},\ \bibinfo {year} {1977})\ pp.\ \bibinfo
  {pages} {XII, 370}\BibitemShut {NoStop}%
\bibitem [{\citenamefont {Sch\"utte}\ and\ \citenamefont {van~der
  Waerden}(1951)}]{vanderWaerden1951}%
  \BibitemOpen
  \bibfield  {author} {\bibinfo {author} {\bibfnamefont {K.}~\bibnamefont
  {Sch\"utte}}\ and\ \bibinfo {author} {\bibfnamefont {B.~L.}\ \bibnamefont
  {van~der Waerden}},\ }\href {\doibase https://doi.org/10.1007/BF02054944}
  {\bibfield  {journal} {\bibinfo  {journal} {Mathematische Annalen}\ }\textbf
  {\bibinfo {volume} {123}},\ \bibinfo {pages} {96 } (\bibinfo {year}
  {1951})}\BibitemShut {NoStop}%
\bibitem [{\citenamefont {Toth}(2013)}]{Toth2013}%
  \BibitemOpen
  \bibfield  {author} {\bibinfo {author} {\bibfnamefont {L.~F.}\ \bibnamefont
  {Toth}},\ }\href
  {https://books.google.ru/books?id=JQ-nBgAAQBAJ&dq=fejes+toth+lagerungen+auf+der+ebene&lr=&source=gbs_navlinks_s}
  {\emph {\bibinfo {title} {Lagerungen in der Ebene auf der Kugel und im
  Raum}}},\ \bibinfo {series} {Grundlehren der mathematischen Wissenschaften},
  Vol.~\bibinfo {volume} {65}\ (\bibinfo  {publisher} {Springer-Verlag},\
  \bibinfo {year} {2013})\ p.\ \bibinfo {pages} {240}\BibitemShut {NoStop}%
\bibitem [{\citenamefont {Melnyk}\ \emph {et~al.}(1977)\citenamefont {Melnyk},
  \citenamefont {Knop},\ and\ \citenamefont {Smith}}]{melnykextr1977}%
  \BibitemOpen
  \bibfield  {author} {\bibinfo {author} {\bibfnamefont {T.}~\bibnamefont
  {Melnyk}}, \bibinfo {author} {\bibfnamefont {O.}~\bibnamefont {Knop}}, \ and\
  \bibinfo {author} {\bibfnamefont {W.}~\bibnamefont {Smith}},\ }\href
  {\doibase 10.1139/v77-246} {\bibfield  {journal} {\bibinfo  {journal}
  {Canadian Journal of Chemistry}\ }\textbf {\bibinfo {volume} {55}},\ \bibinfo
  {pages} {1745} (\bibinfo {year} {1977})}\BibitemShut {NoStop}%
\bibitem [{\citenamefont {Sidi}(2003)}]{Sidi2023}%
  \BibitemOpen
  \bibfield  {author} {\bibinfo {author} {\bibfnamefont {A.}~\bibnamefont
  {Sidi}},\ }\href {\doibase https://doi.org/10.1017/CBO9780511546815} {\emph
  {\bibinfo {title} {Practical Extrapolation Methods}}},\ Mathematics,
  Computational Science, Numerical Analysis and Computational Science\
  (\bibinfo  {publisher} {Cambridge University Press},\ \bibinfo {year}
  {2003})\BibitemShut {NoStop}%
\bibitem [{\citenamefont {Katoot}\ and\ \citenamefont
  {Oberacker}(1989)}]{katoot1989microscopic}%
  \BibitemOpen
  \bibfield  {author} {\bibinfo {author} {\bibfnamefont {M.~W.}\ \bibnamefont
  {Katoot}}\ and\ \bibinfo {author} {\bibfnamefont {V.}~\bibnamefont
  {Oberacker}},\ }\href@noop {} {\bibfield  {journal} {\bibinfo  {journal}
  {Journal of Physics G: Nuclear and Particle Physics}\ }\textbf {\bibinfo
  {volume} {15}},\ \bibinfo {pages} {333} (\bibinfo {year} {1989})}\BibitemShut
  {NoStop}%
\bibitem [{\citenamefont {Zagrebaev}\ and\ \citenamefont
  {Greiner}(2006)}]{zagrebaev2006low}%
  \BibitemOpen
  \bibfield  {author} {\bibinfo {author} {\bibfnamefont {V.}~\bibnamefont
  {Zagrebaev}}\ and\ \bibinfo {author} {\bibfnamefont {W.}~\bibnamefont
  {Greiner}},\ }\href@noop {} {\bibfield  {journal} {\bibinfo  {journal}
  {Journal of Physics G: Nuclear and Particle Physics}\ }\textbf {\bibinfo
  {volume} {34}},\ \bibinfo {pages} {1} (\bibinfo {year} {2006})}\BibitemShut
  {NoStop}%
\bibitem [{\citenamefont {Reid~Jr}(1968)}]{reid1968local}%
  \BibitemOpen
  \bibfield  {author} {\bibinfo {author} {\bibfnamefont {R.~V.}\ \bibnamefont
  {Reid~Jr}},\ }\href@noop {} {\bibfield  {journal} {\bibinfo  {journal}
  {Annals of Physics}\ }\textbf {\bibinfo {volume} {50}},\ \bibinfo {pages}
  {411} (\bibinfo {year} {1968})}\BibitemShut {NoStop}%
\bibitem [{\citenamefont {Blocki}\ \emph {et~al.}(1977)\citenamefont {Blocki},
  \citenamefont {Randrup}, \citenamefont {Swiatecki},\ and\ \citenamefont
  {Tsang}}]{Blocki1977}%
  \BibitemOpen
  \bibfield  {author} {\bibinfo {author} {\bibfnamefont {J.}~\bibnamefont
  {Blocki}}, \bibinfo {author} {\bibfnamefont {J.}~\bibnamefont {Randrup}},
  \bibinfo {author} {\bibfnamefont {W.~J.}\ \bibnamefont {Swiatecki}}, \ and\
  \bibinfo {author} {\bibfnamefont {C.~F.}\ \bibnamefont {Tsang}},\ }\href
  {\doibase https://doi.org/10.1016/0003-4916(77)90249-4} {\bibfield  {journal}
  {\bibinfo  {journal} {Ann. of Phys.}\ }\textbf {\bibinfo {volume} {105}},\
  \bibinfo {pages} {427} (\bibinfo {year} {1977})}\BibitemShut {NoStop}%
\bibitem [{\citenamefont {Brown}\ and\ \citenamefont
  {Jackson}(1976)}]{Brown1976}%
  \BibitemOpen
  \bibfield  {author} {\bibinfo {author} {\bibfnamefont {G.~E.}\ \bibnamefont
  {Brown}}\ and\ \bibinfo {author} {\bibfnamefont {A.~D.}\ \bibnamefont
  {Jackson}},\ }\href
  {https://inis.iaea.org/search/search.aspx?orig_q=RN:7264729} {\emph {\bibinfo
  {title} {The nucleon-nucleon interactions}}}\ (\bibinfo  {publisher}
  {North-Holland Publishing Company},\ \bibinfo {year} {1976})\ pp.\ \bibinfo
  {pages} {1--250}\BibitemShut {NoStop}%
\end{thebibliography}%

\end{document}